%% file: cbc_main.tex
\documentclass[fleqn,usenatbib]{mnras}

\usepackage{newtxtext,newtxmath}
\usepackage[T1]{fontenc}
\usepackage{ae,aecompl}

\usepackage{graphicx}	% Including figure files
\usepackage{amsmath}	% Advanced maths commands
\usepackage{amssymb}	% Extra maths symbols
\usepackage{csvsimple,longtable,booktabs,dcolumn} % include csv into tables
\usepackage{url}
\usepackage{array}
\usepackage{xspace}
\usepackage{multirow}

\newcommand{\afe}{[$\alpha$/Fe]}
\newcommand{\teff}{$T_{\rm eff}$}
\newcommand{\feh}{[Fe/H]}
\newcommand{\logg}{$\log g$}
\newcommand{\thet}{$\theta$}
\newcommand{\mychi}{$\chi^2$}
\newcommand{\mrd}{$\tilde\Delta$}
\newcommand{\deltacolor}{$\Delta {\rm \sf colour}$}
\newcommand{\deltaidx}{$\Delta {\rm \sf idx}$}
\newcommand{\adev}{{\rm \sf adev}}

\newcommand{\syncomil} {\textsc{SynCoMiL}\xspace}

\newcommand{\galaxev}  {\textsc{galaxev}\xspace}
\newcommand{\miles}    {{MILES}\xspace}
\newcommand{\pmiles}   {\textsc{sps-m}\xspace}
\newcommand{\smiles}   {\textsc{sps-s}\xspace}
\newcommand{\cmiles}   {\textsc{sps-c}\xspace}
\newcommand{\rmiles}   {\textsc{sps-r}\xspace}

\DeclareUrlCommand\UScore{\urlstyle{rm}}
\newcommand{\expUScore}{%
  \expandafter\expandafter\expandafter
  \UScore
  \expandafter\expandafter\expandafter
}

\defcitealias{coelho14}{C14}

\usepackage{comment}
\includecomment{comment}
\specialcomment{notes}{\begingroup\sffamily\color{magenta}}{\endgroup}

\title[Theoretical vs. semi-empirical SPS models]{To use or not to use synthetic stellar spectra in population synthesis models?}
\author[P. Coelho et al.]{
Paula R. T. Coelho,$^{1}$\thanks{E-mail: pcoelho@usp.br (PC)}
Gustavo Bruzual,$^{2}$
St\'ephane Charlot$^{3}$
\\
$^{1}$Universidade de S\~ao Paulo, Instituto de Astronomia, Geof\'isica e Ci\^encias Atmosf\'ericas, Rua do Mat\~ao 1226, 05508-090, S\~ao Paulo, Brazil\\
$^{2}$Instituto de Radioastronom{\'i}a y Astrof{\'i}sica, UNAM, Campus Morelia, Michoacan, M{\'e}xico, C.P. 58089, M{\'e}xico\\
$^{3}$Sorbonne Universit\'e, CNRS, UMR7095, Institut d'Astrophysique de Paris, F-75014, Paris, France
}

\date{Accepted XXX. Received YYY; in original form ZZZ}

\pubyear{2018}

\begin{document}   
\label{firstpage}
\pagerange{\pageref{firstpage}--\pageref{lastpage}}
\maketitle

\begin{abstract}
%Abstract: authors must provide an abstract (except for Errata, which do not have abstracts), normally of not more than 250 words for Main Journal
%papers or 200 words for Letters. The abstract should be presented as a single paragraph and briefly summarize the goals, methods, and new results presented in the paper.
Stellar population synthesis (SPS) models are invaluable to study star clusters and galaxies. They provide means to extract stellar masses, stellar ages, star formation histories,
chemical enrichment and dust content of galaxies from their integrated spectral energy distributions, colours or spectra. As most models, they contain
uncertainties which can hamper our ability to model and interpret observed spectra. This work aims at studying a specific source of model uncertainty:
the choice of an empirical vs. a synthetic stellar spectral library. Empirical libraries suffer from limited coverage of parameter
space, while synthetic libraries suffer from modelling inaccuracies. Given our current inability to have both ideal stellar-parameter coverage with ideal
stellar spectra, what should one favour: better coverage of the parameters (synthetic library) or better spectra on a star-by-star basis (empirical library)?
To study this question, we build a synthetic stellar library mimicking the coverage of an empirical library, and SPS models with different choices of 
stellar library tailored to these investigations.
Through the comparison of model predictions
and the spectral fitting of a sample of nearby galaxies, we learned that: predicted colours are more affected by the coverage effect than the choice of a
synthetic vs. empirical library; the effects on predicted spectral indices are multiple and defy simple conclusions; derived galaxy ages are virtually
unaffected by the choice of the library, but are underestimated when SPS models with limited parameter coverage are used; metallicities are robust against limited HRD coverage, but
are underestimated when using synthetic libraries. 
\end{abstract}


%%%%%%%%%%%%%%%%% BODY OF PAPER %%%%%%%%%%%%%%%%%%

\input{introduction}
\input{syncomil}

\input{methodology}

\input{results}

\input{discussion}
\input{conclusions}

\input{acknowledgements}

%%%%%%%%%%%%%%%%% BIBLIOGRAPHY %%%%%%%%%%%%%%%%%%

\bibliographystyle{mnras}
\bibliography{cbc_references}

%%%%%%%%%%%%%%%%% APPENDIX %%%%%%%%%%%%%%%%%%

%\input{appendix_fig}
\input{online_material}

% Don't change these lines
\bsp    % typesetting comment
\label{lastpage}
\end{document}

%% file: introduction.tex
\section{Introduction}
\label{sec_intro}

Stellar population synthesis (SPS) models are among the most powerful tools developed over the last few decades in astrophysics. Originally based on the seminal work by \citet[][see also \citealt{tinsley78}] {tinsley_gunn76}, they are today the means through which we are able to model (and thus, interpret) the integrated properties of galaxies and unresolved star clusters.
In the current era of large spectroscopic galaxy surveys, SPS models play a key role in inferring massive amounts of information about the origin and evolution of galaxies 
 \citep[see, e.g.,][for examples of the use of SPS models in the interpretation of large samples of galaxies]{kauffmann+03,tremonti+04,gallazzi+05,dacunha+08,cid+09,conroy10,sodre+13,cc16}.%\textcolor{red}{[I think here we could briefly cite work that, employing SPS models, made huge impact in our understanding in galaxy evolution. can you help me?]} 

SPS models can be built in a variety of ways, but the so-called \emph{Evolutionary Stellar Population Synthesis} models
are among the ones with more predictive and interpretative power, and have been used very frequently in the literature \citep[e.g.][]{bruzual83, arimoto86,guiderdoni87,CB91,bc93, BC03, bressan94,fritze94, worthey94, vazdekis96,vazdekis+15, fioc97, fioc19, kodama97, maraston98, maraston05, coelho+07, leitherer+99, conroy10, conroy_dokkum12}.  
%\footnote{These five papers together have been cited over 7700 times}. %\textcolor{red}{[do you think i can truly claim that?]}. 
Such models are built upon a set of key ingredients, namely: (i) stellar evolutionary tracks or isochrones; (ii) stellar flux libraries; (iii) initial mass functions; and (iv) star formation histories \citep[see e.g. the review by][]{conroy13}.

A model can only be as good as its ingredients, and there are published studies devoted to the investigation of the uncertainties affecting SPS models \citep[e.g.][]{charlot+96,conroy+09,percival_salaris09}. These studies have shown that limitations of SPS models can arise from uncertainties in both stellar evolution theory and stellar spectral libraries.

In this work, we take a new look at quantifying the uncertainties affecting SPS models by studying a particular source of error: the impact of choosing between an empirical (i.e. observational) or a theoretical library of individual stellar spectra to describe stellar fluxes. By `stellar spectral library', we mean a compilation of homogeneous spectra (either observed or modelled), of which a plethora is available in the literature.\footnote{A comprehensive list has been maintained over the years by David Montes, see \url{https://webs.ucm.es/info/Astrof/invest/actividad/spectra.html}} For stellar population studies, a library 
should ideally provide complete coverage of the
Hertzsprung--Russell diagram (HRD hereafter), accurate and precise atmospheric parameters 
(effective temperature, \teff, surface gravity, {\logg}, abundances, [Fe/H] and [Mg/Fe], rotation, micro- and macro-turbulent velocities, etc.), as well as a good compromise between wavelength coverage, spectral resolution and signal-to-noise ratio (SNR).
Both empirical and theoretical spectral libraries have been largely used in the literature, each presenting advantages and disadvantages.

In a theoretical library \citep[e.g., in recent years,][]{leitherer+10,palacios+10,sordo+10,kirby11,delaverny+12,husser+13,coelho14}, a stellar spectrum has well-defined atmospheric parameters, does not suffer from low SNR or flux calibration problems, and covers a larger wavelength range at a higher spectral resolution than any empirical library. The drawback is that current limitations in our  knowledge of the physics of stellar atmospheres and in the databases of atomic and molecular opacities make theoretical spectra suffer from limited ability to reproduce observations accurately  \citep[e.g.][]{bessell+98,kucinskas+05,kurucz06_lines,MC07,bertone+08,coelho09a_proc,plez11proc,lebzelter+12b,sansom+13,coelho14, knowles+19}. 

In an empirical library, in contrast, all spectral features are accurate (modulo any observational and data reduction problems). Several such libraries have been proposed with different coverages in wavelength, resolution and stellar parameters \citep[e.g., in recent years,][]{ayres10,blanco-cuaresma+14,chen+14,depascale+14,lebzelter+12a,liu+15,villaume+17,worley+12,worley+16,lamost-dr1,mastar}. The strongest limitation of an empirical library is that it is virtually impossible to fully sample the parameter space -- in terms of atmospheric parameters, including abundance ratios -- needed to probe the full range of galaxy evolution studies: stellar populations exist in other galaxies, which are  not represented by the stars we harbour in our Galaxy, such as young metal-poor stars expected to dominate the spectra of high-redshift galaxies \citep[e.g.,][]{stark16} and chemical mixtures different from those tracing the specific history of the solar neighbourhood. A classical example of the latter is the over-abundance of $\alpha$-process over iron-peak elements in high-metallicity populations found in elliptical galaxies \citep[e.g.][]{worthey+92, thomas+05}. Moreover, even to date, assigning atmospheric parameters to observed spectra remains challenging, to the point that different groups often derive different parameters for the same stars. This is a concern for the community, which triggered the creation of an IAU Working Group on Stellar Spectral Libraries.\footnote{\url{https://www.iau.org/science/scientific_bodies/working_groups/306/}} Meanwhile, impressive work is being achieved towards deriving accurate and precise parameters for stars, while exploring and understanding the sources of deviant parameter estimates \citep[e.g.][and references therein]{gaia-eso,smiljanic+14,jofre+14,jofre+15,jofre+17}. 

SPS models incorporating stellar spectral libraries can be roughly classified into 4 types: 

%\begin{notes}
%{Gustavo, can you please place your own work in the 4 groups below as you consider better? add any other work you remember of course. i'm %only trying to restrict to spectral models (i mean, not indices).}
%\end{notes}

\begin{enumerate}
\item  models relying purely on empirical stellar libraries, which we refer to as \emph{semi-empirical SPS models}\footnote{Semi-empirical models are still based on theoretical prescriptions for stellar evolution.} \citep[e.g.,][]{maraston+09,vazdekis+10,vazdekis+16};

\item  models relying purely on theoretical stellar libraries, which we refer to as \emph{fully-theoretical SPS models} \citep[e.g.,][]{leitherer+99,delgado+05,maraston05,coelho+07,percival+09,buzzoni+09proc,lee+09,leitherer+14};

\item models combining empirical and theoretical libraries to sample the parameter space and widen the wavelength coverage \citep[e.g.][]{BC03,maraston_stromback11};

\item models using an empirical library as a base (typically for solar abundances), from which predictions are computed differentially (often via theoretical spectra), which we refer to as \emph{differential SPS models}  \citep[][]{prugniel+07proc,walcher+09,conroy_dokkum12,vazdekis+15}. 
%Such models represent a key recent development, which enabled measurements of different abundance patterns in galaxies at high spectral resolution.

\end{enumerate}

Traditionally, the trend has been to prefer empirical libraries to study stellar absorption features in the spectra of old stellar populations \citep[e.g.][]{vazdekis+16}, while models for young and UV-bright stellar populations heavily rely on theoretical libraries \citep[e.g.][]{leitherer+14}. The recent work by \citet{martins+19} probed different spectral libraries to reproduce the integrated spectra of Galactic star clusters, concluding that modern theoretical libraries are competitive also for modelling old populations. 

Our focus in the present work is on the comparison between semi-empirical and fully-theoretical SPS models. In particular, we aim at investigating how the strengths and caveats of empirical versus theoretical libraries of stellar spectra impact the integrated properties of Simple Stellar Populations (SSPs). The questions we seek to answer are:

\vspace{0.3cm}
\begin{enumerate}
\item \emph{How do the uncertainties identified in theoretical stellar libraries affect integrated colours and spectral indices of model stellar populations?}
\item \emph{How does the non-ideal coverage of the HR diagram by empirical libraries affect the predictions of integrated properties of stellar populations?}
\item \emph{How do random errors in the stellar atmospheric parameters translate into the models?}
\item \emph{To what extent does the choice of an empirical or a theoretical library  affect age and metallicity estimates from integrated light of galaxies?}
\end{enumerate}

\vspace{0.3cm}

To address these questions, we compute SPS models for different choices of  empirical and theoretical stellar libraries, isolating the effects introduced by the use of synthetic versus empirical spectra from those due to the HRD coverage. 
These SPS models are tailored to the specific tests performed in this paper and are not expected to be useful for the general user.

The structure of the paper is as follows:  
we start by detailing a new synthetic stellar library in Section~\ref{sec_syncomil};
in Section~\ref{sec_method} we describe the SPS models built to address the questions outlined above; Section~\ref{sec_results} shows results from the model comparison.
The discussion and conclusions follow in Sections~\ref{sec_discussion} and~\ref{sec_conclusions}.
We provide ancillary information in an online appendix.
%Appendices~\ref{app_modelmodel}--\ref{app_tabs}. 
All models computed for this work are available at \url{http://specmodels.iag.usp.br}.

%% file: syncomil.tex
\begin{figure*}
\begin{center}
\includegraphics[width=0.9\linewidth, trim=0 0 0 0]{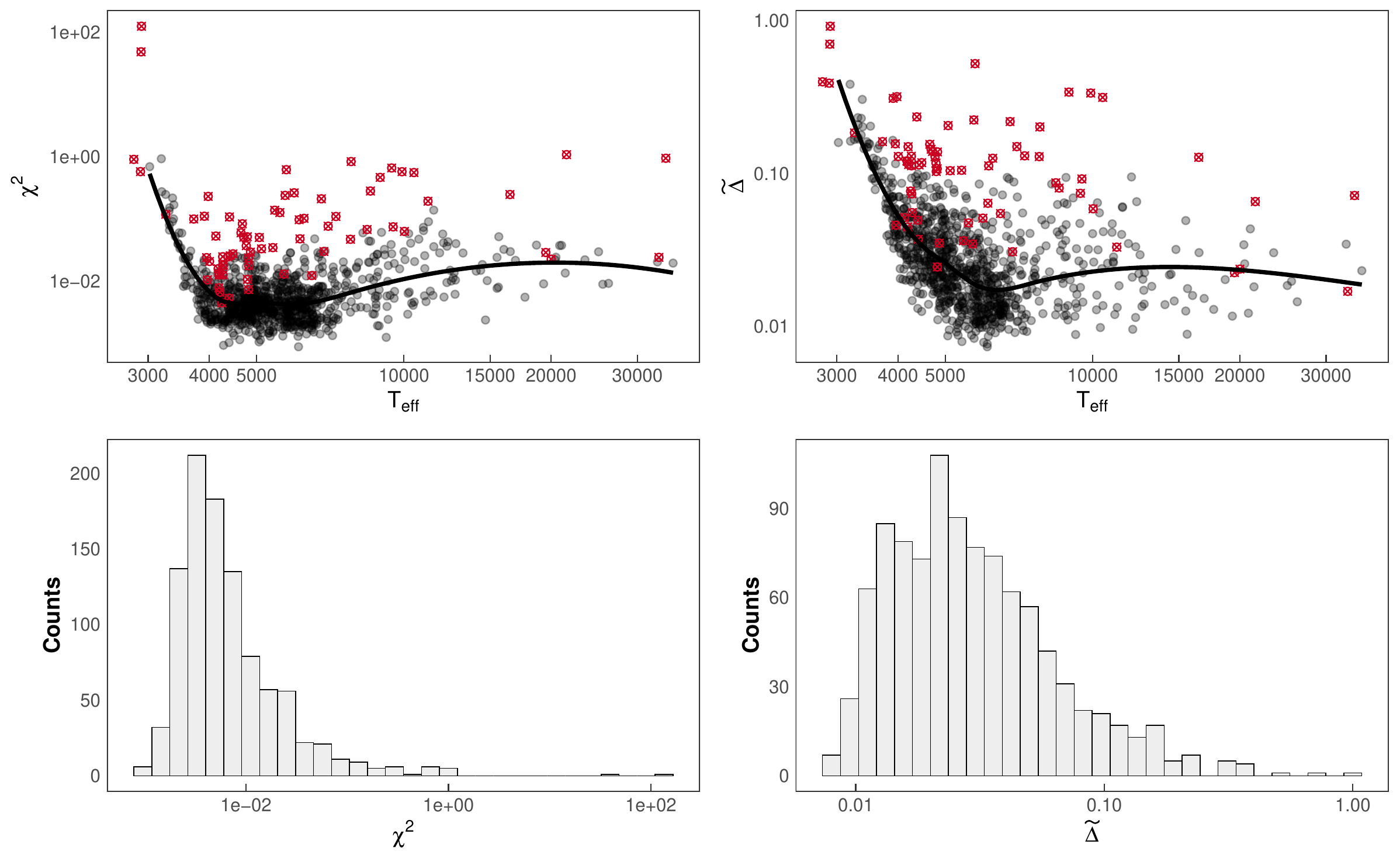}
\caption{\label{fig_metrics} \textit{Top panels:} The \mychi\ and \mrd\ metrics as a function of \teff\ for the \syncomil\ stars.
Red crosses indicate stars which have not been used in the SPS models (see text in Section \ref{sec_syncomil}). The smooth line
corresponds to a LOESS (locally estimated scatter plot smoothing) regression to the points, plotted to aid the eye. \textit{Bottom panels:}
Distribution of \mychi\ and \mrd\ values for the \syncomil\ library.}
\end{center}
\end{figure*}

\section{\syncomil: A synthetic counterpart to the MILES library}
\label{sec_syncomil}

For the present work we built a \textit{Synthetic Counterpart to the \miles Library} (\syncomil\ hereafter), a theoretical spectral grid which mimics the \miles\ stellar library in terms of its wavelength and HRD coverage, and spectral resolution.
\miles \citep{MILES1,MILES2} is a carefully flux-calibrated empirical spectral library widely used in stellar population modelling \citep[e.g.][]{vazdekis+10,vazdekis+12,vazdekis+15,vazdekis+16,martin-hernandez+10proc,maraston_stromback11,rock2016}. It contains spectra for close to 1000 stars covering $\lambda\lambda$ 3540 -- 7410\,\AA\ at FWHM $\sim$ 2.5 \AA\ spectral resolution \citep{fbarroso+11}. Atmospheric parameters for the stars in \miles were first compiled from the literature by \citet{MILES2}. \citet{prugniel+11} and \citet{sharma+16} re-derived these parameters, providing error estimates for the majority of the \miles stars. 
\miles was designed to have an optimal coverage of the HRD in terms of the effective temperature \teff, surface gravity \logg, and metallicity \feh\ of the stars in the library. 
The [Fe/H] vs. \afe\ relation for the \miles stars, characterised by \citet{milone+11}, made the library particularly suitable for comparison with stellar spectral models, where both the abundance pattern and the global metallicity need to be defined.
% making \miles the only empirical library so far with a characterisation of its abundance pattern. \pc{is this claim really true?}

It is well known that atomic and molecular opacities play an essential role in stellar atmosphere models and corresponding synthetic spectra. 
An analysis of libraries from the literature shows different predictions for spectral indices, depending on the exact combination of opacities and grid  \citep[e.g.][]{MC07}. The authors illustrate a trend with \teff\ (hot stars are better reproduced on average than cool ones) and wavelength (redder spectral indices being on average better reproduced than bluer ones).  
The recent work by \citet[][see their fig.~16]{knowles+19} shows that modern grids still show evidence for a light trend with wavelength, and their differential predictions also differ.
Nonetheless, and even though improving the opacities is a slow and time-consuming process, considerable progress has been achieved over time \citep[e.g.][]{peterson+15,kurucz17,2018ApJ...862..146F}. The recent work by \citet{martins+19} compared the ability of different grids (both empirical and synthetic) in reproducing the integrated spectra of globular clusters. They show that current synthetic grids are competitive, despite still showing discrepancies when compared to observations.
\citet{martins+17} discuss current needs regarding synthetic libraries, in particular for applications to stellar population studies.

%Update grid by \citet[][C14 hereafter]{coelho14} (see her table~6 and fig.~10) confirmed the trend with \teff, in the sense that hot stars can be better modelled than cooler stars. 

The computation of \syncomil\ is largely based on the work by \citet{coelho14} (C14 hereafter). Comparisons with MILES library stars discussed in \citetalias{coelho14} (see her table~6 and fig.~10) show trends with \teff\ and wavelength similar to the ones mentioned above. The ingredients for these models are summarised below, and we refer the reader to \citetalias{coelho14} for technical details.

\begin{enumerate}

\item {\it Opacity distribution functions} (ODFs) were adopted from \citetalias{coelho14} for iron abundances \feh\  = --1.3, --1.0, --0.8, --0.5, and from \citet{ATLASODFNEW}\footnote{As made available by F. Castelli; downloaded on April 29 2016 from \url{http://wwwuser.oats.inaf.it/castelli/odfnew.html}} for \feh\  = --3.0, --2.5, --2.0, --1.5, --0.3, --0.2, --0.1, +0.0, +0.2, +0.3, +0.4, +0.5, +1.0. 
A new set of ODFs was computed with \feh\ = --0.5 and \afe\ = +0.2. % $\alpha$-elements to iron ratio

\item {\it Model atmospheres} were computed with 
ATLAS9 \citep{ATLAS1970,ATLAS_LINUX} for stars with \teff\ $\geq$ 3500\,K and the ODFs described above, 
%adopting the ODF closest to the iron abundance \feh\ of each star. We 
adopting the atmospheric parameters of the \miles stars and the  convergence criteria as in \citet{meszaros+12}; For stars with \teff\ $<$ 3500\,K we used existing MARCS atmosphere models\footnote{Available at \url{http://marcs.astro.uu.se}} \citep{MARCS08} to compute a small grid of cool stars.
In all MARCS models we adopted the {\it standard} chemical composition class, with {\it spherical} model geometry for \logg\ $\le$ 1.5, and {\it plane-parallel} for \logg\ $\ge$ 4.5.
% For these MARCS models we adopted the "Standard" chemical composition class, adopted was "Standard composition", "Spherical" model geometry for \logg $\le$ 1.5, and "Plane-parallel" for \logg $\ge$ 4.5;

\item {\it Synthetic stellar spectra} were computed with the SYNTHE code \citep{kurucz_avrett81,ATLAS_LINUX}, based on the ATLAS9 and MARCS models. The opacities are as in \citetalias{coelho14} with one update, the inclusion of the molecular transition C$_2$ D-A from \citet{2013JQSRT.124...11B}\footnote{As made available by R. Kurucz; downloaded on Dec 2016 from \url{http://kurucz.harvard.edu/molecules.html}} to correct for the problem identified by \citet{knowles+19}. 
For stars with \teff\ $\geq$ 3500\,K the synthetic spectrum was computed based on the ATLAS9 models. Cooler star spectra were computed from the MARCS model grids, then interpolated to achieve the atmospheric parameters of \syncomil\ stars. Interpolation was performed linearly in $\theta=5040$/\teff, \logg\ and \feh, with the flux in logarithmic scale. We adopted this scheme after different tests 
comparing synthetic spectra with the interpolated ones. 

\item The synthetic spectra were then corrected for {\it the effect of predicted lines}, interpolating linearly the coefficients listed in Table B1 of \citetalias{coelho14} to the exact atmospheric parameters of the \syncomil\ stars.

\item The spectra were convolved, rebinned and trimmed to match the resolution, dispersion and wavelength range of MILES \citep{fbarroso+11}.
\end{enumerate}

We adopt the \miles\ atmospheric stellar parameters mainly from \citet{prugniel+11}, with the revision for cool stars provided by \citet{sharma+16}. 
For 26 stars we use the parameters from \citet{MILES2}, either because \citet{prugniel+11} do not provide an independent determination, or because we concluded by visual comparison that the \citet{MILES2} parameters permit a closer match between observed and model spectra. 
For stars HD001326B and HD199478 we modified slightly the reported parameters, as we could not obtain converged models for the nominal parameters. 
The changes in \teff\ and \logg\ are nevertheless small -- the smallest needed to achieve convergence -- and are within the reported errors.
We list in Table~\ref{goodstars} the atmospheric parameters adopted for each star in \miles\ and \syncomil, along with their respective sources. To mimic the abundance pattern of the \miles library, we follow table 7 of \citetalias{coelho14}, based on the work by \citet{milone+11}. 
%The adopted errors in \teff, \logg, and \feh\ are listed in Table~\ref{tab_paramerrors}.

\input{abridged_tables}

\syncomil\ spectra were compared with the empirical ones from MILES both visually and quantitatively. Quantitative differences between model and observations were obtained via a $\chi^2$ metrics 

\begin{equation}
\chi^2 = \sum_{\lambda}{ \frac{\left(f^\textrm{syn}_{\lambda} - {f^\textrm{obs}_{\lambda}}\right)^2}{\sigma^{2}_{\lambda}}},
\label{eq1}
\end{equation}

\noindent and a median deviation $\tilde\Delta$ defined as

\begin{equation}
\tilde\Delta = {\rm median} \left({\frac{ | f^\textrm{syn}_{\lambda} - f^\textrm{obs}_{\lambda} | }{f^\textrm{obs}_{\lambda}}}\right),
\label{eq2}
\end{equation}

\noindent where $f^\textrm{syn}_{\lambda}$ and $f^\textrm{obs}_{\lambda}$ are the synthetic and observed fluxes at a given $\lambda$, respectively.
The standard deviation $\sigma_{\lambda}$ appearing in Eq.~(\ref{eq1}) was estimated for each star as follows. P. S\'anchez-Bl\'azquez (priv. comm.) kindly provided us with the error spectra for each star in \miles,
which was used in turn to compute the signal-to-noise ratio SNR($\lambda$) for each star.
Since very deviant values of SNR were obtained for some of the stars, we chose to use the 
median[SNR($\lambda$)] of the distribution of SNR at each $\lambda$ as a fiducial value to
compute $\sigma_{\lambda} = f^\textrm{obs}_{\lambda}$/median[SNR($\lambda$)]. This fiducial SNR($\lambda$)
ranges from $\approx 8$ to $\approx 190$ over the \miles wavelength range.

Fig. \ref{fig_metrics} shows the resulting distributions of $\chi^2$ and $\tilde\Delta$, as well as the corresponding values for each star plotted vs. \teff. 
We found that \teff\ is the only atmospheric parameter which correlates with the metrics in Eqs.~\ref{eq1} and \ref{eq2}.
The patterns in Fig. \ref{fig_metrics} are in agreement with previous work: the larger values of $\chi^2$ and 
$\tilde\Delta$ for cool stars are in agreement with, e.g., \citet{MC07} and \citetalias{coelho14},
and the larger values for \teff\ $\gtrapprox$ 7000K are consistent with the larger errors in \teff\ for these stars reported by \citet{prugniel+11}.
%(see Table \ref{tab_paramerrors}, obtained from \citetalias{coelho14}). 
The red symbols in \ref{fig_metrics} correspond to 71 stars which we opted not to use in the remaining of this work (e.g., bottom panel in Fig. \ref{fig_synmiles_stars}). 
Spectra in \miles which are not suitable for stellar population modelling have been identified previously in the literature \citep[e.g.,][R. Peletier, priv. comm.]{prugniel+11,barber+14}. 
In the present work, a star is not used in the SPS models if any of the conditions below applies.

\begin{enumerate}
\item The observed spectrum shows:
	\begin{itemize}
	\item emission lines,
	\item excessive noise or corrupted pixels,
	\item distortions in the continuum,
	\item $E(B-V)$ > 0.3, as inferred by \citet[][]{prugniel+11},
	\item peculiar spectral features (e.g. HD055496\footnote{This star shows unusually strong molecular bands and is classified as a peculiar star in the SIMBAD Database.}).
	\end{itemize}
\item $\chi^2$ and \teff\ fulfill any of:
	\begin{itemize}
	\item \teff\ < 4000\,K and $\chi^2 > 1$,
   	\item 4000 $\leq$ \teff\ $\leq$ 7000K and $\chi^2 > 0.05$,
	\item \teff\ > 7000\,K and $\chi^2 > 0.15$.
	\end{itemize}
\item $\tilde\Delta$ and \teff\ fulfill any of:
\begin{itemize}
	\item \teff\ $<$ 4000K and $\tilde\Delta > 0.4 $,
   	\item \teff\ $\geq$ 4000K and $\tilde\Delta > 0.1$.
\end{itemize}
\end{enumerate}

Table \ref{badstars} lists the discarded stars.
Most of the discarded stars satisfy more than one of these criteria.
Stars which fall into the quantitative cuts listed above show strong continuum mismatches between model and observations.
We hypothesise that these are due to either large errors in \teff, or problems with flux calibration, reddening, or star identification. The results of \citet{prugniel+11} and \citet{martins+19} also show evidence of residual flux calibration problems in \miles. 
For illustration purposes, in Fig. \ref{fig_synmiles_stars} we compare model and observed spectra for three stars.

%% file: abridged_tables.tex
\begin{table}
\begin{center}
\caption{\label{goodstars}Atmospheric parameters used as input values in the computation of \syncomil (abridged; the full table is available as supplementary online material; see Table \ref{tab_atmpars}). 
Unlisted MILES stars are not used in the SPS models in this work and are reported separately in Table~\ref{badstars} (see Section~\ref{sec_syncomil}).}
\begin{tabular}{clrrrl}
\hline
%MILES ID & Star & T$_{\rm eff}$ & log g & [Fe/H] & Notes \\
MILES ID & Star & \teff & \logg & \feh & Notes \\
\hline
001 & HD224930  & 5411 & 4.19 & -0.78 & b \\
002 & HD225212  & 4117 & 0.68 &  0.14 & c \\
012 & HD001326b & 3571 & 4.81 & -0.57 & d \\
096 & HD016901  & 5345 & 0.85 &  0.00 & a \\
149 & HD027371  & 4995 & 2.76 &  0.15 & b \\
581 & HD143807  &10727 & 3.84 & -0.01 & b \\
\hline
\multicolumn{6}{l}{Atmospheric parameters adopted from: $^a$\citet{MILES2};}\\ 
\multicolumn{6}{l}{$^b$\citet{prugniel+11}; $^c$\citet{sharma+16}; $^d$Used different}\\ %\multicolumn{6}{l}{parameters than}\\
\multicolumn{6}{l}{parameters than proposed in the literature to ensure model}\\ 
\multicolumn{6}{l}{convergence.}\\
%Visit \url{http://specmodels.iag.usp.br}}\\
%\multicolumn{6}{l}{for the full table.}\\
\end{tabular}
\end{center}
\end{table}

%\begin{table}
%\begin{center}
%\caption{\label{tab_paramerrors}Adopted errors in the stellar atmospheric parameters$^e$}
%\begin{tabular}{cccc}
%\hline
%\teff\ interval & $\sigma($\teff$)$ & $\sigma($\logg$)$ &$\sigma$\feh\\
%\hline
 %3000 -- 4000   &   120   &    0.3  & 0.15\\
 %4000 -- 5000   &   120   &    0.2  & 0.15\\
 %5000 -- 7000   &   120   &    0.1  & 0.15\\
 %7000 -- 9000   &   250   &    0.1  & 0.15\\
 %9000 -- 10000  &   250   &    0.2  & 0.15\\
%10000 -- 13000  &   400   &    0.2  & 0.15\\
%13000 -- 16000  &   650   &    0.2  & 0.15\\
%16000 -- 18000  &   850   &    0.2  & 0.15\\
%18000 -- 21000  &  1000   &    0.2  & 0.15\\
%21000 -- 23000  &  1400   &    0.2  & 0.15\\
%above 23000     &  3000   &    0.2  & 0.15\\
%\hline
%\multicolumn{4}{l}{$^e$\citet{coelho14}.}
%\end{tabular}
%\end{center}
%\end{table}

\begin{table}
\begin{center}
\caption{\label{badstars}\miles\ stars unsuitable for SPS modelling (abridged; the full table is available as supplementary online material; see Table \ref{tab_discarded}).}
\begin{tabular}{clrrrl}
\hline
MILES ID & Star & \teff & \logg & \feh & Notes\\
\hline
029 & HD004395 & 5444  & 3.43 & -0.27 & b, 1, 6    \\
044 & HD006474 & 6781  & 0.49 &  0.26 & b, 6, 7    \\
045 & HD006497 & 4401  & 2.55 &  0.00 & c, 1, 6    \\
104 & HD018391 & 5750  & 1.20 & -0.13 & b, 5, 6, 7 \\
140 & HD281679 & 8542  & 2.50 & -1.43 & a, 6       \\
204 & HD041117 & 20000 & 2.40 & -0.12 & b, 3       \\
212 & HD043042 & 6480  & 4.18 &  0.06 & b, 2       \\
246 & HD055496 & 4858  & 2.05 & -1.48 & b, 4       \\
780 & HD199478 & 11200 & 1.90 &  0.00 & d, 3, 6    \\
\hline
\multicolumn{6}{l}{Atmospheric parameters adopted from: $^a$\citet{MILES2};}\\
\multicolumn{6}{l}{$^b$\citet{prugniel+11}; $^c$\citet{sharma+16}; $^d$same as $^a$ but}\\
\multicolumn{6}{l}{unknown \feh\ assumed to be solar. Star discarded due to:}\\
\multicolumn{6}{l}{$^1$Excessive noise or corrupted spectrum; $^2$Visible continuum}\\
\multicolumn{6}{l}{distortions; $^3$Visible emission lines; $^4$Peculiar features;}\\
\multicolumn{6}{l}{$^5E(B-V)>0.3$ from \citet{prugniel+11}; $^6$Removed by cut}\\ 
\multicolumn{6}{l}{in $\chi^2$; $^7$Removed by cut in $\tilde\Delta$ (Section~\ref{sec_syncomil}).}\\ 
%\multicolumn{6}{l}{Visit \url{http://specmodels.iag.usp.br} for the full table.}\\
\end{tabular}
\end{center}
\end{table}

%% file: methodology.tex
\begin{figure}
\begin{center}
\includegraphics[width=8.5cm, trim=0 20 0 0, clip]{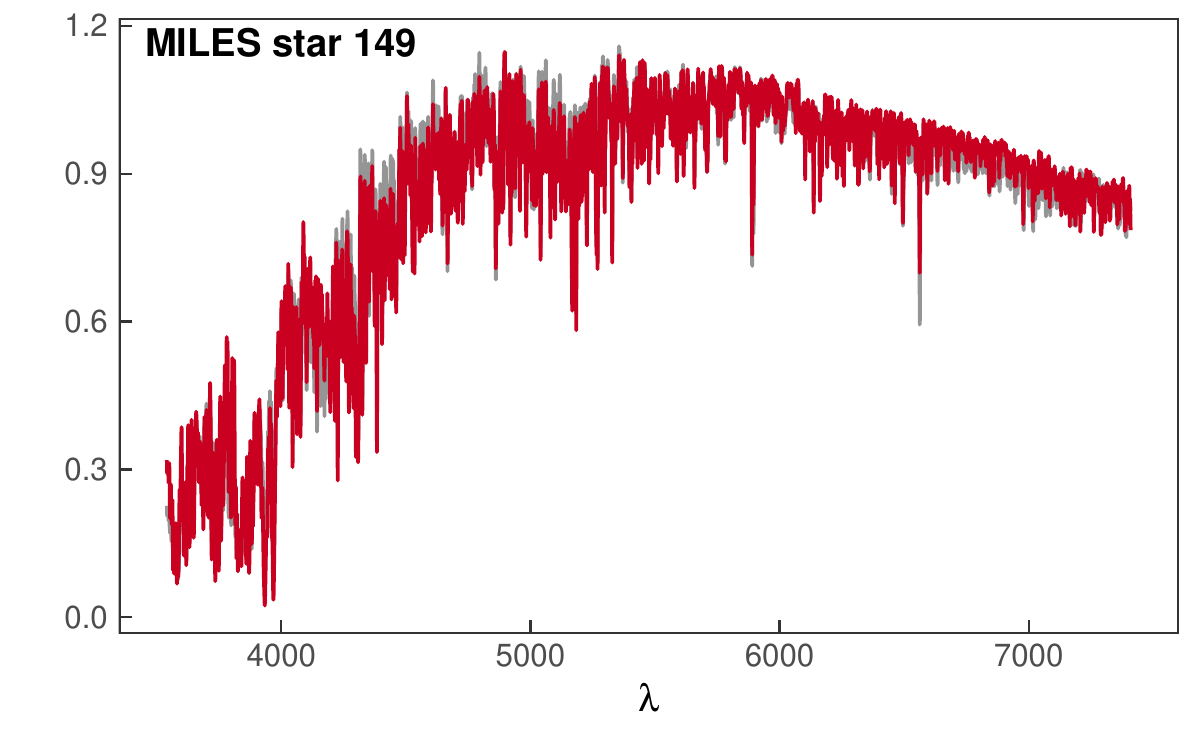}
\includegraphics[width=8.5cm, trim=0 20 0 0, clip]{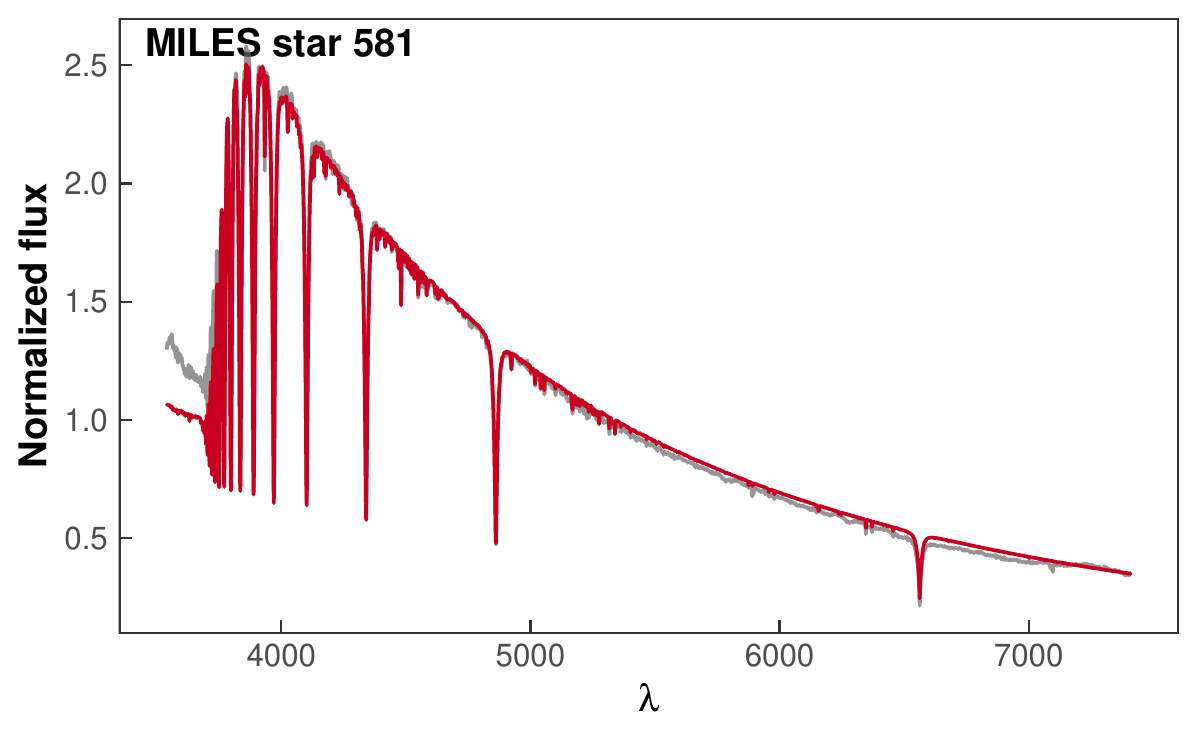}
\includegraphics[width=8.5cm]{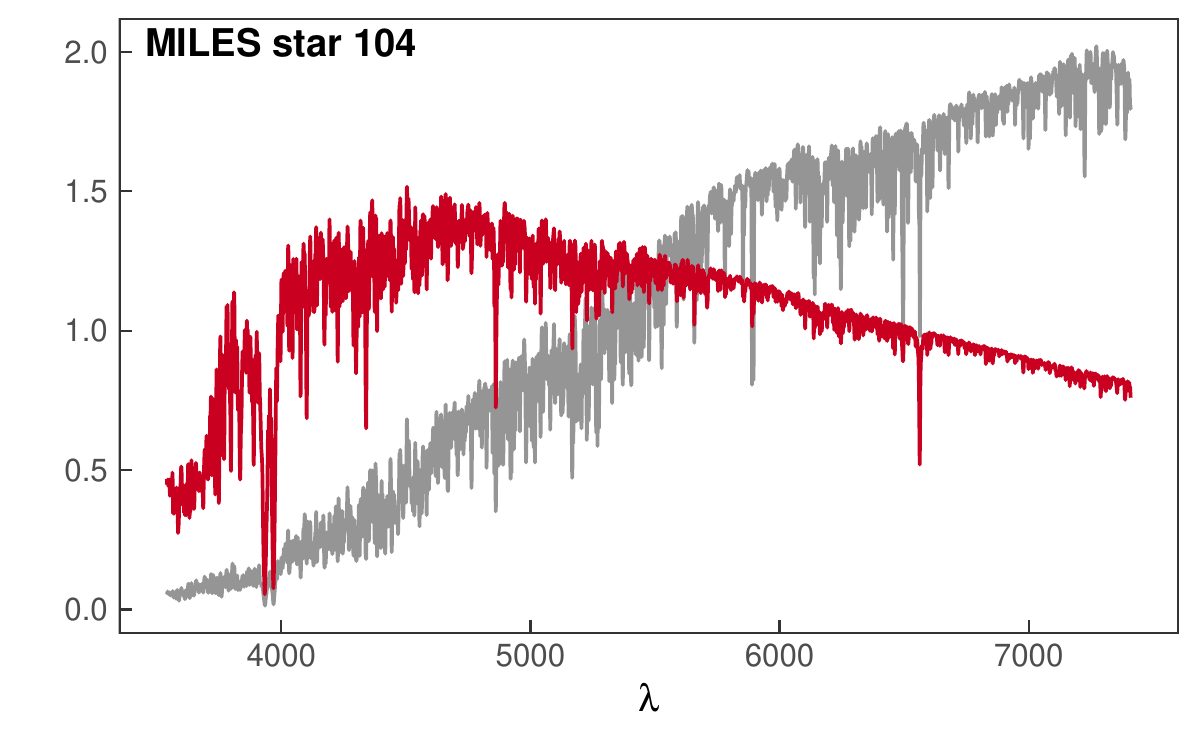}
\caption{\label{fig_synmiles_stars}
Comparison of \miles and \syncomil\ spectra for 3 stars in the \miles library.
{\it (top)}    star 149, HD027371, \teff\,=\,4995\,K,  \logg\,=\,2.76, \feh\,=\,0.15,
{\it (middle)} star 581, HD143807, \teff\,=\,10727\,K, \logg\,=\,3.84, \feh\,=\,-0.01, and
{\it (bottom)} star 104, HD018391, \teff\,=\,5750\,K,  \logg\,=\,1.20, \feh\,=\,-0.13.
The indicated atmospheric parameters are taken from \citet{prugniel+11}.
\miles spectra are shown in gray and \syncomil\ spectra are shown in red.
Fluxes have been normalised to $\int_{3540}^{7410}{F_\lambda\,d\lambda} = 1$. The top and middle panels show \syncomil\ stars around the 25th and 50th percentiles of the \mrd\ distribution in Fig. \ref{fig_metrics}. The bottom panel shows a
`peculiar` case, identified as a Cepheid variable by \citet{prugniel+11}, discarded from the remaining of the present work.
See Section~\ref{sec_syncomil} and Tables~\ref{goodstars} and \ref{badstars}.}
\end{center}
\end{figure}
\section{Methodology}
\label{sec_method}

\begin{figure}
\includegraphics[width=9.5cm, trim=50 180 0 100]{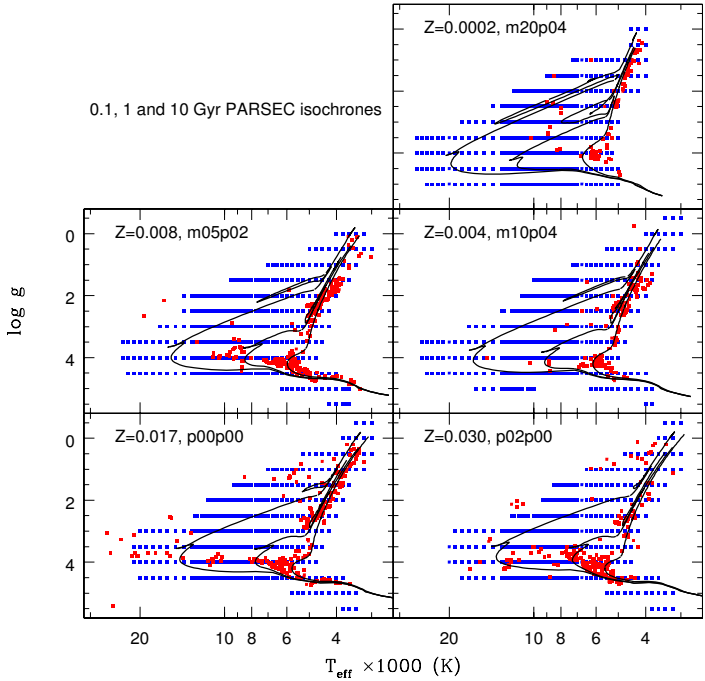}
\caption{\label{fig_tefflogg}Coverage of the stellar flux libraries in the \logg\ vs. \teff\ diagram.  Red symbols indicate the \miles (and \syncomil) stars, and the blue symbols represent the stars from \citetalias{coelho14}. Isochrones for ages 0.1, 1 and 10 Gyr are plotted as black lines. The chemical mixtures are indicated in each panel (see Table \ref{tab_metallicities}).
}
\end{figure}

\begin{table}
\begin{center}
\caption{\label{tab_metallicities}Metallicity regimes explored in the present work}
\begin{tabular}{lcccc}
\hline
\multirow{2}{*}{Bin} & Isochrone &[Fe/H]  & \# \miles    &  \citetalias{coelho14}$^a$\\
                     &  $Z$      & range  &   stars      &         mix    \\
\hline
Sub Solar 3     & 0.0002   & $< -1.3$           &  86   & m20p04 \\
Sub Solar 2     & 0.004     & $[-1.3,-0.6]$    & 144   &  m10p04 \\
Sub Solar 1     & 0.008     & $[-0.6,-0.1]$    & 262   & m05p02 \\
Solar               & 0.017     & $[-0.1,+0.1]$    & 224   & p00p00  \\
Super Solar 1  & 0.030    & $>+0.1$           & 198  & p02p00  \\
\hline
\multicolumn{5}{l}{$^a$ Extended for the present work by adding the mixtures }\\
\multicolumn{5}{l}{(\feh, \afe): (--2.0, 0.4) and (--0.5, 0.2), corresponding to}\\
\multicolumn{5}{l}{$Z = 0.0002$ and $0.008$.}\\
\end{tabular}
\end{center}
\end{table}

To address the questions outlined in Section \ref{sec_intro}, we compute four different sets of SPS models in which all the ingredients are the same except for the choice of the stellar flux library.
The SPS models are computed using the \galaxev code \citep[][and recent updates]{BC03}. This is a flexible code which provides SPS models for a variety of stellar evolutionary tracks, stellar spectral libraries, chemical abundances, initial mass functions, and star formation histories, and probes ideal towards our goal of computing models which differ only in the stellar spectral library. We adopt the PARSEC stellar evolutionary tracks \citep{bressan+12,chen+15} to describe the evolution of stellar populations of the five metallicities listed in Table~\ref{tab_metallicities}. From the evolutionary tracks the \galaxev code builds isochrones for the required age and metallicity. 

Fig.~\ref{fig_tefflogg} shows three isochrones for each stellar metallicity in Table~\ref{tab_metallicities}, 
together with the coverage in the HRD of the \miles, \syncomil, and \citetalias{coelho14} libraries.
This coverage is complete for \citetalias{coelho14} but very sparse for \miles.
The sparse coverage of the HRD by the \miles and other empirical libraries is what forces us to supplement these libraries with synthetic stellar spectra in the \citet[][and recent updates]{BC03} models. One of the goals of the present paper is to quantify the effects on the SPS models of assigning stellar spectra with (\teff,\,\logg,\,\feh) 
relatively far from the true values.

We consider four different sets of SPS models\footnote{All SPS models were computed for the \citet{chabrier03} IMF.} for each mixture in Table~\ref{tab_metallicities}, denoted \pmiles, \smiles, \cmiles\ and \rmiles, which differ only in the stellar spectral library, as follows:
% \begin{enumerate}
% \item \pmiles: the empirical \miles\ library.
% \item \smiles: \ the synthetic \syncomil\ library (Section \ref{sec_syncomil}). 
% \item \cmiles: the synthetic \citetalias{coelho14} library\footnote{The \citetalias{coelho14} spectra were convolved, rebinned and trimmed to match the spectral resolution, dispersion and wavelength range of \miles.}.
% \item \rmiles: 10 realisations of \smiles\,(details below). 
% \end{enumerate}

\vspace{0.2cm}
\pmiles: the empirical \miles\ library.

\smiles: \ the synthetic \syncomil\ library. % (Section \ref{sec_syncomil}). 

\cmiles: \,the synthetic \citetalias{coelho14} library\footnote{The \citetalias{coelho14} spectra were convolved, rebinned and trimmed to match the resolution, dispersion and wavelength range of \miles.}.

\rmiles: \,10 realisations of \smiles\,(details below). 
\vspace{0.2cm}

\noindent To each star along an isochrone, the \galaxev code assigns a spectrum drawn from the selected library. The stellar spectrum is assigned on the basis of the proximity of the stellar parameters of the library stars to the corresponding parameters of the problem star in the HRD, interpolating between neighbouring spectra 
when required.
For the \cmiles models, each problem star in the isochrone is bracketed by four \citetalias{coelho14} stellar models (see Fig.~\ref{fig_tefflogg}), characterized each by (\thet,\,\logg), where \thet = $5040/T_{\rm eff}$. In this case, we interpolate the stellar models logarithmically in flux, first in \thet\ at constant \logg\ and then in \logg.
This interpolation scheme is possible only in very few instances when using the \miles library in the \pmiles models.
In some cases, we can interpolate in \teff\ two \miles spectra of the required \logg, but in most cases we use the \miles spectrum closest in (\teff,\,\logg) to the problem star on the isochrone.

For the \smiles models, we use the same spectral assignment as in the \pmiles models, but draw the corresponding stellar spectra from the \syncomil instead of the \miles library. 

For the \rmiles models, we modify the stellar parameters of each \syncomil\ star by adding to each parameter a Gaussian random error. Then we replace each star in the \smiles models with the star with closest parameters in the modified table. 
This exercise is repeated 10 times for each set of SPS models listed in Table~\ref{tab_metallicities}.
We adopt errors for \teff\ and \logg\ as compiled by \citetalias{coelho14} (see her table 5): 
for \teff\ the error ranges from 120\,K for cool stars to 3000\,K for hot stars; 
for \logg\ the errors range from 0.1 to 0.3 depending on \teff. 
For \feh\ we adopt conservative errors of 0.15 \citep[e.g.,][]{soubiran+98}.
% with $\sigma$ listed in Table \ref{tab_paramerrors}; 

%% file: results.tex
\begin{figure*}
\includegraphics[width=0.85\linewidth]{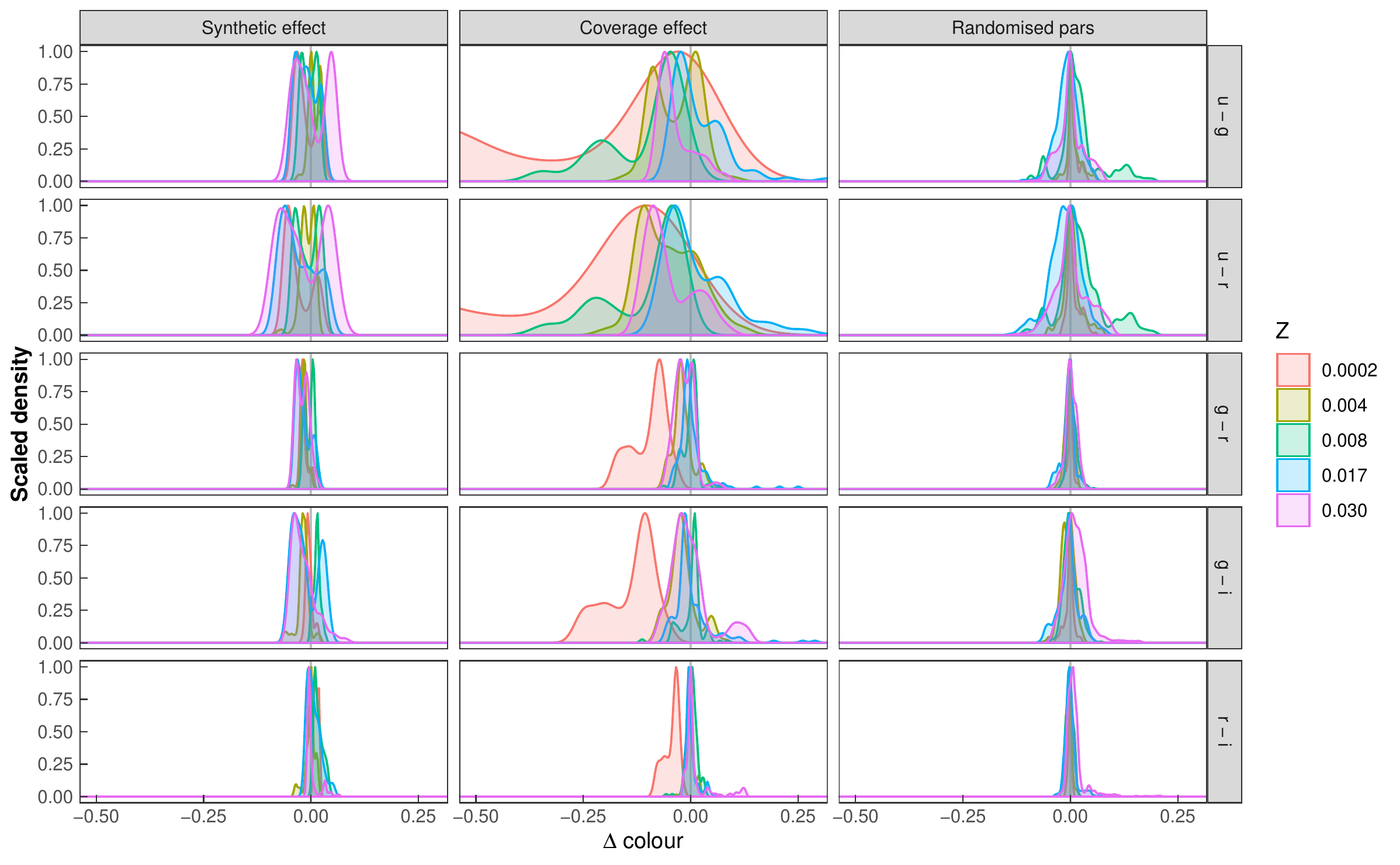}
\caption{\label{fig_colors_density} Density plots showing the distributions of colour differences (\deltacolor, defined in Section \ref{sec_colours}) for the different combinations of SDSS-based colours (in rows) and the model effects (in columns).  Colours indicate the metallicities Z, as indicated in the label.
%The different colours correspond to different metallicity regimes, as indicated in the label.
}
\end{figure*}

\begin{figure}
\includegraphics[width=\linewidth]{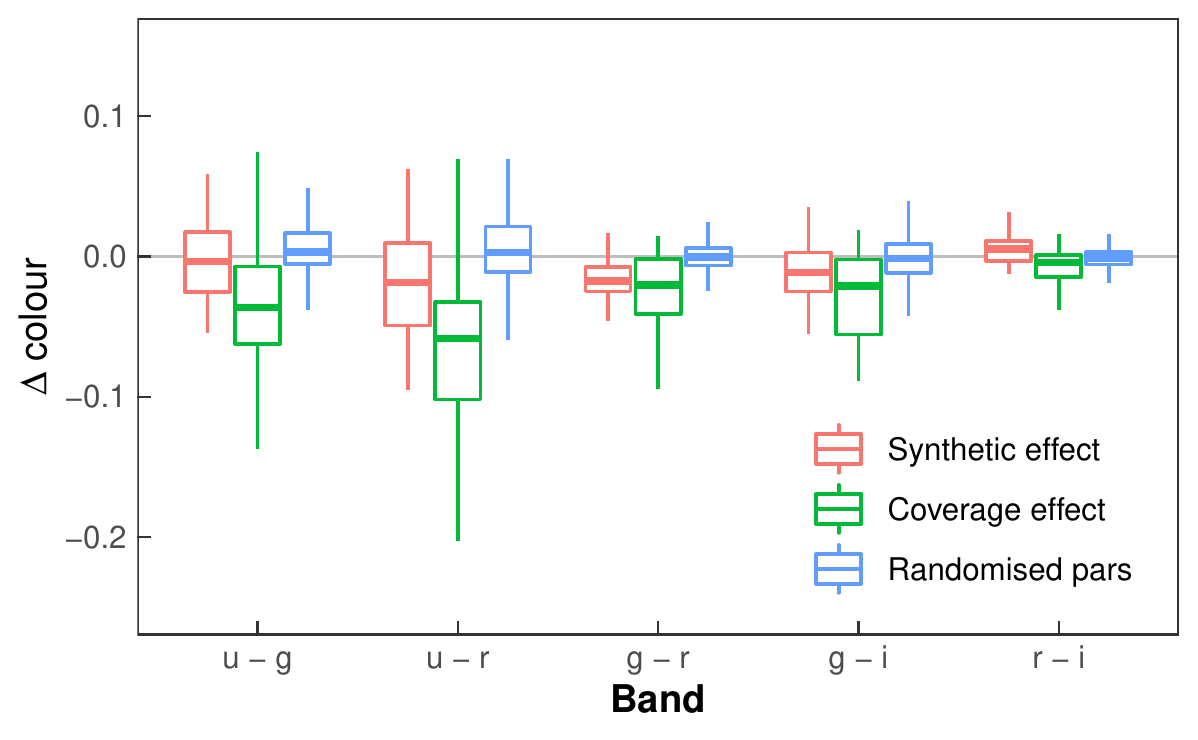}
\caption{\label{fig_colors_boxplot}The values of \deltacolor\ are given for each color (x-axis) and effect (coloured labels).
Each box indicates the range from the 25\% to the 75\% percentiles, with the horizontal line indicating the median value.
The whiskers have length equal to $\pm 1.5 \times$IQR. Outliers are not shown.}
\end{figure}

\begin{figure*}
\includegraphics[width=0.85\linewidth]{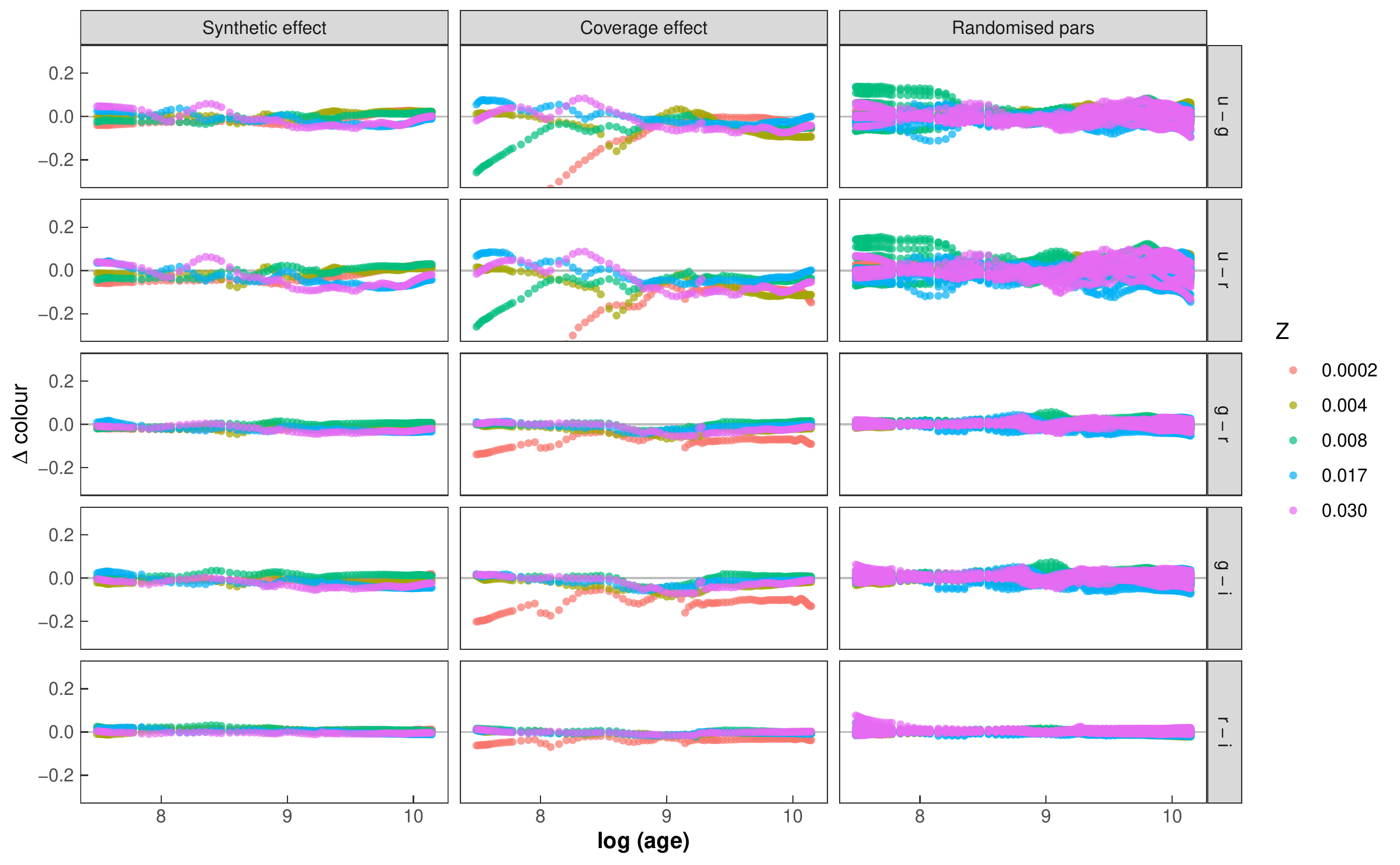}
\caption{\label{fig_colors_distributions}\deltacolor\ differences for the three effects (columns) described in Section \ref{sec_modelmodel}, {plotted as a function of stellar population age (in yr)}. Different colours are shown in rows,
and different metallicities are identified as indicated in the label.}
\end{figure*}
% \begin{figure*}
% \includegraphics[width=87mm]{figures/cbc_fig8a.png}
% \includegraphics[width=87mm]{figures/cbc_fig8b.png}
% \includegraphics[width=87mm]{figures/cbc_fig8c.png}
% \caption{\label{fig_idxscatter}
% Spectral indices predicted by different models at the age indicated by the auxiliary axis.
% ({\it top left}) \pmiles (y-axis) vs. \smiles (x-axis);
% ({\it top right}) \cmiles (y-axis) vs. \smiles (x-axis);
% ({\it bottom}) \rmiles (y-axis) vs. \smiles (x-axis).}
% \end{figure*}

%=================================================

\section{Results}
\label{sec_results}

The different SPS models described in Section \ref{sec_method} were compared to each other in two ways: 
(1) via direct model -- model comparisons (Section \ref{sec_modelmodel}), and
(2) using the models to derive stellar population parameters via spectral fits to a sample of galaxy spectra (Section \ref{sec_specfit}).

\subsection{\label{sec_modelmodel}Model -- model comparisons}

From the direct comparison of the colours and spectral line indices predicted by the different models we aim at understanding the following three effects:

\emph{Synthetic effect:} comparing the predictions of the \pmiles and \smiles models we can assess the consequences of using theoretical instead of empirical stellar spectra for a fixed coverage of the HRD.

\emph{Coverage effect:} comparing the predictions of the \smiles and \cmiles models we can isolate the consequences of limited vs. complete HRD spectral coverage on the observables.

\emph{Random error effect:} comparing the \rmiles and \smiles models, we can assess the effects of random errors in the stellar atmospheric parameters on the model predictions.

%%%%%%%%%%% RESULTS COLORS{}

\subsubsection{Broad-band colors}
\label{sec_colours}

For each set of models, we compute 5 colours, $u-g$, $u-r$, $g-r$, $g-i$, and $r-i$, using the SDSS $ugri$ filter response functions
\citep{doi+10}\footnote{The \miles spectra extend from 3540 to 7410\,\AA\ and do not cover the full widths of the $u$ and $i$ bands.
The $u$ band extends from 2980 to 4130\,\AA, with $\lambda_\mathrm{eff}(u) = 3560$\,\AA\ inside the \miles range,
while the $i$ band extends from 6430 to 8630\,\AA, with $\lambda_\mathrm{eff}(i) = 7500$\,\AA\ not far from the \miles edge.
The stellar flux is considered to be zero for $\lambda < 3540$\,\AA\ and $\lambda > 7410$\,\AA\ when computing
the $u$ and $i$ magnitudes.
Given that all our SPS models cover the same wavelength range as the \miles library, we consider that the use of the $u$ and $i$ bands is still useful and informative.}.
For each age and metallicity of the SPS models we compute the following colour differences in direct relation to the effects listed above:
\begin{subequations}\label{eq:colors}
\begin{align}
\Delta{\rm \sf colour}_{\rm Synthetic}= {c}_{\pmiles}-{c}_{\smiles},\\
\Delta{\rm \sf colour}_{\rm Coverage} = {c}_{\cmiles}-{c}_{\smiles},\\
\Delta{\rm \sf colour}_{\rm RanError} = {c}_{\rmiles}-{c}_{\smiles},
\end{align}
\end{subequations}
\noindent where $c$ is a colour. Results are shown in Figs.~\ref{fig_colors_density}, \ref{fig_colors_boxplot} and \ref{fig_colors_distributions}, and listed in Table \ref{tab_colors}. 

Fig.~\ref{fig_colors_density} shows the distributions of \deltacolor\ for the different colours (rows) and effects (columns). We notice that the distributions are not always symmetric, and that they are typically broader for the coverage effect than for the synthetic and randomised-parameter effects. 
%The highest and lowest metallicities are more affected, as expected, since for these regimes the coverage of the empirical library is poorer.
%\footnote{We illustrate the HR coverage of the 5 intervals in metallicity in section \ref{online_hrcoverage} of the online material.}. 
%The large coverage effect seen for the most metal poor regime is in agreements with the $Q_n$ parameter by \citet{vazdekis+10}.
These results are illustrated in a condensed manner as a boxplot in Fig. \ref{fig_colors_boxplot}, indicating the corresponding median and interquartile range (IQR)\footnote{IQR is equal to the difference between the 75th and 25th percentiles. For a normal distribution, IQR $=1.35 \times \sigma$.} given in Table \ref{tab_colors}. In %Appendix
Fig.~\ref{fig_colors_distributions}, we show \deltacolor\ as a function of stellar population age for the different colours and effects. A large variance in colours involving the $u$ band arises at ages younger than 1\,Gyr, %$\log(\mathrm{age/yr})=9$, 
especially at low metallicity, which is not surprising given the limited number of hot metal poor stars in empirical libraries.

In summary, the coverage effect dominates the systematics and variance in  \deltacolor. The most affected colours are those involving the $u$ band, and the largest \deltacolor\ is $\Delta(u-r)_{\rm Coverage}=0.06$.

% \vspace{.5cm}
%The lessons learned are: 
%\begin{enumerate}
%\item \deltacolor\ are typically small, the largest being 0.06 %corresponding to the {\sf coverage effect} on the u--r colour; 
%\item the most affected colours are those that use the {\sf u} band, and; 
%\item the {\sf coverage effect} dominate the systematics and the variances.
%\end{enumerate}

%Median colour differences \deltacolor\
%, for the three model comparisons performed in this work. \pc{complete}
\begin{table}
\begin{center}
\caption{\label{tab_colors}Median and IQR values of \deltacolor\ in magnitude units.}
\begin{tabular}{crrrrrr}
\hline
\multirow{4}{*}{Colour} & \multicolumn{2}{c}{Synthetic}             & \multicolumn{2}{c}{Coverage }              & \multicolumn{2}{c}{Random error} \\
                        & \multicolumn{2}{c}{effect}                & \multicolumn{2}{c}{effect}                 & \multicolumn{2}{c}{effect} \\
                        & \multicolumn{2}{c}{(\pmiles\ --\smiles)}  & \multicolumn{2}{c}{(\cmiles\ -- \smiles)}  & \multicolumn{2}{c}{(\rmiles\ -- \smiles)} \\
                        & Median & IQR                              & Median  & IQR                              & Median & IQR \\
\hline
$u-g$ & --0.002 & 0.043 & -0.038 & 0.068 &   0.003 & 0.021\\
$u-r$ & --0.018 & 0.059 & -0.061 & 0.085 &   0.001 & 0.031\\
$g-r$ & --0.017 & 0.017 & -0.018 & 0.043 &   0.001 & 0.011\\
$g-i$ & --0.009 & 0.035 & -0.017 & 0.062 & --0.001 & 0.020\\
$r-i$ &   0.007 & 0.019 & -0.003 & 0.017 & --0.001 & 0.009\\
\hline
\end{tabular}
\end{center}
\end{table}

%%%%%%%%%%% RESULTS INDICES

\subsubsection{Spectral indices}
\label{sec_specidx}

\begin{table}
\begin{center}
\caption{\label{tab_indices}Spectral indices measured in the present work.}
\begin{tabular}{ll}
\hline
Indices  & Reference \\
\hline
H10, H9, H8                     & \citet{marcillac+06} \\
HK                              & \citet{brodie+86}   \\
B4000                           & \citet{kauffmann+03} \\
H$\delta$A, H$\delta$F, H$\gamma$A, H$\gamma$F & \citet{worthey+97} \\
\hline
CN1, CN2, Ca4227, G4300,        & \citet{trager+98} \\ 
Fe4383, Ca4455, Fe4531,         & \\ 
Fe4668, H$_\beta$, Fe5015, Mg1, & \\ 
Mg2, Mgb, Fe5270, Fe5335,       & \\ 
Fe5406, Fe5709, Fe5782, NaD,    & \\ 
TiO1, TiO2                      & \\ 
\hline
\end{tabular}
\end{center}
\end{table}

\begin{figure*}
\includegraphics[width=0.85\linewidth]{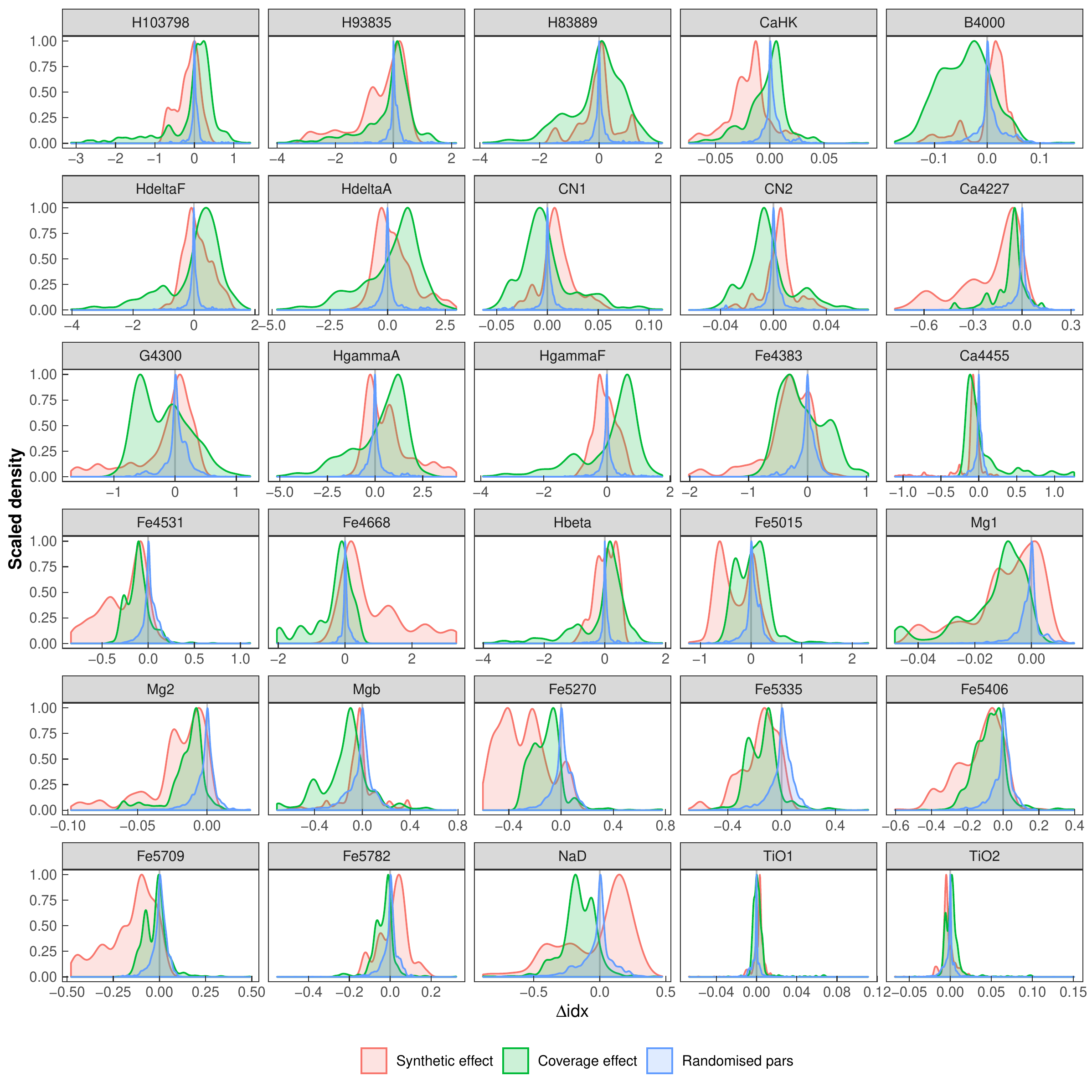}
\caption{\label{fig_idx_distributions}Density plots showing the distributions of index differences (\deltaidx, as defined
in Section \ref{sec_specidx}) for the indices listed in Table \ref{tab_indices}. The different effects are shaded as indicated in the label.}
\end{figure*}

\begin{figure*}
\includegraphics[width=0.85\linewidth]{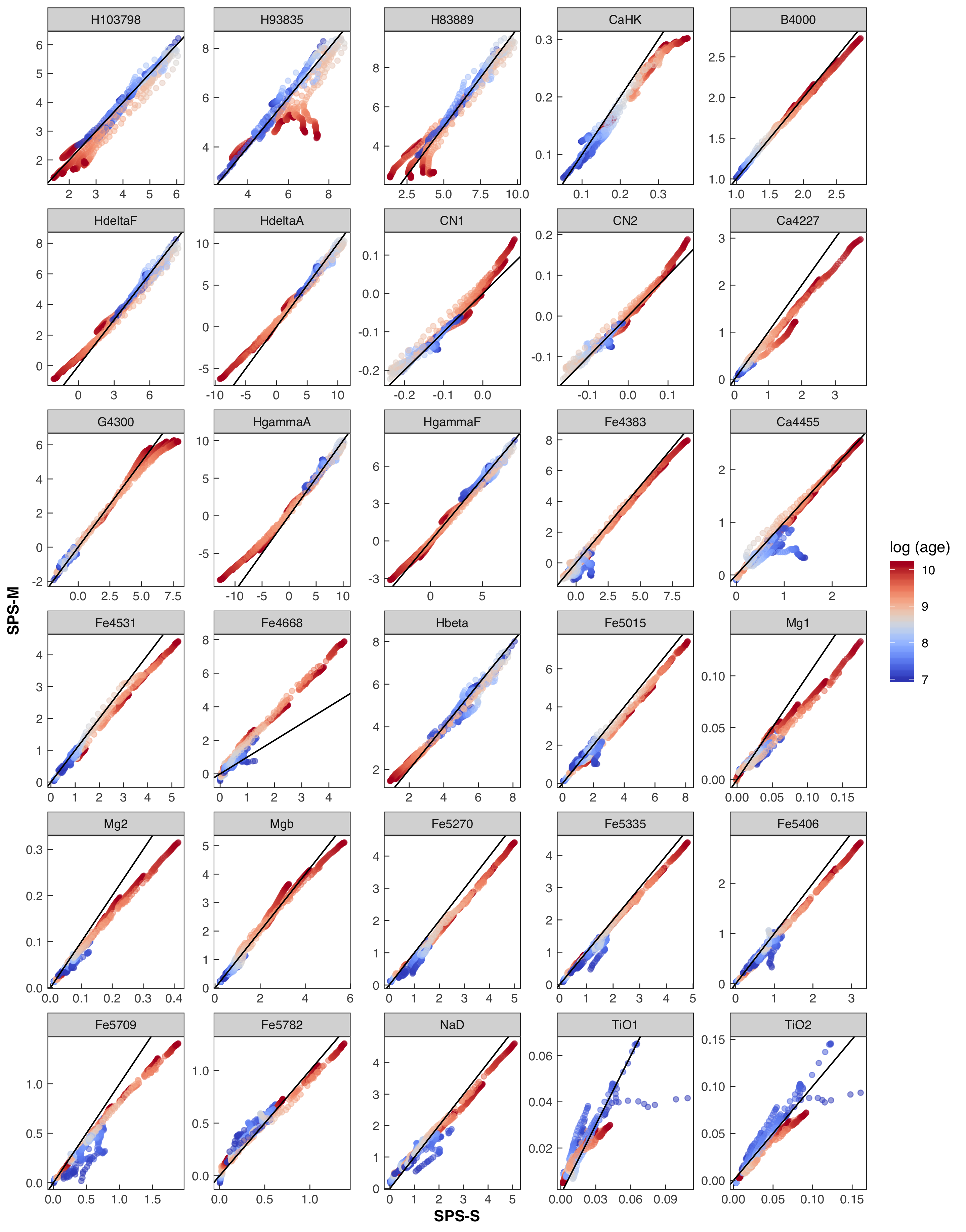}
\caption{\label{fig_idxscattera}
Spectral indices predicted by \pmiles (y-axis) vs. \smiles (x-axis) models. The points are colour-coded by the age of the population, as indicated by the auxiliary axis.}
\end{figure*}

\begin{figure*}
\includegraphics[width=0.85\linewidth]{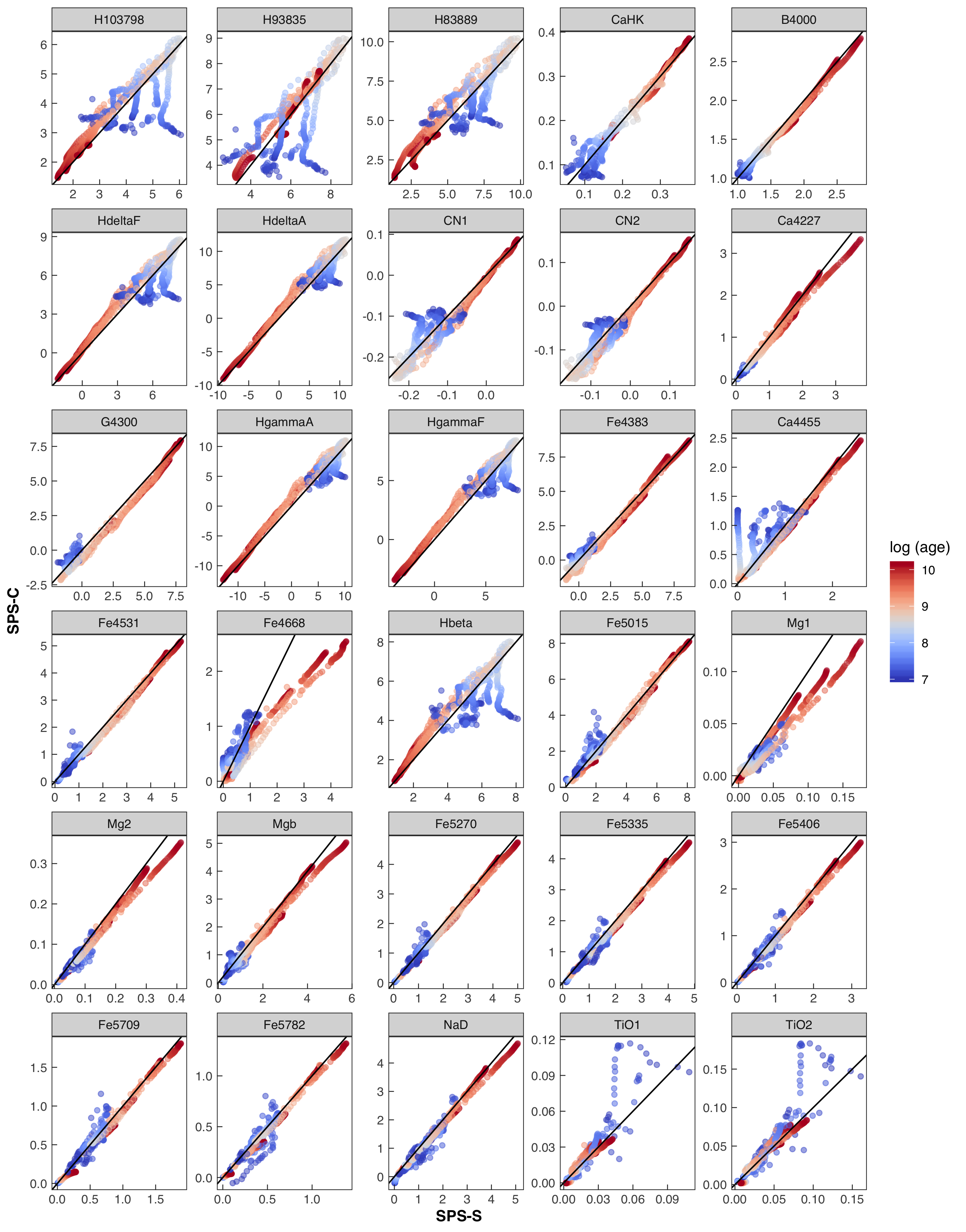}
\caption{\label{fig_idxscatterb}
Spectral indices predicted by \cmiles (y-axis) vs. \smiles (x-axis) models. The points are colour-coded by the age of the population, as indicated by the auxiliary axis.
}
\end{figure*}

\begin{figure*}
\includegraphics[width=0.85\linewidth]{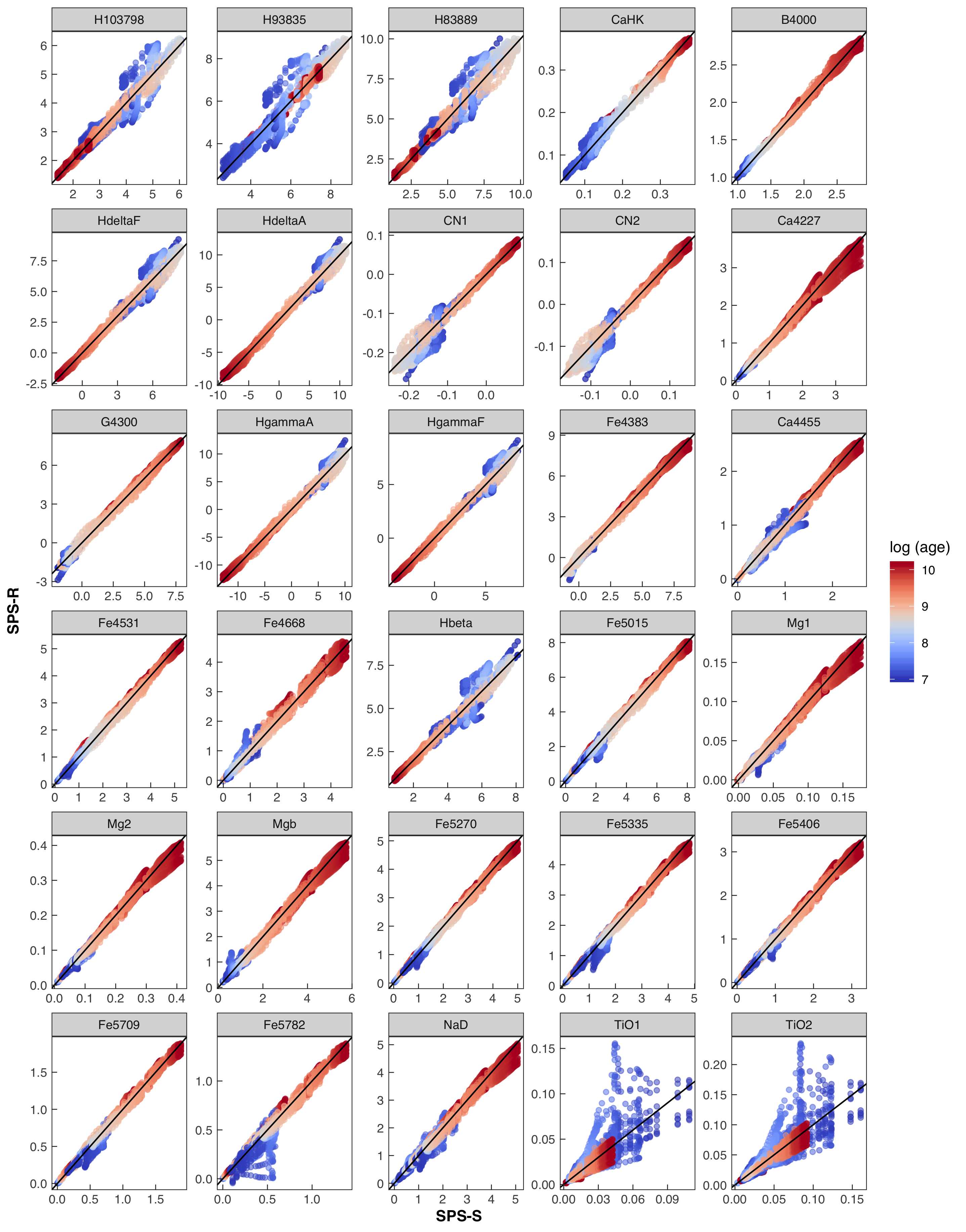}
\caption{\label{fig_idxscatterc}
Spectral indices predicted by \rmiles (y-axis) vs. \smiles (x-axis) models. The points are colour-coded by the age of the population, as indicated by the auxiliary axis.}
\end{figure*}

\begin{figure*}
\includegraphics[width=0.85\textwidth]{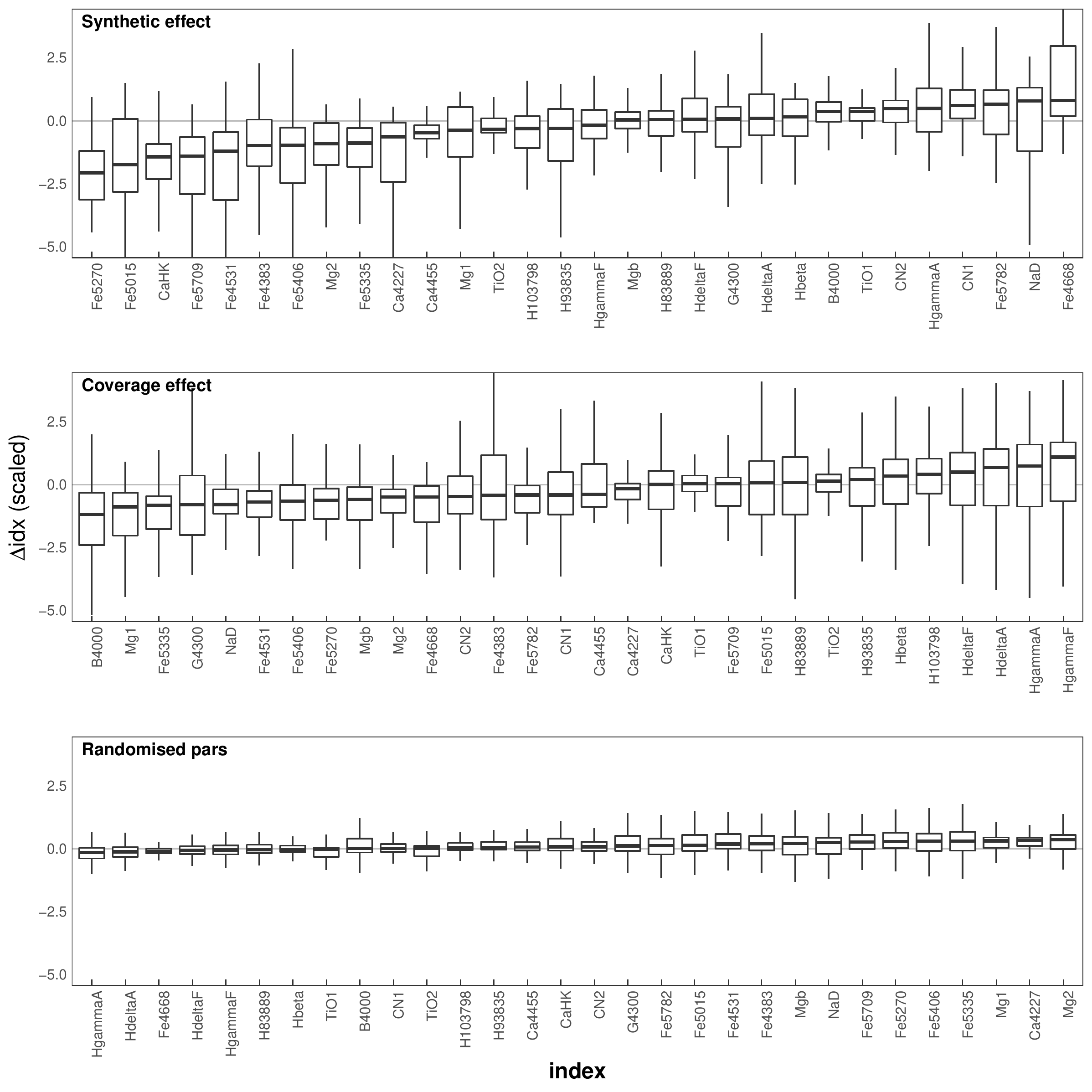}
\caption{\label{fig_idxboxplot}Boxplots of \deltaidx\ are shown for the synthetic (top), coverage (middle) and randomised parameters (bottom) effects. In each row, \deltaidx\, are sorted in crescent order to facilitate identification of most affected indices in each case. The values of \deltaidx\ have been scaled to their standard score to facilitate comparison
(see Section \ref{sec_specidx}).}
\end{figure*}

The spectral indices listed in Table \ref{tab_indices} were measured in all SPS models.
We define \deltaidx\ as the index difference between two sets of models:
\begin{subequations}\label{eq:idx}
\begin{align}
\Delta {\rm \sf idx}_{\rm Synthetic} = {I}_{\pmiles} - {I}_{\smiles},\\
\Delta {\rm \sf idx}_{\rm Coverage}  = {I}_{\cmiles} - {I}_{\smiles},\\
\Delta {\rm \sf idx}_{\rm RanError}  = {I}_{\rmiles} - {I}_{\smiles},
\end{align}
\end{subequations}
\noindent where $I$ is any spectral index. 
We measure $\Delta{\rm \sf idx}$ for all ages and metallicities. Results are shown in Figs. \ref{fig_idx_distributions} -- \ref{fig_idxboxplot},
and listed in Table~\ref{tab_deltaidx}.
% \ref{fig_idxscattera}, \ref{fig_idxscatterb}, \ref{fig_idxscatterc} 
%$(Section \ref{online_modelmodel}, Figs. 

Fig. \ref{fig_idx_distributions} shows the distributions of \deltaidx\ for
each index. 
The impact of the three effects on the indices is complex and defies simple conclusions. In all cases, the effect of the random errors on the atmospheric parameters is the least important.
The indices for which the synthetic effect is most prominent are: CaHK, Fe4668, Fe5270, Fe5709 and NaD.
%\smiles models underestimate the Fe4668 and NaD indices, and overestimates the CaHK, Fe5270 and Fe5709 indices.
The impact of HRD coverage in the indices is not negligible and dominates over the synthetic effect in the case of, e.g., H8, B4000, and G4300.
For many indices the synthetic and coverage effects are comparable and may introduce different systematic effects. 
Table \ref{tab_deltaidx} lists the median and IQR values of \deltaidx\ for each index and effect.

%Section \ref{app_modelmodel}, 

%The lessons learned are: 
%\begin{enumerate}
%\item in all cases, the effect of randomising the atmospheric parameters is the least important;
%\item the indices were the synthetic effect is most prominent are: CaHK, Fe4668, Fe5270, Fe5709 and NaD. \synlib\ underestimate the values for indices Fe4668 and NaD, and overestimates the values for CaHK, Fe5270 and Fe5709; 
%\item the impact of HR coverage in the indices is not negligible, and dominates over the synthetic effect in the case of H8, B4000, G4300, and;
%\item for the other indices the Synthetic and Coverage effects are comparable, and may introduce different systematic effects. 
%\end{enumerate}

%For each spectral fit parameter inferred in Section \ref{s_specfit}, this table shows the median and IQR values of the Synthetic effect (parameters obtained from MILES-based SSPs minus parameters obtained from \synlib-based SSPs), and Coverage effect (parameters obtained from Theoretical-based SSPs minus parameters obtained from \synlib-based SSPs).

\begin{table}
\begin{center}
\caption{\label{tab_deltaidx}Median and IQR values of \deltaidx.}
\begin{tabular}{lcccccc}
\hline
\multirow{4}{*}{Index} & \multicolumn{2}{c}{Synthetic}             & \multicolumn{2}{c}{Coverage }              & \multicolumn{2}{c}{Random error} \\
                       & \multicolumn{2}{c}{effect}                & \multicolumn{2}{c}{effect}                 & \multicolumn{2}{c}{effect} \\
                       & \multicolumn{2}{c}{(\pmiles\ --\smiles)}  & \multicolumn{2}{c}{(\cmiles\ -- \smiles)}  & \multicolumn{2}{c}{(\rmiles\ -- \smiles)} \\
                       & Median & IQR                              & Median  & IQR                              & Median & IQR \\
\hline
 H103798    & -0.093 & 0.365 &  0.116 & 0.399 &  0.005 & 0.081\\
 H93835     & -0.154 & 0.997 &  0.088 & 0.730 &  0.017 & 0.151\\
 H83889     &  0.046 & 0.472 &  0.068 & 1.095 &  0.003 & 0.159\\
 CaHK       & -0.018 & 0.018 &  0.000 & 0.020 &  0.001 & 0.006\\
 B4000      &  0.014 & 0.027 & -0.038 & 0.071 &  0.002 & 0.019\\
 H$\delta$A &  0.138 & 1.068 &  0.520 & 1.467 & -0.006 & 0.250\\
 H$\delta$F &  0.052 & 0.568 &  0.242 & 0.903 & -0.003 & 0.134\\
 H$\gamma$A &  0.458 & 1.285 &  0.653 & 1.820 & -0.010 & 0.308\\
 H$\gamma$F & -0.058 & 0.473 &  0.478 & 0.977 & -0.003 & 0.148\\
 CN1        &  0.009 & 0.016 & -0.006 & 0.024 &  0.000 & 0.005\\
 CN2        &  0.005 & 0.010 & -0.006 & 0.017 &  0.000 & 0.004\\
 Ca4227     & -0.104 & 0.257 & -0.052 & 0.070 &  0.001 & 0.037\\
 G4300      &  0.014 & 0.439 & -0.225 & 0.647 &  0.025 & 0.163\\
 Fe4383     & -0.288 & 0.450 & -0.150 & 0.624 &  0.003 & 0.144\\
 Ca4455     & -0.071 & 0.072 & -0.057 & 0.227 &  0.002 & 0.045\\
 Fe4531     & -0.181 & 0.370 & -0.109 & 0.143 &  0.012 & 0.080\\
 Fe4668     &  0.409 & 1.204 & -0.150 & 0.622 &  0.011 & 0.081\\
 H$\beta$   &  0.074 & 0.550 &  0.146 & 0.661 &  0.003 & 0.095\\
 Fe5015     & -0.378 & 0.631 &  0.019 & 0.462 &  0.034 & 0.138\\
 Mg1        & -0.006 & 0.015 & -0.009 & 0.013 &  0.000 & 0.003\\
 Mg2        & -0.016 & 0.021 & -0.011 & 0.012 & -0.000 & 0.007\\
 Mgb        & -0.026 & 0.091 & -0.112 & 0.184 & -0.001 & 0.101\\
 Fe5270     & -0.294 & 0.248 & -0.109 & 0.155 &  0.008 & 0.079\\
 Fe5335     & -0.131 & 0.179 & -0.123 & 0.153 &  0.008 & 0.086\\
 Fe5406     & -0.101 & 0.179 & -0.074 & 0.113 &  0.004 & 0.056\\
 Fe5709     & -0.114 & 0.162 & -0.011 & 0.080 &  0.005 & 0.040\\
 Fe5782     &  0.031 & 0.090 & -0.025 & 0.056 &  0.003 & 0.032\\
 NaD        &  0.084 & 0.393 & -0.163 & 0.151 & -0.001 & 0.102\\
 TiO1       &  0.003 & 0.004 &  0.000 & 0.005 &  0.000 & 0.003\\
 TiO2       & -0.004 & 0.007 &  0.001 & 0.008 & -0.000 & 0.005\\
\hline
\end{tabular}
\end{center}
\end{table}

%\pc{Email from GB regarding the different D4000 indices:}
%\textcolor{magenta}{This are different ways to measure the 4000 A break.}\\
%\textcolor{magenta}{D4000 = original definition = ratio of flux(4050:4250) to flux(3750:3950) measured in Fnu units.}\\
%\textcolor{magenta}{B4VN =B4000 = very narrow definition of 4000 A break = flux(4000:4100) over (3850:3950)}\\
%\textcolor{magenta}{D4000vn = is a fitting function version of the 4000 A break by Gorgas, Cardiel, Pedraz, Gonzalez (1998)}\\
%\textcolor{magenta}{I will stick to any of the first 2. I think the narrow version is preferred by SDSS people which have more or
%les the same resolution as miles.}\\

In Figs. \ref{fig_idxscattera}, \ref{fig_idxscatterb} and \ref{fig_idxscatterc} we plot the indices computed from 
\pmiles, \cmiles\ and \rmiles against those of \smiles\ models, respectively. 
The dependence with the age of the population is apparent: as expected, the largest deviations from the 1-to-1 line 
occur for the old population in the case of the synthetic effect, and for the young population in the case of the coverage effect. 
In these plots we can locate opacity related problems, e.g., Fe4668 (see discussion in \citealt{knowles+19}),
or the high sensitivity to HRD coverage of the Balmer line indices.
The effect of randomising the stellar parameters is to increase the dispersion of the data points but they do not deviate significantly from the 1-to-1 line except for young populations: most likely this is induced by the coverage effect due to the scarcity of hot stars in the HRD.

The boxplot in Fig. \ref{fig_idxboxplot} is convenient to easily rank the sensitivity of each index to the effect being explored.
To facilitate visualisation, \deltaidx\ has been scaled to its standard score (or z-value), defined as:
\begin{equation}
z = \frac{x - \mu}{\sigma},
\end{equation}  
\noindent where $x$ is the raw \deltaidx\ value, $\mu$ is the mean value of $x$ for the population, and $\sigma$ its standard deviation.
In each panel (\emph{top}: synthetic effect; \emph{middle}: coverage effect; \emph{bottom}: randomised parameters) the indices are sorted in increasing order of \deltaidx. 
Deviations from the zero-line indicate systematic effects and the size of the box indicates the IQR. %It is easy to see that: T
The further away from the zero-line the midline of a box is, the more the index is affected by systematics. 

In summary, the consequences of the synthetic and the coverage effects on the indices are multiple and difficult to summarise in simple terms. The outcome of randomising the stellar parameters is small compared to the other effects.
% the indices most affected by the synthetic effect are Fe5270 (\smiles\ overestimates it) and Fe4668 (\smiles\ underestimates it);
%the indices most affected by systematics from the coverage effect are B4000 (\pmiles\ overestimates it) and H$\gamma_F$ (\pmiles\ underestimates it).

%=================================================

\subsection{\label{sec_specfit}Spectral fitting of galaxies}

The model -- model comparisons of the previous section illustrate interesting differences between the sets of SPS models, but cannot be easily translated into the question that matters the most: how does the adoption of different spectral libraries change the age and metallicity of galaxies derived from their integrated light?

To address this question, we use the \pmiles, \smiles, and \cmiles models to derive stellar population parameters from the 
spectra of $\sim$1000 nearby galaxies ($0.04 < z < 0.06$).
We use the sample of \citet{gadotti09} which encompasses galaxies of different morphology: ellipticals, spirals (with classical or pseudo-bulges, with and without bars), and bulgeless discs. 
The spectra were obtained from the SDSS database \citep{sdss-dr2} and processed as in \citet{cg11}. 
We use the \textsc{Starlight} spectral fitting code \citep{cid+05} to infer the stellar population parameters, processing the
sample three times, once for each set of SPS models.
We refer the reader to the quoted work for details on the sample and technique.

We show in Fig.~\ref{fig_stresults} the results for two parameters related to the fit-quality and three parameters related to the stellar population. The parameters related to the fit-quality are \mychi (as in equation 1) and an average relative deviation \adev, defined as
\begin{equation}
{\rm \sf adev} = \frac{1}{N} \sum_{\lambda}
\left(\frac{|f^\textrm{model}_\lambda-f^\textrm{obs}_\lambda|}{f^\textrm{obs}_\lambda}\right),
\label{eqadev}
\end{equation}  
\noindent where N is the number of wavelength points in the spectrum, $f^\textrm{model}_\lambda$ is the model spectrum fitted by \textsc{Starlight}, and $f^\textrm{obs}_\lambda$ is the observed galaxy spectrum.

The synthetic effect can be inspected in the left hand side column of Fig.~\ref{fig_stresults}, which compares the results from \pmiles\ (y-axis) vs. those from \smiles\ (x-axis). We highlight that spectral fits performed with \smiles models tend to have larger
$\chi^2$ and \adev\ than those obtained with \pmiles models, even though there are no noticeable differences in the retrieved values of A$_V$ and  log (age). The values of \feh\ obtained with the \smiles models
are systematically lower than those obtained with the \pmiles models.
The differences in \feh\ correlate with \feh, lower metallicities showing larger differences.
%The combined impact of the Synthetic effect can be summarised as:
%\begin{enumerate}
%\item spectral fits performed with \synlib\-based SSPs tend to have larger $\chi^2$ and \adev\ than those obtained with MILES-based SSPs;
%\item there are no noticeable differences in the retrieved values of A$_V$;
%\item there are no noticeable differences in the retrieved mean ages of the galaxies, and;
%\item $[$Fe/H$]$ obtained from \synlib\-based SSPs are systematically lower than those obtained with MILES-based SSPs, and there is an evidence of correlation with [Fe/H] (lower metallicities having larger differences).
%\end{enumerate}

The coverage effect can be inspected in the right hand side column of Fig.~\ref{fig_stresults}, which shows the results from \cmiles\ (y-axis) vs. results from \smiles\ (x-axis) models. We notice that spectral fits performed with the \smiles and \cmiles models
have similar \adev, but $\chi^2$ is slightly larger for the \cmiles fits.
There are no noticeable differences in the retrieved values of Av.
Whereas the values of log(age) obtained from the \cmiles models are larger than for the \smiles models, there are no systematic differences in the retrieved mean [Fe/H], even though there are some outliers from the 1--to--1 line, which remain to be investigated.
%\pc{Paula, do you have information on the galaxy mass obtained from the \starlight fits? We could comment on their similarity or if any difference exixsts.} PC: not easily. 
%The combined impact of the Coverage effect can be summarised as:
%\begin{enumerate}
%\item spectral fits performed with \synlib\- and \textsc{Theoretical}-based SSPs are similar in \adev\, but $\chi^2$ are slightly larger in \textsc{Theoretical}-based fits;
%\item there are no noticeable differences in the retrieved values of Av;
%\item ages obtained from \textsc{Theoretical}-based SSPs are larger than the ones from \synlib\-based models, and;
%\item there are no systematic differences in the retrieved mean [Fe/H] of the galaxies, but there are some strong outliers to the 1--to--1 line, which remain to be investigated.
%\end{enumerate}

Table \ref{tab_stresults} shows the median residuals and IQRs for the parameters obtained from
\pmiles vs. \smiles fits (synthetic effect), and \cmiles vs.s \smiles fits (coverage effect).
In all cases, the systematic differences are inside the IQR ranges. 
We note that the median residuals on \feh\ and log(age) are of the same order of the IQR for the synthetic effect and the coverage effect, respectively.

\begin{table}
\begin{center}
\caption{\label{tab_stresults}Median residuals and IQR for spectral fits}
\begin{tabular}{lrrrr}
\hline
\multirow{3}{*}{Parameter} & \multicolumn{2}{c}{Synthetic effect}       & \multicolumn{2}{c}{Coverage effect} \\
                           & \multicolumn{2}{c}{(\pmiles\ -- \smiles)}  & \multicolumn{2}{c}{(\cmiles\ -- \smiles)} \\
                           & median & IQR  & median & IQR   \\
\hline
$\chi^2$ 	           & --0.05 & 0.13 &   0.04 & 0.06 \\	
\adev\                 & --0.16 & 0.29 &   0.09 & 0.10 \\
A$_V$	               &   0.00 & 0.04 &   0.01 & 0.05 \\
mean $\log{\rm (age)}$ &   0.00 & 0.12 &   0.11 & 0.14 \\
mean [Fe/H]	           &   0.13 & 0.13 & --0.01 & 0.12 \\
\hline
\end{tabular}
\end{center}
\end{table}

%% file: discussion.tex
\begin{figure*}
\includegraphics[width=0.8\linewidth]{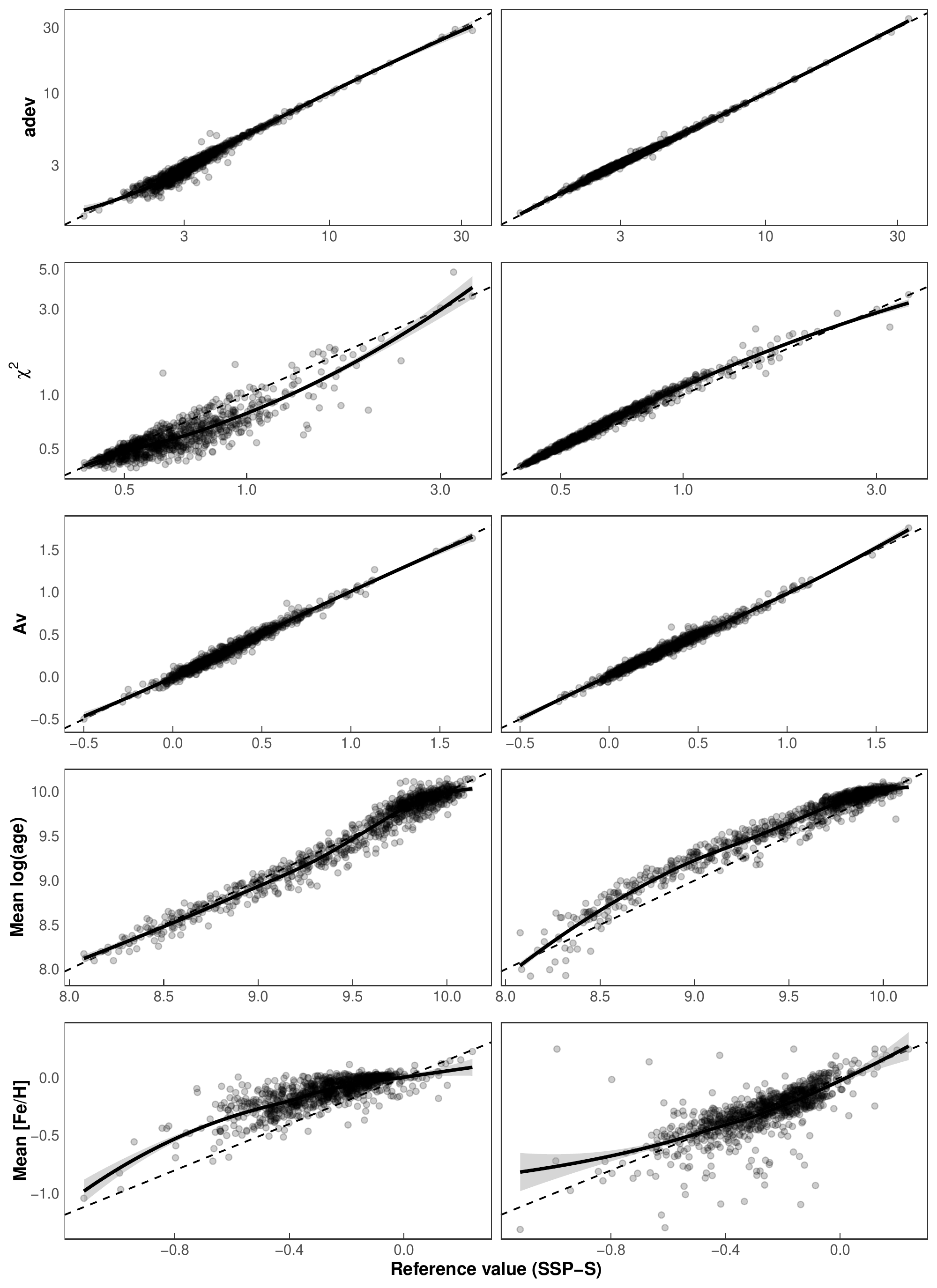}
\caption{\label{fig_stresults}Results from the spectral fitting of a sample of nearby galaxies. 
The left-side column shows the results obtained from \pmiles\ (y-axis) vs. \smiles\ (x-axis) models.
The right-side column shows the results obtained from \cmiles\ (y-axis) vs. \smiles\ (x-axis) models.
The different parameters
inferred are given in rows: $\chi^2$, \adev\ (Eq.~\ref{eqadev}), reddening Av, light-weighted log(age), and light-weighted \feh.
The smooth line corresponds to a LOESS (locally estimated scatterplot smoothing) regression to the points and is shown to guide the eye.}
\end{figure*}

\section{Discussion}
\label{sec_discussion}

\subsection{Semi-empirical vs. theoretical SPS models}
\emph{How do uncertainties in theoretical stellar libraries affect integrated colours and line indices measured in SPS models?}

\vspace{0.2cm}
Several works in the literature compare synthetic stellar spectra to empirical ones on a star-by-star basis \citep[e.g.][]{MC07,bertone+08}.
Recent work by \citet{martins+19} provides insight into how synthetic vs. empirical spectra compare at the level of the integrated properties of stellar populations.
These authors model integrated spectra of clusters combining CMDs and stellar libraries from different sources. 
Their conclusions depend on the wavelength range.
From 3900 to 6300\,\AA, or when considering specific spectral features, empirical libraries do better.
In the range 3525 to 6300\,\AA, a theoretical library outperformed the empirical ones in 70\% of their tests, 
essentially because the shape of the continuum is more in agreement with the observations in the theoretical case.

Our results add more information to the theoretical vs. empirical library debate. The SPS models built in this work are largely based on the
theoretical library by \citetalias{coelho14}.
Her fig. 10 illustrates the systematic differences between synthetic spectra for given {\teff} and \logg, and \miles spectra averaged over all stars in the library with the same atmospheric parameters (within uncertainties).
The two main possible reasons for these differences are: either they result from abundance patterns unaccounted for in the synthetic grid (such as variations of C and N due to dredge-up in giants stars), or they reflect true deficiencies in the spectral modelling, related to either the physics of the stellar atmosphere or to the adopted atomic opacities. In the present analysis we assume that all differences are due to the opacities. 

In Section \ref{sec_modelmodel}, colours and spectral indices from the \pmiles\ and \smiles models were compared to quantify how the inaccuracies in the opacities translate into the integrated light of the stellar population. 
The results from  Table \ref{tab_colors} show that the systematic effect on broad-band colours ranges from $-0.002$ in $u-g$ to $-0.018$ in $u-r$. The IQR of the distributions range from 0.017 in $g-r$ to 0.059 in $u-r$. For comparison, the reported accuracy (global rms dispersion) of the flux calibration for the \miles library is 0.013 mag in $B-V$. 

Comparing the spectral indices we get a closer view of the role played by the opacities. Figs.~\ref{fig_idx_distributions}, \ref{fig_idxscattera} and \ref{fig_idxboxplot} show a complex and index-dependent behaviour.
One may argue that the best performance of the \smiles\ models occurs for the indices H10, B4000, H$\delta_F$, CN$_2$, G4300, H$\gamma_F$, Fe4383, H$\beta$, Mg$_b$ and Fe5782, whereas the indices that deserve most attention from modellers are Fe5270, Fe5015, CaHK, Fe5709, Fe4383 and Fe4668 (the latter had already been pointed out by \citealt{knowles+19}).

It is unclear to what extent the results presented here can be safely applied to other theoretical libraries. 
Given that the broad-band colours depend mostly on averaged opacities and on the physics of the stellar atmosphere, we expect the results for the colours to be similar for other modern libraries of statistical fluxes \citep[e.g.][]{MARCS08, ATLASODFNEW}, or in high-resolution spectral libraries as long as the effect of predicted lines has been taken into account (see discussion in section 3 of \citetalias{coelho14}). 
On the other hand, the results for the spectral indices depend on the details of the specific opacities, and it is less likely that they can be reliably adopted for other synthetic libraries.
As such, the reader should see our results on the spectral indices
%present results on the spectral indices 
as guidelines for the use of the \citetalias{coelho14} library. 

%\url{http://www.physics.csbsju.edu/stats/box2.html} 

%=================================================
\subsection{The effect of the HRD coverage}
\emph{How does the non-ideal or poor coverage of the HRD by empirical libraries affect the predictions of SPS models?} 

\vspace{0.2cm}
Empirical libraries cannot cover the HRD homogeneously and completely due to observational constrains. 
We do not harbour in our Galaxy massive metal-poor stars, which are likely to be present in high redshift galaxies, nor can we cover the abundance patterns of galaxies with diverse star formation and chemical enrichment histories. By construction, with semi-empirical SPS models we can reproduce to a better degree the properties of populations similar to the solar neighbourhood than others. 

\citet{vazdekis+10} discuss this issue in the context of three empirical libraries \miles, STELIB \citep{stelib}, and Lick/IDS \citep{worthey+94}. They define the quality parameter $Q_n$ to quantify the reliability of their semi-empirical SPS models as a function of age, metallicity and IMF (see their Section 3.2). The authors conclude that the minimum age at which a model is reliable ranges from 60 Myr around solar metallicity to 10 Gyr in the most metal-poor regime, with a varying degree of quality $Q_n$. 

Here we retake this discussion comparing our \smiles\ and \cmiles models: the first mimics the HRD coverage of an empirical library, and the second covers the HRD completely and homogeneously for populations older than 30 Myr (cf. Fig.~\ref{fig_tefflogg}). Differences found in this experiment will be due to the incomplete HRD coverage of the empirical library, since \syncomil\ and \citetalias{coelho14} share the same codes and opacities.

The effects that the different HRD coverage have on colours can be seen
in Figs.~\ref{fig_colors_density}, \ref{fig_colors_boxplot} and \ref{fig_colors_distributions}, and are summarised in Table~\ref{tab_colors}. 
For all colours, except $r-i$, the coverage effect introduces systematic differences and variances larger than the synthetic effect. The largest difference occurs in $u-r$, with a systematic value of $-0.061$ and IQR of 0.085. The effect is largest in the most metal-poor regime, in accordance with the results of \citet{vazdekis+10}. This is expected since the coverage of the HRD is poorest at the lowest metallicities.

The effects on spectral indices are shown in Figs. \ref{fig_idx_distributions}, \ref{fig_idxscatterb}, 
and \ref{fig_idxboxplot}, and are difficult to generalise.
Some indices seem to be little affected (at least in comparison to the other effects) such as Fe5709 and Fe5782, while in other cases, such as B4000, the difference is prominent. 
Fig. \ref{fig_idxboxplot} shows that the most affected indices are,
at one extreme, B4000 and Mg1 (\smiles models, on average, underestimate the indices), and, at the other extreme, H$\delta$ and H$\gamma$ (\smiles models, on average, overestimate the indices). 

%=================================================
\subsection{Errors in the atmospheric parameters}
\emph{How do errors in the stellar atmospheric parameters translate into SPS models?}

\vspace{0.2cm}
\citet{percival_salaris09} performed an interesting investigation on the possible impact of systematic uncertainties in the atmospheric parameters on the integrated spectra of stellar populations. These authors simulated a systematic difference between the {\teff} scale of the isochrones and that of the stellar flux libraries, and considered errors within the typical offsets found in the literature (100\,K in \teff, 0.25 in \logg\ and 0.15 in \feh). 
Their results raised a caution, by showing that small systematic differences between the atmospheric parameter scales can mimic non-solar abundance ratios or multiple populations in the analysis of integrated spectra. If this result is confirmed, much of what the community is concluding in terms of abundance patterns in galaxies can be an artefact, due to offsets between parameter scales and not truly a tracer of different chemical evolution.

At the suggestion of the referee, we compare our results to \citet{percival_salaris09}. See their tables~1 and 2, where these authors show the impact of their tests on Lick indices for 2 choices of SSPs.
In Table~\ref{tab_compare_ps09}, we list $\tilde\Delta$idx and IQR($\Delta$idx) from our experiments for the same choices of age, metallicity and indices as in \citet{percival_salaris09}.
The effect of random errors on the stellar parameters is typically smaller than the effect introduced by adding a systematic difference of 100\,K in the \teff\ scale, and of comparable magnitude to the effect of adding systematic differences in \logg\ and \feh\ of 0.25 and 0.15, respectively.
The exact reason why a systematic difference in the \teff\ scale has a larger impact than adding random errors to the stellar parameters is difficult to trace. We can hypothesise that random errors tend to cancel out, while a systematic difference in the \teff\ scale is equivalent to selecting a younger (hotter) or older (cooler) isochrone when building an SSP model without changing the selection of stellar spectra.

In any case, a look at table~3 of \citet{percival_salaris09} shows that the deviations introduced by tampering with the atmospheric parameters is comparable to (or larger than) typical observational errors on the spectral indices. As such, these effects cannot be safely neglected.
Here we complement this investigation with the \rmiles models, simulating the effect of \emph{random} rather than systematic variations in the atmospheric parameters. Our tests reveal that both in colours (Figs.  \ref{fig_colors_boxplot} and \ref{fig_colors_distributions}, Table \ref{tab_colors}) and spectral indices (Fig. \ref{fig_idx_distributions} and \ref{fig_idxboxplot}), the effect of randomising the atmospheric parameters is small compared to the other effects. We conclude then that random errors in the stellar parameters (within the uncertainties adopted here, see Section \ref{sec_method}) are not a major source of concern for current SPS models. 

\begin{table}
\caption{\label{tab_compare_ps09}$\tilde\Delta$idx and IQR($\Delta$idx) for Lick indices (Z = 0.017)}
\begin{center}
\begin{tabular}{lrcrc}
\hline
      & \multicolumn{2}{c}{Age = 4 Gyr}      & \multicolumn{2}{c}{Age = 14 Gyr}     \\
Index & $\tilde\Delta$idx & IQR($\Delta$idx) & $\tilde\Delta$idx & IQR($\Delta$idx) \\
\hline
 H$_{\delta F}$ &  0.101 & 0.071 &  0.041 & 0.059\\
 H$_{\gamma F}$ &  0.131 & 0.086 &  0.090 & 0.196\\
 CN1            & -0.003 & 0.007 &  0.002 & 0.006\\
 CN2            & -0.002 & 0.008 &  0.006 & 0.005\\
 Ca4227         & -0.040 & 0.137 &  0.082 & 0.370\\
 G4300          &  0.019 & 0.025 &  0.114 & 0.147\\
 Fe4383         & -0.166 & 0.316 &  0.174 & 0.264\\
 Ca4455         & -0.038 & 0.069 &  0.047 & 0.043\\
 Fe4531         & -0.004 & 0.087 &  0.051 & 0.033\\
 H$_\beta$      &  0.045 & 0.035 &  0.012 & 0.085\\
 Fe5015         &  0.025 & 0.163 &  0.164 & 0.182\\
 Mg1            & -0.001 & 0.009 & -0.002 & 0.018\\
 Mg2            & -0.004 & 0.014 &  0.005 & 0.031\\
 Mgb            & -0.154 & 0.121 &  0.153 & 0.387\\
 Fe5270         & -0.037 & 0.141 &  0.085 & 0.119\\
 Fe5335         & -0.018 & 0.140 &  0.054 & 0.227\\
 Fe5406         & -0.021 & 0.125 &  0.034 & 0.130\\
 Fe5709         &  0.005 & 0.066 &  0.035 & 0.006\\
 Fe5782         &  0.009 & 0.061 &  0.017 & 0.059\\
 NaD            & -0.009 & 0.137 &  0.001 & 0.445\\
 TiO1           &  0.001 & 0.004 & -0.001 & 0.006\\
 TiO2           &  0.001 & 0.004 &  0.000 & 0.007\\
\hline
\end{tabular}
\end{center}
\end{table}

\subsection{Inferring stellar-population parameters from integrated light}
\emph{To what extent does the choice of an empirical or a theoretical library in the SPS model affect age, metallicity and reddening estimates from integrated light?}

\vspace{0.2cm}
Possibly, the most important result of the experiments performed in this work is to estimate to what extent stellar population parameters -- age, metallicity, reddening -- derived from SPS models vary if we adopt in the model an empirical or a theoretical library. 
Given our current inability to have both complete coverage of the HRD with high quality stellar spectra, what should one favour: complete coverage of the HRD (theoretical library) or accurate spectra on a star-by-star basis (empirical library)?

There are several ways proposed in the literature to derive the stellar population parameters from integrated light, but for the purpose of the present paper we choose to investigate results obtained from spectral fitting. To that end we adopt the widely used code \textsc{Starlight} \citep{cid+05} to fit a sample of nearby galaxies \citep{gadotti09, cg11}. In Section \ref{sec_modelmodel} we show that the synthetic and coverage effect impact both colours and spectral features. 
A code such as \textsc{Starlight}, which fits the continuum and the spectral features together, is a good option to evaluate in a global manner how the stellar library of choice will influence the derived
galaxy properties.

Results are shown in Fig. \ref{fig_stresults} and summarised in Table \ref{tab_stresults}. The first noticeable feature is that the use of synthetic spectra tends to increase \mychi\ and \adev\ for the fit, i.e., the fits tend to be statistically worse.
%(left-side column, panels \adev\ and \mychi). 
This is not surprising, and reminds us that atomic and molecular opacities still need improvement. 
Nevertheless, and to some extent a surprising result, there are no important differences in the reddening or the ages derived using \pmiles\ and \smiles\ models.
%(left-side column, panel log(age)). 
There is only a hint that for intermediate ages (9 $\lesssim$ log(age) $\lesssim$ 9.5), the values obtained with \smiles\ are slightly lower than for \pmiles. 
On the other hand, the effect of HRD coverage on inferred ages is more significant: %(right-side column, panel log(age)): 
the use of models with non-optimal HRD coverage
%(incomplete coverage of the HRD) 
underestimates virtually all ages, the effect being more pronounced for log(age) $\sim$ 9.0.

The effects on the derived metallicities are shown in the bottom panels of Fig. \ref{fig_stresults}. \smiles\ models 
%(left-side column)
recover lower metallicites than \pmiles\ models, with a median difference of 0.13. This result is consistent with the top-panel of Fig. \ref{fig_idxboxplot}, which shows more indices below the zero-line %(\smiles\ predictions are stronger than the ones from \pmiles) 
than around or above it. 
This is also in agreements with results from the literature that show
that in general synthetic stellar spectra are stronger-lined than observed stellar spectra \citep[e.g.][]{MC07, coelho14}. 
There is an indication that the difference is stronger towards lower-metallicities, but the origin of this tendency is unclear. 
%On the other hand, 
Having a complete coverage of the HRD 
%(\cmiles\ vs \smiles\ results, right-side bottom panel) 
does not seem to affect the derived mean metallicities, although the dispersion increases.

%How to bin in ages? 

%Cid+05 (sec 2.2.2) uses young (logage < 8), intermediate ([8,9]), old (>9). "These age ranges were defined on the basis of simulations, by seeking which combinations of the xj produce smaller input minus output residuals".

%Walcher+15 (sec 4.1) uses young (age < 1.5Gyr), intermediate ([1.5, 9.5]), old (> 9.5Gyr).
%\end{comment}

%% file: conclusions.tex
\section{Conclusions}
\label{sec_conclusions}

It has been traditionally accepted that SPS models tailored to study young populations favour the use of theoretical stellar libraries, 
due to their better HRD and wavelength coverage, 
while models targeting intermediate and old populations favour empirical stellar libraries, whose detailed spectral features are more reliable than in synthetic spectra. 
%Models for a wide range of applications resort to combining both types of libraries for a better sampling of parameter space. 

In this paper we perform experiments with especially built SPS models to investigate and quantify the impact of the choice of stellar library type on: 
{\it (i)} the predicted colours and magnitudes of evolving simple stellar populations, and 
{\it (ii)} the inferred age, metallicity and redenning of a galaxy from spectral fits to its integrated spectrum. 

We build a new synthetic stellar library which mimics the HRD coverage of a widely used empirical library (Section \ref{sec_syncomil}). During this exercise we identified 71 \miles stars (Table \ref{badstars}) whose spectra we consider not suitable for stellar population modelling, and we recommend not to use these spectra in SPS models.

We build SPS models using different stellar libraries (Section \ref{sec_method}). We name \emph{synthetic effect} the differences introduced in the SPS model predictions by using a theoretical vs. an empirical library, for identical HRD coverage.
Analogously, we name \emph{coverage effect} the differences introduced in the SPS models by using libraries with a limited vs. a complete HRD coverage. 
The results of our tests are given in Section \ref{sec_results}, and further discussed in Section \ref{sec_discussion}. 
The lessons learned are as follows.

In the majority of the cases the coverage effect is responsible for the larger deviations in the predicted colours, especially those involving the $u$ band, for which the lack of hot stars in the empirical library is more
noticeable.

For spectral indices, the coverage and synthetic effects are comparable for most features in the wavelength range considered. 
Some indices are more sensitive to the synthetic effect (e.g. CaHK, Ca4227, Fe4668, Fe5270), indicating spectral regions that deserve more attention from the modellers to improve the theoretical grids.  
Other indices are more sensitive to the coverage effect (H8, B4000, G4300), and we warn users of semi-empirical SPS models to take the predictions for these indices with caution. 

We test the effects that random errors in the atmospheric parameters of the stars have on the SPS model predictions, and conclude that, for the typical errors in atmospheric parameters adopted in our experiments,
this effect is minor in comparison to the other effects.

We use different SPS models to infer the stellar population parameters of a sample of nearby galaxies by spectral fitting. 
The synthetic effect is null for the mean light-weighted log(age) of the galaxies, but metallicity is underestimated by an average of $\sim$ 0.13. %The difference is smaller at solar metallicity, and larger towards lower metallicities.
The coverage effect results in galaxy ages being underestimated (for all ages but more strongly around log(age) $\sim$ 9), but has little impact on the inferred metallicities other than increasing the dispersion. 
The inferred reddening is virtually unaffected by either effect.

Strictly speaking, our results are valid for the specific HRD coverage of the \miles library and the synthetic grid of \citetalias{coelho14}. Nonetheless, we believe that our conclusions on the coverage effect will not change much in the near future, given that \miles already has an optimal coverage of the HRD for solar neighbourhood stars. 
More stars from different populations (LMC, SMC, galactic bulge) need to be introduced in the library to possibly produce a significant change, at the expense of lowering the spectral resolution and/or the SNR, due to current observational constraints. We expect that the synthetic effect on optical colours 
will be similar for most modern theoretical libraries currently in use. 
The effect on the spectral indices is more sensitive to the specific choice of the synthetic grid and atomic and molecular opacities.

Overall, we conclude that in several instances a sparse coverage of the HRD can introduce larger errors than the inaccuracies of the synthetic spectra. As such, one has to decide with care which kind of SPS models -- semi-empirical or fully theoretical -- to favour, depending on the application. As of now, SPS models built on current theoretical grids of synthetic spectra are very competitive, for all ages.

%We conclude that current theoretical grids of synthetic spectra are quite competitive. SPS models based on these libraries can be used to study stellar populations of all ages. 
% I am unsure to write the sentence above because it claims for theoretical grids in general, while we only tested mine. So i tried a milder concluding line.

%% file: acknowledgements.tex
\section*{Acknowledgements}
PC acknowledges support from Conselho Nacional de Desenvolvimento Cient\'ifico e Tecnol\'ogico
(CNPq 310041/2018-0). 
PC and GB acknowledge support from Funda\c c\~ao de Amparo \`a Pesquisa do Estado de S\~ao Paulo
 through projects FAPESP 2017/02375-2 and 2018/05392-8. PC and SC acknowledge support from USP-COFECUB 2018.1.241.1.8-40449YB.
GB acknowledges financial support from the National Autonomous University of Mexico (UNAM) through grant DGAPA/PAPIIT IG100319 and from CONACyT through grant CB2015-252364. 

PC thanks A. Ederoclite for his patience and support during the development of this work. 
PC thanks P. Prugniel for innumerous discussions about stellar parameters and spectra, and for providing his derived E(B-V) values for MILES spectra.
The authors thank P. S\'anchez-Bl\'azquez for providing the error spectra for MILES, and
R. Peletier for providing information on caveats on some MILES spectra.

Funding for the SDSS and SDSS-II has been provided by the Alfred P. Sloan Foundation, the Participating Institutions, the National Science Foundation, the U.S. Department of Energy, the National Aeronautics and Space Administration, the Japanese Monbukagakusho, the Max Planck Society, and the Higher Education Funding Council for England. The SDSS Web Site is
\url{http://www.sdss.org/}.

The SDSS is managed by the Astrophysical Research Consortium for the Participating Institutions. The Participating Institutions are the American Museum of Natural History, Astrophysical Institute Potsdam, University of Basel, University of Cambridge, Case Western Reserve University, University
of Chicago, Drexel University, Fermilab, the Institute for Advanced Study, the Japan Participation Group, Johns Hopkins University, the Joint Institute for Nuclear Astrophysics, the Kavli Institute for Particle Astrophysics and Cosmology, the Korean Scientist Group, the Chinese Academy of Sciences
(LAMOST), Los Alamos National Laboratory, the Max-Planck-Institute for Astronomy (MPIA), the Max-Planck-Institute for Astrophysics (MPA), New Mexico State University, Ohio State University, University of Pittsburgh, University of Portsmouth, Princeton University, the United States Naval Observatory, and the University of Washington.

%% file: online_material.tex
\appendix
\onecolumn
\section{Online-only material}
\label{online_tabs}

\begin{longtable}{clrrrc}
\caption{\label{tab_atmpars}Atmospheric parameters used as input values in the computation of \syncomil (Section 2). References for the source of the stellar parameters are given in column (f): 
$^a$\citet{MILES2}; 
$^b$\citet{prugniel+11}; 
$^c$\citet{sharma+16}; 
$^d$Computed with different parameters than proposed in the literature to ensure model convergence. % (the values in parenthesis indicate the original determination). 
Stars with MILES ID not listed in this table were not used in the SPS models of this work and are reported separately in Table~\ref{tab_discarded}.
}\\

\toprule\bfseries 
MILES ID & \bfseries Star & \bfseries T$_{\rm eff}$ & \bfseries log g & \bfseries [Fe/H] & \bfseries Reference \\
(a) & (b) & (c) & (d) & (e) & (f) \\ 

\midrule 
\endfirsthead
\caption{continued.}\\
\hline
MILES ID & \bfseries Star & \bfseries T$_{\rm eff}$ & \bfseries log g & \bfseries [Fe/H] & \bfseries Reference \\
(a) & (b) & (c) & (d) & (e) & (f) \\ 
\hline
\endhead
\bottomrule \endfoot   

%    \csvreader[
%    	head to column names,
%    	late after line=\\,
%    	late after last line=,
%    	before reading={\catcode`\#=12},after reading={\catcode`\#=6}
%    	]{ok_stars.csv}{}
%    	{\milesid & \expUScore{\star} & \teff & \logg & \feh & \ref}

  1 & HD224930 & 5411 & 4.19 & -0.78 & b\\
  2 & HD225212 & 4117 & 0.68 & 0.14 & c\\
  3 & HD225239 & 5559 & 3.72 & -0.51 & b\\
  4 & HD000004 & 6779 & 3.87 & 0.21 & b\\
  5 & HD000249 & 4731 & 2.83 & -0.31 & c\\
  6 & HD000319 & 8641 & 4.29 & -0.35 & b\\
  7 & HD000400 & 6190 & 4.15 & -0.22 & b\\
  8 & HD000245 & 5749 & 4.13 & -0.57 & b\\
  9 & HD000448 & 4770 & 2.61 & 0.02 & c\\
  10 & BD+130013 & 5000 & 3.00 & -0.75 & b\\
  11 & HD000886 & 20454 & 3.79 & -0.03 & b\\
  12 & HD001326b & 3571 & 4.81 & -0.57 & d\\
  13 & HD001461 & 5666 & 4.21 & 0.19 & b\\
  14 & HD001918 & 4888 & 2.44 & -0.40 & b\\
  15 & HD002628 & 7335 & 3.95 & -0.09 & b\\
  16 & HD002665 & 4986 & 2.28 & -1.96 & b\\
  17 & HD002796 & 4837 & 1.78 & -2.23 & b\\
  18 & HD002857 & 8000 & 2.70 & -1.50 & b\\
  19 & HD003008 & 4364 & 0.68 & -1.83 & c\\
  20 & HD003369 & 16005 & 3.71 & 0.04 & b\\
  21 & HD003360 & 20375 & 3.80 & -0.04 & b\\
  22 & HD003567 & 6094 & 4.18 & -1.14 & b\\
  23 & HD003546 & 4945 & 2.36 & -0.66 & b\\
  24 & HD003574 & 4019 & 1.13 & 0.01 & c\\
  25 & HD003651 & 5211 & 4.48 & 0.21 & b\\
  26 & HD003795 & 5345 & 3.72 & -0.63 & b\\
  27 & HD003883 & 7616 & 3.81 & 0.68 & b\\
  28 & HD004307 & 5773 & 3.97 & -0.24 & b\\
  30 & HD004539 & 25000 & 5.40 & 0.00 & b\\
  31 & HD004628 & 4964 & 4.65 & -0.23 & b\\
  32 & HD004656 & 3934 & 1.67 & -0.13 & c\\
  33 & HD004744 & 4590 & 2.32 & -0.74 & c\\
  34 & HD004906 & 5157 & 3.58 & -0.66 & b\\
  35 & HD005268 & 4904 & 2.35 & -0.57 & b\\
  36 & HD005384 & 3933 & 1.79 & 0.18 & c\\
  37 & HD005395 & 4845 & 2.45 & -0.43 & c\\
  38 & HD005780 & 3917 & 1.64 & -0.71 & c\\
  39 & HD005916 & 4954 & 2.31 & -0.75 & b\\
  40 & HD006186 & 4865 & 2.36 & -0.35 & b\\
  41 & HD006203 & 4506 & 2.20 & -0.41 & c\\
  42 & HD006268 & 4571 & 1.13 & -2.63 & c\\
  43 & HD006229 & 5181 & 2.50 & -1.14 & b\\
  46 & HD006582 & 5323 & 4.33 & -0.79 & b\\
  47 & HD006805 & 4505 & 2.48 & 0.07 & c\\
  48 & HD005848 & 4451 & 2.25 & 0.09 & c\\
  49 & HD006834 & 6482 & 4.22 & -0.58 & b\\
  50 & HD006755 & 5097 & 2.53 & -1.58 & b\\
  51 & HD006833 & 4502 & 1.78 & -0.84 & c\\
  52 & HD007106 & 4678 & 2.55 & -0.02 & c\\
  53 & HD007351 & 3619 & 0.36 & -0.35 & c\\
  54 & HD007374 & 12247 & 4.16 & 0.16 & b\\
  55 & HD007595 & 4327 & 1.82 & -0.68 & c\\
  56 & HD007672 & 4939 & 2.78 & -0.42 & b\\
  57 & HD008724 & 4792 & 1.76 & -1.63 & c\\
  58 & HD008829 & 7129 & 4.10 & -0.17 & b\\
  59 & HD009138 & 4041 & 1.89 & -0.50 & c\\
  60 & HD009356 & 6800 & 4.24 & -0.80 & b\\
  61 & HD009562 & 5766 & 3.89 & 0.14 & b\\
  63 & HD009826 & 6139 & 4.06 & 0.11 & b\\
  65 & HD010380 & 4154 & 1.85 & -0.24 & c\\
  66 & HD010307 & 5875 & 4.28 & 0.06 & b\\
  67 & HD010700 & 5348 & 4.39 & -0.46 & b\\
  69 & HD010780 & 5406 & 4.63 & 0.15 & b\\
  70 & HD010975 & 4843 & 2.44 & -0.23 & c\\
  71 & HD011257 & 7103 & 4.08 & -0.27 & b\\
  72 & HD011397 & 5526 & 4.24 & -0.58 & b\\
  73 & HD011964 & 5272 & 3.85 & 0.05 & b\\
  75 & HD012438 & 4937 & 2.35 & -0.73 & b\\
  76 & HD013043 & 5823 & 4.11 & 0.06 & b\\
  77 & BD+290366 & 5666 & 4.25 & -0.95 & b\\
  78 & HD013267 & 15500 & 2.57 & -0.10 & b\\
  79 & HD013555 & 6515 & 4.07 & -0.16 & b\\
  80 & HD013520 & 4023 & 1.61 & -0.27 & c\\
  81 & BD-010306 & 5723 & 4.28 & -0.89 & b\\
  82 & HD013783 & 5516 & 4.37 & -0.49 & b\\
  83 & HD014221 & 6619 & 4.07 & -0.17 & b\\
  84 & HD014802 & 5777 & 3.89 & -0.07 & b\\
  85 & HD014829 & 8750 & 3.15 & -1.57 & b\\
  86 & HD014938 & 6275 & 4.22 & -0.25 & b\\
  88 & HD015798 & 6527 & 4.07 & -0.12 & b\\
  89 & HD016031 & 6039 & 4.09 & -1.63 & b\\
  90 & HD016234 & 6225 & 4.18 & -0.19 & b\\
  91 & HD016232 & 6314 & 4.29 & 0.11 & b\\
  92 & HD016673 & 6260 & 4.30 & 0.00 & b\\
  93 & HD016784 & 5782 & 4.08 & -0.68 & b\\
  94 & BD+460610 & 5889 & 4.13 & -0.86 & b\\
  95 & G004-036 & 6073 & 4.20 & -1.66 & b\\
  96 & HD016901 & 5345 & 0.85 & 0.00 & a\\
  97 & HD017081 & 12722 & 4.20 & 0.28 & b\\
  98 & HD017361 & 4630 & 2.53 & 0.02 & c\\
  99 & HD017491 & 3258 & 0.65 & -0.15 & c\\
  100 & HD017382 & 5339 & 4.64 & 0.17 & b\\
  101 & HD017548 & 6013 & 4.20 & -0.53 & b\\
  102 & HD017378 & 8477 & 1.25 & 0.00 & b\\
  103 & HD018191 & 3199 & 0.78 & -0.05 & c\\
  105 & HD018907 & 5069 & 3.43 & -0.65 & b\\
  106 & HD019445 & 5900 & 4.20 & -2.07 & b\\
  108 & HD019373 & 5947 & 4.15 & 0.11 & b\\
  109 & HD019994 & 6051 & 4.02 & 0.16 & b\\
  110 & HD020041 & 11509 & 2.01 & 0.23 & b\\
  111 & HD020512 & 5267 & 3.81 & -0.13 & b\\
  112 & HD020619 & 5710 & 4.47 & -0.18 & b\\
  113 & HD020630 & 5733 & 4.45 & 0.12 & b\\
  114 & HD020893 & 4363 & 2.26 & 0.07 & c\\
  115 & BD+430699 & 4736 & 4.72 & -0.38 & c\\
  116 & HD021017 & 4419 & 2.67 & 0.07 & c\\
  117 & HD021197 & 4376 & 4.50 & 0.13 & c\\
  118 & HD021581 & 4825 & 2.00 & -1.70 & a\\
  119 & BD+660268 & 5300 & 4.20 & -2.00 & b\\
  120 & HD022049 & 5115 & 4.72 & 0.05 & b\\
  121 & HD022484 & 5987 & 4.07 & -0.05 & b\\
  122 & HD021910 & 4798 & 2.48 & -0.45 & c\\
  123 & HD022879 & 5870 & 4.23 & -0.80 & b\\
  124 & HD023249 & 5020 & 3.73 & 0.08 & b\\
  125 & HD023261 & 5165 & 4.56 & 0.24 & b\\
  126 & HD023194 & 8031 & 4.00 & -0.17 & b\\
  127 & HD023439a & 5181 & 4.47 & -0.90 & b\\
  128 & HD023439b & 4786 & 4.63 & -1.09 & c\\
  129 & HD023607 & 7586 & 3.97 & -0.03 & b\\
  130 & HD023841 & 4306 & 2.05 & -0.66 & c\\
  131 & HD023924 & 7776 & 3.94 & 0.07 & b\\
  132 & HD024616 & 5014 & 3.16 & -0.71 & b\\
  133 & HD024341 & 5405 & 3.71 & -0.62 & b\\
  134 & HD024421 & 6168 & 4.20 & -0.29 & b\\
  135 & HD024451 & 4418 & 4.57 & -0.09 & c\\
  136 & HD025329 & 4964 & 4.60 & -1.58 & c\\
  137 & HD025532 & 5600 & 2.50 & -1.35 & b\\
  138 & HD025673 & 5112 & 4.54 & -0.40 & b\\
  139 & HD026297 & 4497 & 1.11 & -1.79 & c\\
  141 & HD026322 & 7008 & 3.94 & 0.13 & b\\
  143 & HD284248 & 6113 & 4.14 & -1.55 & b\\
  144 & BD-060855 & 5442 & 4.60 & -0.69 & b\\
  145 & HD026965 & 5114 & 4.41 & -0.26 & b\\
  146 & HD285690 & 4971 & 4.70 & 0.18 & c\\
  147 & HD027126 & 5425 & 4.14 & -0.38 & b\\
  148 & HD027295 & 11034 & 3.99 & -0.11 & b\\
  149 & HD027371 & 4995 & 2.76 & 0.15 & b\\
  150 & HD027771 & 5285 & 4.59 & 0.27 & b\\
  151 & HD027819 & 7871 & 3.89 & -0.06 & b\\
  152 & HD028305 & 4964 & 2.72 & 0.20 & b\\
  153 & HD285773 & 5348 & 4.56 & 0.25 & b\\
  154 & HD028946 & 5314 & 4.55 & -0.10 & b\\
  155 & HD028978 & 8864 & 3.42 & -0.26 & b\\
  156 & HD029065 & 4034 & 1.69 & -0.35 & c\\
  157 & HD029139 & 3851 & 1.62 & -0.13 & c\\
  158 & BD+501021 & 5081 & 4.48 & -0.65 & b\\
  159 & BD+450983 & 5155 & 4.45 & -0.22 & b\\
  160 & HD030743 & 6484 & 4.16 & -0.34 & b\\
  161 & HD030504 & 4022 & 1.75 & -0.50 & c\\
  162 & HD030649 & 5791 & 4.21 & -0.48 & b\\
  163 & HD031128 & 5949 & 4.18 & -1.45 & b\\
  164 & HD030959 & 3562 & 0.37 & -0.09 & c\\
  165 & HD030834 & 4194 & 1.61 & -0.35 & c\\
  166 & HD031295 & 8822 & 4.11 & -0.73 & b\\
  167 & HD031767 & 4367 & 1.50 & -0.02 & c\\
  168 & HD032147 & 4650 & 4.58 & 0.16 & c\\
  169 & HD032655 & 7114 & 3.47 & 0.23 & b\\
  170 & HD033256 & 6477 & 4.15 & -0.27 & b\\
  171 & HD033276 & 7223 & 3.80 & 0.22 & b\\
  172 & HD293857 & 5628 & 4.38 & 0.10 & b\\
  173 & HD033608 & 6461 & 4.03 & 0.21 & b\\
  174 & HD034538 & 4870 & 2.96 & -0.36 & b\\
  175 & MS0515.4-0710 & 5206 & 4.41 & 0.05 & c\\
  176 & HD034411 & 5842 & 4.16 & 0.08 & b\\
  177 & HD035155 & 3637 & 0.09 & -0.53 & c\\
  178 & HD035179 & 4942 & 2.48 & -0.60 & c\\
  179 & HD035369 & 4915 & 2.49 & -0.24 & b\\
  180 & HD035296 & 6171 & 4.31 & 0.01 & b\\
  182 & HD036003 & 4378 & 4.58 & -0.15 & c\\
  183 & HD036395 & 3579 & 4.72 & -0.05 & c\\
  184 & HD037160 & 4754 & 2.64 & -0.64 & c\\
  185 & HD037792 & 6509 & 4.17 & -0.54 & b\\
  186 & HD037536 & 3775 & 0.22 & 0.14 & c\\
  187 & HD037828 & 4505 & 1.36 & -1.41 & c\\
  188 & HD037394 & 5279 & 4.60 & 0.20 & b\\
  189 & HD037984 & 4445 & 2.15 & -0.52 & c\\
  190 & HD038392 & 4941 & 4.75 & -0.02 & c\\
  191 & HD038393 & 6316 & 4.23 & -0.09 & b\\
  192 & HD038007 & 5705 & 3.98 & -0.31 & b\\
  193 & HD038545 & 8673 & 3.68 & -0.48 & b\\
  194 & HD038751 & 4776 & 2.71 & 0.11 & c\\
  195 & HD038656 & 4943 & 2.55 & -0.15 & b\\
  196 & HD039364 & 4660 & 2.46 & -0.74 & c\\
  197 & HD039853 & 3858 & 1.58 & -0.61 & c\\
  198 & HD039833 & 5869 & 4.39 & 0.18 & b\\
  199 & HD039801 & 3666 & 0.20 & 0.07 & c\\
  200 & HD039970 & 12006 & 2.13 & 0.19 & b\\
  201 & HD040657 & 4264 & 1.81 & -0.73 & c\\
  202 & HD250792 & 5554 & 4.33 & -1.01 & b\\
  203 & HD041312 & 4044 & 1.77 & -0.75 & c\\
  205 & HD251611 & 5382 & 3.40 & -1.44 & b\\
  206 & HD041692 & 14800 & 3.30 & -0.01 & b\\
  207 & HD041636 & 4688 & 2.42 & -0.29 & c\\
  208 & HD042182 & 5041 & 4.63 & 0.13 & b\\
  209 & HD041597 & 4607 & 2.01 & -0.54 & c\\
  211 & HD042543 & 3707 & 0.17 & 0.18 & c\\
  213 & HD043318 & 6330 & 4.04 & -0.07 & b\\
  214 & BD+371458 & 5450 & 3.38 & -2.12 & b\\
  215 & HD043380 & 4521 & 2.40 & -0.05 & c\\
  216 & HD044007 & 4987 & 2.33 & -1.53 & b\\
  217 & HD043378 & 9284 & 4.05 & -0.27 & b\\
  218 & HD043947 & 5983 & 4.28 & -0.27 & b\\
  219 & HD044030 & 4026 & 1.75 & -0.51 & c\\
  222 & HD045282 & 5309 & 3.19 & -1.42 & b\\
  223 & HD045829 & 4499 & 0.56 & 0.11 & c\\
  224 & HD046341 & 5835 & 4.28 & -0.66 & b\\
  225 & HD047205 & 4728 & 3.10 & 0.19 & c\\
  226 & HD046703 & 6250 & 1.00 & -1.50 & b\\
  227 & HD047914 & 3938 & 1.79 & 0.04 & c\\
  228 & HD048329 & 4496 & 0.75 & 0.08 & c\\
  229 & HD048433 & 4464 & 2.01 & -0.22 & c\\
  230 & BD+151305 & 4901 & 4.66 & 0.11 & b\\
  231 & HD048565 & 6030 & 3.94 & -0.63 & b\\
  232 & HD048682 & 6088 & 4.28 & 0.11 & b\\
  233 & HD049161 & 4168 & 1.82 & 0.17 & c\\
  234 & HD049331 & 3830 & 0.44 & 0.13 & c\\
  235 & HD049933 & 6647 & 4.19 & -0.45 & b\\
  236 & HD050778 & 4009 & 1.77 & -0.43 & c\\
  237 & HD050420 & 7319 & 3.75 & 0.11 & b\\
  238 & HD051440 & 4313 & 1.72 & -0.66 & c\\
  239 & HD052005 & 4071 & 0.66 & 0.08 & c\\
  240 & HD052973 & 5657 & 1.12 & 0.09 & b\\
  241 & HD053927 & 4911 & 4.71 & -0.28 & b\\
  242 & HD054605 & 6268 & 0.97 & 0.10 & a\\
  243 & BD+371665 & 5128 & 3.50 & -0.65 & b\\
  244 & HD054810 & 4726 & 2.58 & -0.33 & c\\
  245 & HD054719 & 4405 & 2.14 & 0.13 & c\\
  247 & HD055693 & 5773 & 4.16 & 0.25 & b\\
  248 & HD055575 & 5811 & 4.19 & -0.38 & b\\
  249 & HD056274 & 5769 & 4.40 & -0.53 & b\\
  250 & HD056577 & 3904 & 0.48 & 0.12 & c\\
  253 & HD057264 & 4599 & 2.41 & -0.40 & c\\
  254 & HD058207 & 4806 & 2.55 & -0.11 & c\\
  255 & HD058551 & 6306 & 4.27 & -0.42 & b\\
  257 & HD059374 & 5873 & 4.21 & -0.82 & b\\
  258 & BD+241676 & 6230 & 3.81 & -2.55 & b\\
  259 & HD059984 & 5973 & 4.07 & -0.68 & b\\
  260 & HD059881 & 7623 & 3.64 & 0.15 & b\\
  261 & HD060219 & 5900 & 1.83 & -0.49 & a\\
  262 & HD060179 & 9550 & 3.83 & -0.13 & b\\
  263 & LHS1930 & 5420 & 4.33 & -1.11 & b\\
  264 & HD060522 & 3834 & 1.54 & -0.02 & c\\
  265 & HD061064 & 6646 & 3.69 & 0.27 & b\\
  266 & BD-011792 & 5131 & 3.38 & -0.81 & b\\
  267 & HD061606 & 4890 & 4.67 & 0.05 & b\\
  268 & HD061772 & 4096 & 1.46 & -0.04 & c\\
  269 & HD061603 & 3953 & 1.43 & 0.19 & c\\
  270 & HD061935 & 4802 & 2.57 & -0.06 & c\\
  271 & HD061913 & 3568 & 0.55 & 0.06 & c\\
  272 & BD+002058a & 6096 & 4.17 & -1.22 & b\\
  273 & HD062345 & 5029 & 2.61 & -0.01 & b\\
  274 & HD062301 & 5933 & 4.12 & -0.62 & b\\
  275 & HD062721 & 3913 & 1.81 & -0.36 & c\\
  276 & HD063302 & 4264 & 0.12 & 0.12 & c\\
  277 & HD063352 & 4149 & 1.71 & -0.60 & c\\
  278 & BD-182065 & 4878 & 2.43 & -0.71 & c\\
  279 & HD064332 & 3515 & 0.19 & -0.11 & c\\
  280 & HD064090 & 5405 & 4.19 & -1.65 & b\\
  281 & HD063791 & 4822 & 1.94 & -1.62 & c\\
  282 & HD064606 & 5302 & 4.42 & -0.76 & b\\
  283 & HD064488 & 8837 & 3.65 & -0.36 & b\\
  284 & HD065228 & 5861 & 1.24 & 0.06 & b\\
  285 & HD065583 & 5281 & 4.33 & -0.65 & b\\
  286 & HD065714 & 4983 & 2.50 & 0.18 & b\\
  287 & HD065953 & 3986 & 1.73 & -0.34 & c\\
  288 & HD065900 & 9235 & 3.69 & -0.16 & b\\
  289 & HD066141 & 4265 & 1.98 & -0.51 & c\\
  290 & HD066573 & 5680 & 4.26 & -0.58 & b\\
  291 & HD067523 & 6810 & 3.59 & 0.60 & b\\
  292 & HD067228 & 5732 & 3.84 & 0.12 & b\\
  293 & BD+800245 & 5509 & 3.74 & -1.85 & b\\
  294 & HD068284 & 5945 & 3.97 & -0.52 & b\\
  295 & HD069267 & 4068 & 1.49 & -0.19 & c\\
  296 & HD069611 & 5773 & 4.09 & -0.58 & b\\
  297 & HD069830 & 5412 & 4.49 & 0.04 & b\\
  298 & HD233511 & 6005 & 4.13 & -1.52 & b\\
  299 & HD069897 & 6328 & 4.18 & -0.23 & b\\
  300 & HD070272 & 3921 & 1.46 & -0.06 & c\\
  301 & HD071030 & 6541 & 4.03 & -0.15 & b\\
  302 & HD072184 & 4606 & 2.87 & 0.18 & c\\
  303 & HD072324 & 4881 & 2.43 & 0.00 & b\\
  304 & HD072660 & 9290 & 3.39 & -0.20 & b\\
  305 & HD073471 & 4495 & 2.13 & 0.10 & c\\
  306 & HD072905 & 5919 & 4.47 & 0.01 & b\\
  307 & HD073898 & 4912 & 2.30 & -0.56 & b\\
  308 & HD073665 & 5024 & 2.70 & 0.21 & b\\
  309 & HD073394 & 4612 & 1.48 & -1.49 & c\\
  310 & HD074000 & 6178 & 4.03 & -1.85 & b\\
  311 & HD073593 & 4815 & 2.82 & -0.17 & c\\
  312 & HD074011 & 5795 & 4.08 & -0.56 & b\\
  313 & HD074395 & 5546 & 1.41 & 0.09 & b\\
  314 & BD+251981 & 6668 & 4.28 & -1.20 & b\\
  315 & HD074442 & 4689 & 2.49 & -0.03 & c\\
  316 & HD074377 & 4674 & 4.47 & -0.37 & b\\
  317 & HD074721 & 8900 & 3.38 & -1.32 & b\\
  318 & BD-122669 & 6800 & 4.10 & -1.50 & b\\
  319 & HD074462 & 4747 & 1.83 & -1.40 & c\\
  320 & HD075318 & 5432 & 4.48 & -0.13 & b\\
  321 & HD075691 & 4299 & 2.10 & -0.16 & c\\
  322 & HD075732 & 5260 & 4.35 & 0.43 & b\\
  323 & HD076151 & 5748 & 4.42 & 0.15 & b\\
  324 & HD076292 & 6958 & 3.88 & 0.16 & b\\
  325 & HD076932 & 5908 & 4.09 & -0.82 & b\\
  326 & HD076780 & 5704 & 4.28 & 0.18 & b\\
  327 & BD-052678 & 5492 & 3.85 & -2.02 & b\\
  328 & HD076910 & 6397 & 4.23 & -0.51 & b\\
  329 & BD-032525 & 5869 & 4.09 & -1.60 & b\\
  331 & HD077338 & 5300 & 4.30 & 0.36 & b\\
  332 & HD077236 & 4343 & 1.89 & -0.89 & c\\
  333 & HD078541 & 3917 & 1.45 & -0.37 & c\\
  334 & HD078234 & 6976 & 4.04 & -0.06 & b\\
  335 & HD078558 & 5651 & 4.06 & -0.45 & b\\
  336 & HD078209 & 7519 & 3.77 & 0.55 & b\\
  337 & HD078737 & 6550 & 4.19 & -0.46 & b\\
  338 & HD078732 & 4939 & 2.27 & -0.07 & b\\
  339 & HD079211 & 3846 & 4.63 & -0.17 & c\\
  340 & HD079452 & 4982 & 2.31 & -0.79 & b\\
  341 & HD079765 & 7146 & 4.11 & -0.26 & b\\
  342 & HD079633 & 7223 & 4.06 & -0.26 & b\\
  343 & HD080390 & 3366 & 0.53 & -0.10 & c\\
  344 & HD081009 & 8829 & 3.79 & 0.84 & b\\
  345 & HD081029 & 6714 & 4.14 & -0.08 & b\\
  346 & HD081192 & 4745 & 2.57 & -0.76 & c\\
  348 & BD+092190 & 6270 & 4.11 & -2.86 & a\\
  349 & HD082074 & 5090 & 3.21 & -0.43 & b\\
  350 & HD082590 & 6669 & 4.22 & -0.81 & b\\
  351 & HD082734 & 4906 & 2.56 & 0.20 & c\\
  352 & HD082210 & 5445 & 3.64 & -0.10 & b\\
  353 & HD082885 & 5520 & 4.41 & 0.40 & b\\
  354 & HD083212 & 4472 & 0.99 & -1.64 & c\\
  355 & HD081817 & 4168 & 1.40 & 0.15 & c\\
  356 & HD083425 & 4150 & 1.99 & -0.47 & c\\
  357 & HD083618 & 4244 & 1.88 & -0.20 & c\\
  358 & HD083632 & 4167 & 1.41 & -0.85 & c\\
  359 & HD233666 & 5161 & 2.42 & -1.62 & b\\
  360 & HD083506 & 4875 & 2.36 & 0.12 & c\\
  361 & HD084441 & 5398 & 2.02 & -0.06 & b\\
  362 & HD084737 & 5872 & 4.05 & 0.12 & b\\
  363 & HD084937 & 6211 & 4.00 & -2.05 & b\\
  364 & HD085235 & 8769 & 3.69 & -0.23 & b\\
  366 & HD085503 & 4425 & 2.56 & 0.27 & c\\
  367 & HD085773 & 4345 & 0.66 & -2.48 & c\\
  368 & HD086986 & 8000 & 2.55 & -1.70 & b\\
  369 & HD087141 & 6359 & 3.90 & 0.09 & b\\
  370 & HD087140 & 5092 & 2.48 & -1.70 & b\\
  371 & HD087737 & 10958 & 2.11 & 0.11 & b\\
  372 & HD087822 & 6573 & 4.06 & 0.10 & b\\
  373 & HD088230 & 4017 & 4.67 & -0.01 & c\\
  374 & HD088446 & 5848 & 3.89 & -0.51 & b\\
  375 & HD088725 & 5647 & 4.24 & -0.64 & b\\
  377 & HD088737 & 6106 & 3.89 & 0.20 & b\\
  378 & HD088986 & 5766 & 4.04 & 0.04 & b\\
  379 & HD089010 & 5642 & 3.80 & 0.00 & b\\
  380 & HD089254 & 7166 & 3.83 & 0.31 & b\\
  381 & HD089449 & 6467 & 4.11 & 0.11 & b\\
  382 & HD089484 & 4381 & 1.79 & -0.50 & c\\
  383 & HD089707 & 5937 & 4.25 & -0.47 & b\\
  384 & HD089744 & 6169 & 3.93 & 0.18 & b\\
  385 & HD089995 & 6472 & 4.08 & -0.24 & b\\
  386 & HD089822b & 10182 & 3.85 & 0.07 & b\\
  387 & HD090508 & 5776 & 4.31 & -0.30 & b\\
  388 & HD237903 & 4106 & 4.64 & -0.16 & c\\
  389 & HD091347 & 5887 & 4.26 & -0.44 & b\\
  390 & HD091889 & 6109 & 4.16 & -0.21 & b\\
  391 & HD092523 & 4112 & 1.77 & -0.43 & c\\
  392 & HD093329 & 8400 & 3.10 & -1.20 & b\\
  393 & HD093487 & 5215 & 2.41 & -1.06 & b\\
  394 & HD094028 & 6076 & 4.23 & -1.30 & b\\
  395 & BD-103166 & 5329 & 4.38 & 0.42 & b\\
  396 & HD095128 & 5852 & 4.24 & 0.02 & b\\
  397 & HD095578 & 3849 & 1.26 & -0.01 & c\\
  398 & HD095735 & 3454 & 4.78 & -0.27 & c\\
  399 & BD+442051a & 3628 & 4.87 & -0.48 & c\\
  400 & HD096360 & 3471 & 0.80 & 0.00 & c\\
  401 & BD+362165 & 6144 & 4.18 & -1.45 & b\\
  402 & HD097560 & 5328 & 2.69 & -1.04 & b\\
  403 & HD097633 & 9201 & 3.68 & -0.16 & b\\
  404 & HD097907 & 4307 & 2.16 & -0.19 & c\\
  405 & HD097916 & 6478 & 4.28 & -0.73 & b\\
  406 & HD097855 & 6416 & 4.18 & -0.39 & b\\
  407 & HD098468 & 4528 & 2.06 & -0.36 & b\\
  408 & HD098553 & 5832 & 4.27 & -0.44 & b\\
  409 & HD099109 & 5242 & 4.29 & 0.40 & b\\
  410 & HD233832 & 4970 & 4.49 & -0.59 & b\\
  411 & HD099648 & 4977 & 2.24 & -0.03 & b\\
  412 & HD099747 & 6738 & 4.19 & -0.46 & b\\
  413 & HD099998 & 3979 & 1.53 & -0.37 & c\\
  414 & HD100906 & 5042 & 2.31 & -0.46 & b\\
  415 & HD101227 & 5534 & 4.52 & -0.32 & b\\
  416 & HD101501 & 5535 & 4.52 & 0.03 & b\\
  417 & HD101606 & 6362 & 4.13 & -0.57 & b\\
  418 & HD102224 & 4455 & 2.02 & -0.40 & c\\
  419 & BD+511696 & 5656 & 4.28 & -1.30 & b\\
  420 & HD102328 & 4380 & 2.56 & 0.28 & c\\
  421 & HD102634 & 6281 & 4.15 & 0.22 & b\\
  422 & HD102870 & 6081 & 4.07 & 0.14 & b\\
  423 & HD103095 & 5165 & 4.74 & -1.21 & b\\
  424 & HD103578 & 8509 & 3.80 & -0.19 & b\\
  425 & HD103877 & 7170 & 3.76 & 0.65 & b\\
  426 & HD103932 & 4431 & 4.55 & 0.06 & c\\
  427 & HD104307 & 4423 & 2.24 & -0.07 & c\\
  428 & HD104304 & 5485 & 4.23 & 0.30 & b\\
  429 & HD104833 & 7588 & 3.51 & 0.49 & b\\
  430 & HD105262 & 8500 & 1.50 & -1.87 & b\\
  431 & HD105452 & 7041 & 4.13 & -0.19 & b\\
  432 & HD105546 & 5131 & 2.36 & -1.46 & b\\
  433 & HD105740 & 4791 & 2.78 & -0.54 & c\\
  434 & HD106038 & 6014 & 4.18 & -1.25 & b\\
  435 & CD-2809374 & 4995 & 3.11 & -0.76 & b\\
  436 & HD106516 & 6207 & 4.25 & -0.69 & b\\
  437 & HD107113 & 6543 & 4.19 & -0.42 & b\\
  438 & HD107213 & 6209 & 4.00 & 0.24 & b\\
  439 & BD+172473 & 5283 & 3.37 & -1.04 & b\\
  440 & BD+312360 & 4716 & 2.44 & -0.79 & c\\
  441 & HD108177 & 6278 & 4.19 & -1.41 & b\\
  442 & HD108564 & 4634 & 4.67 & -1.00 & c\\
  443 & HD108915 & 5037 & 3.32 & -0.10 & b\\
  444 & HD109443 & 6758 & 4.17 & -0.60 & b\\
  445 & HD109871 & 3979 & 1.77 & -0.15 & b\\
  446 & HD109995 & 8550 & 2.39 & -1.66 & b\\
  447 & HD110014 & 4425 & 2.34 & 0.25 & c\\
  448 & HD110379 & 6857 & 4.17 & -0.19 & b\\
  449 & HD110897 & 5851 & 4.28 & -0.53 & b\\
  450 & HD110885 & 5545 & 2.99 & -1.06 & b\\
  451 & HD112028 & 9443 & 2.88 & -0.39 & b\\
  452 & HD111631 & 3908 & 4.67 & 0.02 & c\\
  453 & HD111786 & 8080 & 3.88 & -1.50 & b\\
  454 & HD111980 & 5876 & 4.04 & -1.02 & b\\
  455 & HD112127 & 4384 & 2.46 & 0.22 & c\\
  456 & HD112413 & 12303 & 4.09 & 0.90 & b\\
  457 & HD113092 & 4269 & 1.47 & -0.81 & c\\
  458 & HD113022 & 6491 & 4.09 & 0.11 & b\\
  460 & HD114038 & 4567 & 2.33 & -0.02 & c\\
  461 & HD114330 & 9570 & 3.95 & -0.13 & b\\
  462 & HD114606 & 5584 & 4.15 & -0.53 & b\\
  463 & HD114710 & 5994 & 4.35 & 0.05 & b\\
  464 & HD114642 & 6491 & 4.04 & -0.04 & b\\
  465 & HD114946 & 4999 & 3.12 & -0.36 & b\\
  466 & HD115383 & 6047 & 4.24 & 0.17 & b\\
  467 & HD115589 & 5227 & 4.39 & 0.28 & b\\
  468 & HD115617 & 5539 & 4.35 & 0.02 & b\\
  469 & HD115659 & 5104 & 2.64 & 0.04 & b\\
  470 & HD116114 & 8226 & 4.10 & 0.67 & b\\
  471 & HD116316 & 6487 & 4.26 & -0.51 & b\\
  472 & HD116544 & 4417 & 3.31 & 0.15 & c\\
  473 & HD117200 & 6843 & 4.02 & 0.03 & b\\
  474 & HD117176 & 5467 & 3.86 & -0.10 & b\\
  475 & HD117635 & 5175 & 4.48 & -0.42 & b\\
  476 & HD117876 & 4688 & 2.26 & -0.51 & c\\
  479 & HD118244 & 6391 & 4.19 & -0.46 & b\\
  480 & BD+302431 & 16904 & 4.20 & 0.77 & a\\
  481 & HD119228 & 3705 & 1.06 & -0.10 & c\\
  482 & HD119288 & 6595 & 4.18 & -0.23 & b\\
  483 & HD119291 & 4295 & 4.57 & 0.04 & c\\
  484 & HD119667 & 3740 & 0.89 & -0.11 & c\\
  485 & HD120136 & 6386 & 4.15 & 0.24 & b\\
  486 & HD121130 & 3543 & 0.85 & -0.02 & c\\
  487 & HD120933 & 3594 & 1.05 & -0.15 & c\\
  488 & HD121370 & 5967 & 3.78 & 0.28 & b\\
  489 & HD121299 & 4695 & 2.58 & 0.10 & c\\
  490 & HD121258 & 6570 & 4.00 & -0.92 & a\\
  491 & BD+342476 & 6200 & 3.96 & -2.05 & b\\
  492 & HD122106 & 6321 & 3.84 & 0.16 & b\\
  493 & HD122563 & 4618 & 1.32 & -2.67 & c\\
  494 & HD122742 & 5485 & 4.35 & 0.03 & b\\
  495 & HD123299 & 10371 & 3.95 & -0.19 & b\\
  496 & HD122956 & 4734 & 1.59 & -1.68 & c\\
  497 & HD123657 & 3408 & 0.66 & -0.05 & c\\
  498 & HD123821 & 4900 & 2.28 & -0.13 & b\\
  499 & HD124186 & 4384 & 2.56 & 0.27 & c\\
  500 & HD124292 & 5398 & 4.35 & -0.11 & b\\
  502 & HD124850 & 6207 & 3.86 & -0.06 & b\\
  503 & HD125184 & 5536 & 3.85 & 0.24 & b\\
  504 & HD125451 & 6700 & 4.12 & 0.01 & b\\
  505 & BD+012916 & 4375 & 0.77 & -1.86 & c\\
  506 & HD126141 & 6699 & 4.14 & -0.05 & b\\
  507 & HD126053 & 5598 & 4.22 & -0.37 & b\\
  509 & HD126218 & 5137 & 2.70 & 0.24 & b\\
  510 & HD126660 & 6293 & 4.13 & 0.02 & b\\
  511 & HD126614 & 5399 & 4.02 & 0.52 & b\\
  512 & HD126778 & 4832 & 2.41 & -0.52 & b\\
  513 & HD126681 & 5577 & 4.25 & -1.12 & b\\
  514 & HD127243 & 4903 & 2.27 & -0.78 & b\\
  515 & HD127334 & 5579 & 4.10 & 0.20 & b\\
  516 & BD+182890 & 5024 & 2.31 & -1.54 & b\\
  517 & HD128167 & 6777 & 4.18 & -0.39 & b\\
  518 & HD128429 & 6427 & 4.22 & -0.08 & b\\
  519 & HD128801 & 10200 & 3.50 & -1.40 & b\\
  520 & CD-2610417 & 4625 & 4.62 & -0.21 & c\\
  521 & HD128959 & 5857 & 3.84 & -0.51 & b\\
  522 & HD129174 & 12052 & 3.99 & 0.18 & b\\
  523 & HD130095 & 8900 & 3.35 & -1.70 & b\\
  524 & HD130322 & 5391 & 4.52 & 0.12 & b\\
  525 & HD130817 & 6749 & 4.14 & -0.29 & b\\
  526 & HD130705 & 4375 & 2.56 & 0.34 & c\\
  527 & HD130694 & 4093 & 1.74 & -0.77 & c\\
  528 & HD131430 & 4287 & 2.19 & 0.09 & c\\
  529 & HD132142 & 5163 & 4.38 & -0.37 & b\\
  530 & HD131918 & 4118 & 1.56 & -0.10 & c\\
  531 & HD131976 & 3541 & 4.72 & -0.14 & c\\
  532 & HD131977 & 4501 & 4.59 & -0.05 & c\\
  533 & HD132345 & 4403 & 2.54 & 0.31 & c\\
  534 & HD132475 & 5823 & 3.93 & -1.37 & b\\
  535 & HD132933 & 3797 & 1.30 & -0.71 & c\\
  536 & HD133124 & 4006 & 1.76 & -0.01 & c\\
  537 & BD+062986 & 3965 & 4.59 & -0.42 & c\\
  538 & BD+302611 & 4400 & 1.04 & -1.45 & c\\
  539 & HD134063 & 4880 & 2.31 & -0.69 & b\\
  540 & HD134083 & 6573 & 4.17 & -0.01 & b\\
  541 & HD134169 & 5807 & 3.99 & -0.81 & b\\
  542 & HD134440 & 4955 & 4.70 & -1.34 & c\\
  543 & HD134439 & 5172 & 4.68 & -1.27 & b\\
  544 & HD134987 & 5623 & 4.09 & 0.26 & b\\
  545 & HD136064 & 6083 & 3.94 & 0.03 & b\\
  546 & HD135482 & 4530 & 2.32 & -0.04 & c\\
  547 & HD135722 & 4850 & 2.39 & -0.39 & b\\
  548 & HD135485 & 15500 & 4.00 & 0.50 & b\\
  549 & HD136726 & 4176 & 1.85 & -0.09 & c\\
  550 & HD136202 & 6139 & 4.00 & 0.05 & b\\
  551 & HD137071 & 3929 & 1.10 & 0.06 & c\\
  552 & HD136834 & 4856 & 4.50 & 0.23 & c\\
  553 & HD137391 & 7186 & 3.93 & 0.10 & b\\
  554 & HD137759 & 4459 & 2.43 & 0.08 & c\\
  555 & HD137471 & 3793 & 1.13 & -0.04 & c\\
  556 & HD137510 & 5872 & 3.90 & 0.30 & b\\
  557 & HD137704 & 4044 & 1.74 & -0.47 & c\\
  558 & HD137909 & 8466 & 4.06 & 0.96 & b\\
  559 & HD138290 & 6822 & 4.14 & -0.10 & b\\
  560 & HD138481 & 3898 & 1.25 & -0.07 & c\\
  561 & HD139669 & 3930 & 1.44 & 0.12 & c\\
  562 & HD138776 & 5524 & 3.99 & 0.35 & b\\
  563 & HD138764 & 14054 & 3.88 & 0.08 & b\\
  564 & HD139195 & 4946 & 2.64 & -0.13 & b\\
  565 & HD139641 & 4945 & 2.82 & -0.51 & b\\
  566 & HD139446 & 5065 & 2.65 & -0.27 & b\\
  567 & HD140160 & 9557 & 3.66 & 0.35 & b\\
  568 & HD140283 & 5687 & 3.55 & -2.53 & a\\
  569 & BD+053080 & 5034 & 4.46 & -0.44 & b\\
  570 & HD141004 & 5823 & 4.10 & -0.03 & b\\
  571 & HD141714 & 5332 & 3.22 & -0.20 & b\\
  572 & HD141851 & 8246 & 3.89 & -2.00 & a\\
  573 & HD142373 & 5783 & 3.93 & -0.54 & b\\
  574 & HD142575 & 6779 & 4.23 & -0.70 & b\\
  575 & HD142908 & 7038 & 3.98 & -0.02 & b\\
  576 & HD142860 & 6309 & 4.18 & -0.16 & b\\
  577 & HD142703 & 6903 & 4.32 & -1.10 & b\\
  578 & HD143459 & 10498 & 4.00 & -0.39 & b\\
  579 & HD143761 & 5752 & 4.13 & -0.26 & b\\
  580 & MS1558.4-2232 & 4250 & 3.50 & 0.10 & a\\
  581 & HD143807 & 10727 & 3.84 & -0.01 & b\\
  582 & HD144172 & 6432 & 4.14 & -0.38 & b\\
  583 & HD144872 & 4785 & 4.76 & -0.29 & c\\
  584 & HD144585 & 5767 & 4.09 & 0.29 & b\\
  585 & HD144608 & 5363 & 2.62 & 0.03 & b\\
  586 & HD145148 & 4868 & 3.65 & 0.10 & b\\
  587 & HD145675 & 5270 & 4.31 & 0.48 & b\\
  588 & HD145250 & 4530 & 2.28 & -0.33 & c\\
  589 & HD145976 & 6927 & 4.08 & -0.02 & b\\
  590 & HD146051 & 3779 & 1.46 & -0.15 & c\\
  591 & HD147379b & 3873 & 4.68 & 0.06 & c\\
  592 & HD146624 & 9125 & 3.99 & -0.27 & b\\
  593 & HD147923 & 4787 & 4.76 & -0.28 & c\\
  594 & BD-114126 & 4702 & 4.62 & 0.00 & b\\
  595 & HD148112 & 10052 & 3.51 & 0.47 & b\\
  596 & BD+090352 & 6131 & 4.04 & -1.88 & b\\
  597 & HD148513 & 4114 & 2.16 & 0.19 & c\\
  598 & BD+112998 & 5527 & 2.97 & -1.01 & b\\
  599 & HD148816 & 5828 & 4.05 & -0.72 & b\\
  600 & HD148897 & 4278 & 0.99 & -1.20 & c\\
  601 & HD150275 & 4622 & 2.40 & -0.70 & c\\
  602 & HD148786 & 5144 & 2.70 & 0.21 & b\\
  603 & HD149009 & 3862 & 1.20 & 0.09 & c\\
  604 & HD148898 & 9141 & 3.85 & 0.38 & b\\
  605 & HD149121 & 11099 & 3.89 & 0.03 & b\\
  606 & HD149161 & 3915 & 1.64 & -0.26 & c\\
  607 & BD+093223 & 5200 & 2.00 & -2.31 & b\\
  608 & HD149382 & 35500 & 5.70 & -1.30 & b\\
  609 & HD149661 & 5281 & 4.59 & 0.13 & b\\
  610 & HD150012 & 6651 & 3.96 & 0.13 & b\\
  611 & HD150177 & 6190 & 4.08 & -0.58 & b\\
  612 & HD150281 & 5164 & 4.54 & 0.14 & b\\
  613 & HD150453 & 6589 & 4.07 & -0.24 & b\\
  614 & HD151203 & 3570 & 0.91 & 0.03 & c\\
  615 & HD151217 & 3878 & 1.59 & 0.05 & c\\
  616 & HD152601 & 4661 & 2.67 & 0.10 & c\\
  617 & HD152781 & 4969 & 3.55 & 0.08 & b\\
  618 & HD153286 & 7534 & 3.75 & 0.49 & b\\
  619 & HD153882 & 9999 & 3.50 & 0.61 & b\\
  620 & HD154733 & 4227 & 2.20 & -0.09 & c\\
  621 & HD155763 & 12500 & 3.50 & -0.11 & b\\
  622 & HD155358 & 5888 & 4.09 & -0.63 & b\\
  623 & HD155078 & 6508 & 4.00 & 0.03 & b\\
  625 & HD156026 & 4381 & 4.66 & -0.28 & c\\
  626 & HD157373 & 6552 & 4.18 & -0.43 & b\\
  627 & HD157214 & 5621 & 4.05 & -0.41 & b\\
  628 & HD157089 & 5792 & 4.05 & -0.57 & b\\
  629 & HD157910 & 5227 & 2.58 & -0.07 & b\\
  630 & HD157881 & 4023 & 4.64 & -0.01 & c\\
  631 & HD157856 & 6523 & 4.04 & -0.07 & b\\
  632 & HD157919 & 6826 & 3.61 & 0.29 & b\\
  633 & BD+233130 & 5017 & 2.31 & -2.45 & b\\
  634 & HD159332 & 6298 & 4.01 & -0.10 & b\\
  635 & HD159307 & 6395 & 4.19 & -0.54 & b\\
  636 & HD159482 & 5740 & 4.11 & -0.75 & b\\
  637 & HD160933 & 5770 & 3.79 & -0.31 & b\\
  638 & HD160762 & 17678 & 3.69 & 0.02 & b\\
  639 & HD160693 & 5691 & 4.07 & -0.56 & b\\
  640 & HD161074 & 3949 & 1.86 & -0.14 & c\\
  641 & HD161149 & 7015 & 3.70 & 0.33 & b\\
  642 & HD161096 & 4497 & 2.49 & 0.20 & c\\
  644 & HD161227 & 7320 & 3.72 & 0.38 & b\\
  645 & HD161695 & 11506 & 2.23 & 0.11 & b\\
  646 & HD161797 & 5454 & 3.82 & 0.22 & b\\
  647 & HD161817 & 7636 & 2.93 & -0.95 & a\\
  648 & HD162211 & 4486 & 2.45 & -0.06 & c\\
  649 & BD+203603 & 6115 & 4.10 & -2.10 & b\\
  650 & HD164058 & 3963 & 1.51 & 0.07 & c\\
  651 & HD163990 & 3318 & 0.48 & -0.04 & c\\
  652 & HD163993 & 5091 & 2.87 & 0.09 & b\\
  653 & HD164136 & 7140 & 3.87 & 0.00 & b\\
  654 & HD164349 & 4594 & 1.74 & -0.01 & c\\
  655 & HD164353 & 15600 & 2.55 & -0.03 & b\\
  656 & HD164432 & 21371 & 3.81 & -0.02 & b\\
  657 & HD165195 & 4391 & 0.75 & -2.32 & c\\
  658 & HD164975 & 5863 & 1.22 & 0.06 & b\\
  659 & HD165341 & 5349 & 4.58 & 0.16 & b\\
  660 & HD165438 & 4868 & 3.43 & 0.02 & b\\
  661 & HD165908 & 6045 & 4.19 & -0.50 & b\\
  662 & HD166208 & 5107 & 2.73 & 0.15 & b\\
  663 & HD165634 & 4907 & 2.31 & -0.14 & b\\
  664 & HD166620 & 4968 & 4.55 & -0.18 & b\\
  665 & HD166161 & 5201 & 2.33 & -1.25 & b\\
  666 & HD166285 & 6389 & 4.10 & -0.06 & b\\
  667 & HD166460 & 4478 & 2.19 & -0.05 & c\\
  668 & HD167105 & 9000 & 2.36 & -1.50 & b\\
  669 & HD167006 & 3597 & 1.12 & -0.16 & c\\
  670 & HD167768 & 4953 & 2.29 & -0.69 & b\\
  671 & HD169027 & 11030 & 3.89 & -0.08 & b\\
  672 & HD168322 & 4745 & 2.33 & -0.51 & c\\
  673 & HD167665 & 6125 & 4.15 & -0.16 & b\\
  674 & HD168720 & 3788 & 1.45 & -0.08 & c\\
  675 & HD168723 & 4923 & 3.0 & -0.22 & b\\
  676 & HD168608 & 5580 & 1.00 & 0.03 & b\\
  677 & HD170693 & 4396 & 2.1 & -0.49 & c\\
  679 & HD170737 & 5093 & 3.36 & -0.77 & b\\
  681 & HD171391 & 5150 & 2.89 & 0.04 & b\\
  682 & HD171443 & 4220 & 2.04 & -0.10 & c\\
  683 & HD171496 & 4933 & 2.29 & -0.73 & c\\
  684 & HD171999 & 5276 & 4.35 & 0.27 & b\\
  685 & HD172380 & 3364 & 0.42 & -0.06 & c\\
  686 & HD172103 & 6815 & 4.01 & 0.03 & b\\
  687 & HD172365 & 5886 & 1.28 & 0.05 & b\\
  688 & HD172958 & 11464 & 3.97 & 0.02 & b\\
  689 & HD173524 & 11323 & 3.93 & 0.10 & b\\
  690 & HD173740 & 3311 & 5.01 & -0.34 & c\\
  691 & HD172816 & 3318 & 0.72 & -0.14 & c\\
  692 & HD173093 & 6373 & 4.11 & 0.00 & b\\
  693 & HD173648 & 7914 & 3.70 & 0.38 & b\\
  694 & HD173650 & 11832 & 3.71 & 0.64 & b\\
  695 & HD173667 & 6458 & 4.04 & 0.01 & b\\
  696 & HD175305 & 5036 & 2.51 & -1.44 & b\\
  697 & HD173819 & 4392 & -0.25 & -0.67 & c\\
  698 & HD174567 & 10256 & 3.95 & -0.07 & b\\
  699 & HD174912 & 5936 & 4.34 & -0.45 & b\\
  700 & HD175225 & 5286 & 3.70 & 0.20 & b\\
  701 & HD174959 & 14681 & 4.00 & -0.80 & a\\
  702 & HD174704 & 7193 & 3.63 & 0.79 & b\\
  703 & HD175535 & 5197 & 2.85 & -0.07 & b\\
  704 & HD175588 & 3484 & 0.47 & -0.14 & c\\
  705 & HD175865 & 3316 & 0.36 & 0.06 & c\\
  706 & HD175640 & 12077 & 3.94 & 0.17 & b\\
  707 & HD178089 & 6722 & 4.07 & -0.09 & b\\
  708 & HD175892 & 8705 & 4.11 & -0.29 & b\\
  709 & HD176301 & 12667 & 4.18 & 0.18 & b\\
  710 & HD176232 & 8743 & 4.47 & 0.53 & b\\
  711 & HD176437 & 11226 & 4.11 & 0.09 & b\\
  712 & HD177463 & 4611 & 2.30 & -0.23 & c\\
  713 & HD180711 & 4837 & 2.49 & -0.18 & b\\
  714 & HD179761 & 12746 & 4.22 & 0.30 & b\\
  715 & HD180163 & 18663 & 3.69 & 0.04 & b\\
  716 & HD181096 & 6347 & 4.03 & -0.17 & b\\
  717 & HD180928 & 4024 & 1.79 & -0.66 & c\\
  718 & HD181470 & 9802 & 3.91 & -0.13 & b\\
  719 & HD182293 & 4437 & 2.74 & -0.02 & c\\
  721 & HD182572 & 5473 & 3.91 & 0.34 & b\\
  722 & HD183324 & 10325 & 4.17 & -1.24 & b\\
  723 & CD-2415398 & 6269 & 2.93 & -1.17 & b\\
  724 & HD185144 & 5293 & 4.56 & -0.12 & b\\
  725 & HD338529 & 6178 & 3.95 & -2.09 & b\\
  726 & HD184499 & 5743 & 4.07 & -0.54 & b\\
  727 & HD184786 & 3454 & 0.35 & 0.03 & c\\
  728 & HD184406 & 4428 & 2.71 & 0.03 & c\\
  729 & HD185351 & 5045 & 3.27 & 0.08 & b\\
  730 & HD185657 & 4813 & 2.51 & -0.19 & c\\
  732 & HD185859 & 26200 & 3.05 & -0.09 & b\\
  733 & HD186408 & 5731 & 4.15 & 0.08 & b\\
  734 & HD186427 & 5648 & 4.18 & 0.03 & b\\
  735 & HD188119 & 4904 & 2.34 & -0.49 & b\\
  737 & HD187879 & 21004 & 3.14 & 0.05 & b\\
  738 & HD187691 & 6099 & 4.12 & 0.14 & b\\
  739 & HD187921 & 5502 & 0.68 & 0.10 & b\\
  741 & HD188650 & 5764 & 2.90 & -0.40 & a\\
  742 & HD188510 & 5539 & 4.28 & -1.43 & b\\
  743 & HD188512 & 5082 & 3.48 & -0.22 & b\\
  745 & HD188947 & 4828 & 2.62 & 0.06 & b\\
  746 & HD189005 & 5080 & 2.32 & -0.26 & b\\
  747 & HD189558 & 5770 & 3.92 & -1.04 & b\\
  748 & HD189849 & 7804 & 3.89 & -0.01 & b\\
  749 & HD190360 & 5468 & 4.11 & 0.21 & b\\
  750 & HD190404 & 5008 & 4.51 & -0.58 & b\\
  752 & HD190178 & 6263 & 4.05 & -0.66 & b\\
  753 & HD190390 & 6440 & 1.55 & -1.05 & a\\
  754 & HD191026 & 5177 & 3.81 & 0.08 & b\\
  755 & HD191046 & 4438 & 1.67 & -0.75 & c\\
  756 & HD192907 & 10444 & 3.97 & -0.18 & b\\
  757 & HD345957 & 5988 & 4.06 & -1.20 & b\\
  759 & HD192640 & 8774 & 4.42 & -0.80 & b\\
  760 & HD192909 & 3942 & 0.91 & -0.02 & c\\
  761 & HD193281 & 8597 & 4.11 & -0.37 & b\\
  762 & HD194598 & 6090 & 4.24 & -1.04 & b\\
  763 & HD194943 & 6971 & 4.04 & -0.01 & b\\
  764 & HD196502 & 9417 & 3.58 & 0.74 & b\\
  765 & HD195633 & 6119 & 4.09 & -0.52 & b\\
  766 & HD195838 & 6152 & 4.08 & 0.00 & b\\
  767 & HD196544 & 8678 & 3.80 & -0.12 & b\\
  768 & HD196755 & 5582 & 3.64 & -0.02 & b\\
  769 & HD197177 & 4964 & 1.92 & -0.03 & b\\
  770 & HD197572 & 5188 & 0.83 & 0.15 & b\\
  771 & HD197461 & 7334 & 4.02 & -0.07 & b\\
  772 & HD198149 & 4970 & 3.29 & -0.19 & b\\
  773 & HD197989 & 4728 & 2.44 & -0.20 & c\\
  775 & HD198183 & 15630 & 3.81 & 0.05 & b\\
  776 & HD198001 & 9266 & 4.00 & -0.32 & b\\
  777 & BD+044551 & 6089 & 4.21 & -1.26 & b\\
  778 & HD198478 & 16500 & 2.17 & -0.21 & b\\
  779 & HD199191 & 4696 & 2.53 & -0.70 & c\\
  781 & HD199799 & 3387 & 0.12 & -0.25 & c\\
  782 & HD200527 & 3503 & 0.20 & -0.07 & c\\
  783 & HD200580 & 6003 & 4.34 & -0.50 & b\\
  784 & HD200905 & 3977 & 0.79 & 0.10 & c\\
  785 & HD200779 & 4225 & 4.59 & 0.02 & c\\
  786 & HD200790 & 6115 & 3.98 & 0.02 & b\\
  787 & HD201078 & 6151 & 1.85 & 0.09 & b\\
  788 & HD201091 & 4162 & 4.64 & -0.31 & c\\
  789 & HD201601 & 7657 & 3.92 & 0.07 & a\\
  790 & HD201891 & 5881 & 4.18 & -1.05 & b\\
  791 & HD201889 & 5762 & 4.13 & -0.74 & b\\
  792 & HD202109 & 4925 & 2.39 & -0.23 & b\\
  793 & HD202447 & 6277 & 4.01 & 0.26 & b\\
  794 & HD202671 & 14353 & 3.25 & 0.36 & b\\
  795 & HD203638 & 4553 & 2.48 & 0.12 & c\\
  796 & HD204041 & 8737 & 4.45 & -0.44 & b\\
  798 & HD204155 & 5718 & 3.93 & -0.69 & b\\
  799 & HD204613 & 5718 & 3.88 & -0.38 & b\\
  800 & HD205021 & 25500 & 3.70 & -0.10 & b\\
  801 & HD204381 & 5081 & 2.76 & -0.09 & b\\
  802 & HD204754 & 12610 & 4.20 & 0.30 & b\\
  803 & HD204543 & 4590 & 1.17 & -1.97 & c\\
  804 & HD204587 & 4111 & 4.61 & -0.11 & c\\
  805 & HD205435 & 5069 & 2.86 & -0.12 & b\\
  806 & HD205153 & 6005 & 4.02 & 0.07 & b\\
  807 & HD205512 & 4703 & 2.57 & 0.03 & c\\
  808 & HD206078 & 4741 & 2.54 & -0.59 & c\\
  809 & HD206165 & 19300 & 2.65 & -0.27 & b\\
  810 & HD206952 & 4643 & 2.61 & 0.15 & c\\
  811 & HD206453 & 5026 & 2.34 & -0.41 & b\\
  812 & HD207130 & 4741 & 2.65 & 0.08 & c\\
  813 & HD206826 & 6490 & 4.09 & -0.11 & b\\
  816 & HD207076 & 3022 & 0.74 & -0.12 & c\\
  817 & HD207330 & 20815 & 3.69 & 0.04 & b\\
  818 & HD207222 & 9230 & 4.01 & -0.36 & b\\
  821 & HD208906 & 6048 & 4.27 & -0.68 & b\\
  822 & HD209369 & 6632 & 4.06 & -0.04 & b\\
  823 & HD209459 & 11015 & 3.99 & -0.07 & b\\
  824 & HD209975 & 32983 & 3.40 & 0.06 & b\\
  825 & HD210295 & 4864 & 2.30 & -1.26 & c\\
  826 & HD210424 & 12771 & 4.21 & 0.22 & b\\
  827 & HD210745 & 4286 & 0.65 & 0.11 & c\\
  828 & HD210595 & 6639 & 4.11 & -0.47 & b\\
  829 & HD210705 & 6939 & 4.16 & -0.18 & b\\
  830 & HD211075 & 4318 & 1.84 & -0.42 & c\\
  831 & HD212454 & 14466 & 3.33 & 0.35 & b\\
  832 & HD212943 & 4634 & 2.62 & -0.31 & c\\
  833 & HD213119 & 3914 & 1.42 & -0.14 & c\\
  835 & HD213042 & 4505 & 4.49 & 0.12 & c\\
  837 & HD214080 & 23445 & 3.28 & 0.04 & b\\
  839 & HD214567 & 4997 & 2.62 & -0.23 & b\\
  840 & HD214714 & 5224 & 2.03 & -0.67 & b\\
  841 & HD214994 & 9373 & 3.73 & -0.14 & b\\
  843 & HD215648 & 6243 & 4.03 & -0.21 & b\\
  844 & HD216228 & 4745 & 2.49 & -0.03 & c\\
  845 & HD216174 & 4371 & 1.82 & -0.61 & c\\
  846 & HD216131 & 4999 & 2.70 & -0.07 & b\\
  847 & HD216143 & 4638 & 1.38 & -2.09 & c\\
  848 & HD216219 & 5637 & 3.10 & -0.36 & b\\
  849 & HD216385 & 6323 & 4.06 & -0.15 & b\\
  850 & HD217382 & 4105 & 1.96 & 0.09 & c\\
  851 & HD216640 & 4612 & 3.11 & 0.17 & c\\
  852 & HD216831 & 13207 & 3.05 & 0.05 & b\\
  854 & HD217014 & 5674 & 4.14 & 0.18 & b\\
  855 & HD217107 & 5523 & 4.11 & 0.33 & b\\
  856 & HD217754 & 7089 & 3.80 & 0.27 & b\\
  857 & HD218031 & 4713 & 2.42 & -0.17 & c\\
  858 & HD218235 & 6463 & 4.06 & 0.23 & b\\
  859 & HD218329 & 3796 & 1.46 & 0.19 & c\\
  860 & HD218502 & 6167 & 4.11 & -1.75 & b\\
  861 & HD218640 & 5799 & 3.26 & 0.36 & b\\
  862 & HD218804 & 6493 & 4.17 & -0.13 & b\\
  863 & HD218857 & 5057 & 2.43 & -1.93 & b\\
  864 & HD219134 & 4759 & 4.63 & 0.04 & c\\
  866 & BD+384955 & 5270 & 3.50 & -2.23 & b\\
  867 & HD219449 & 4666 & 2.52 & 0.00 & c\\
  868 & HD219623 & 6138 & 4.24 & 0.07 & b\\
  869 & HD219617 & 5941 & 4.12 & -1.36 & b\\
  870 & HD219615 & 4890 & 2.42 & -0.57 & b\\
  871 & HD219734 & 3665 & 0.99 & -0.04 & c\\
  872 & HD219916 & 5070 & 2.84 & -0.04 & b\\
  874 & HD220009 & 4296 & 1.90 & -0.80 & c\\
  875 & HD220575 & 12241 & 4.09 & 0.27 & b\\
  876 & BD+592723 & 5987 & 3.98 & -1.89 & b\\
  877 & HD220825 & 10228 & 3.71 & 0.78 & b\\
  878 & HD220933 & 10515 & 3.72 & -0.06 & b\\
  879 & HD220954 & 4784 & 2.61 & 0.06 & c\\
  880 & HD221170 & 4608 & 1.29 & -2.05 & c\\
  881 & HD221148 & 4588 & 3.19 & 0.34 & c\\
  882 & HD221345 & 4662 & 2.33 & -0.34 & c\\
  883 & HD221377 & 6553 & 4.20 & -0.60 & b\\
  885 & HD221756 & 8833 & 4.21 & -0.64 & b\\
  886 & HD221830 & 5719 & 4.09 & -0.40 & b\\
  887 & HD222404 & 4758 & 3.16 & 0.10 & c\\
  888 & HD222368 & 6231 & 4.14 & -0.08 & b\\
  889 & HD222451 & 6698 & 4.07 & 0.10 & b\\
  891 & HD223047 & 5002 & 1.26 & 0.04 & b\\
  893 & HD223524 & 4560 & 2.41 & 0.02 & c\\
  894 & HD223640 & 12429 & 3.93 & 0.73 & b\\
  895 & HD224458 & 4809 & 2.27 & -0.44 & c\\
  896 & BD+612575 & 6222 & 1.97 & 0.35 & a\\
  897 & NGC288-77 & 4238 & 1.12 & -1.32 & c\\
  898 & HD020902 & 6690 & 1.31 & -0.05 & b\\
  899 & Mel22\_0296 & 5196 & 4.25 & -0.03 & b\\
  900 & Mel22\_2462 & 5219 & 4.47 & -0.03 & b\\
  901 & HD025825 & 6005 & 4.38 & 0.13 & b\\
  902 & HD026736 & 5772 & 4.39 & 0.13 & b\\
  903 & HD284253 & 5283 & 4.47 & 0.13 & b\\
  904 & HD027383 & 6091 & 4.34 & 0.13 & b\\
  905 & HD027524 & 6580 & 4.14 & 0.13 & b\\
  906 & HD027561 & 6682 & 4.14 & 0.13 & b\\
  907 & HD027962 & 8809 & 3.80 & 0.13 & b\\
  908 & HD028483 & 6455 & 4.25 & 0.13 & b\\
  909 & HD028546 & 7490 & 3.85 & 0.13 & b\\
  910 & HD029375 & 7240 & 3.93 & 0.13 & b\\
  911 & HD030034 & 7446 & 3.91 & 0.13 & b\\
  912 & HD030210 & 7694 & 3.66 & 0.13 & b\\
  913 & HD030676 & 6104 & 4.37 & 0.13 & b\\
  914 & HD031236 & 7262 & 3.93 & 0.13 & b\\
  917 & M79\_223 & 4170 & 0.68 & -1.60 & c\\
  918 & NGC2420-140 & 4421 & 1.90 & -0.31 & c\\
  919 & M67\_F-108 & 4235 & 2.21 & 0.00 & c\\
  920 & HD107276 & 7969 & 3.99 & -0.05 & b\\
  921 & HD107513 & 7360 & 3.99 & -0.05 & b\\
  922 & HD109307 & 8162 & 3.91 & -0.05 & b\\
  923 & M3\_IV-25 & 4442 & 1.02 & -1.50 & c\\
  924 & M3\_III-28 & 4306 & 0.74 & -1.50 & c\\
  925 & M3\_398 & 4580 & 1.15 & -1.50 & c\\
  926 & M5\_III-03 & 4174 & 0.51 & -1.29 & c\\
  928 & M5\_II-76 & 5974 & 2.44 & -1.11 & a\\
  930 & M5\_IV-19 & 4251 & 0.90 & -1.29 & c\\
  933 & M4\_LEE-2303 & 6748 & 2.51 & -1.19 & a\\
  934 & M13\_A-171 & 4266 & 1.35 & -0.80 & c\\
  935 & M13\_B-786 & 4114 & 0.43 & -1.53 & c\\
  936 & M92\_IV-114 & 4726 & 1.44 & -2.31 & c\\
  937 & M92\_III-13 & 4226 & 0.38 & -2.31 & c\\
  938 & HD170764 & 5802 & 0.99 & 0.17 & b\\
  939 & HD170820 & 4499 & 1.44 & 0.03 & c\\
  940 & NGC6791-R4 & 3382 & 0.20 & 0.42 & c\\
  941 & NGC6791-R5 & 3193 & 1.12 & 0.42 & c\\
  942 & NGC6791-R16 & 3890 & 1.81 & 0.42 & c\\
  943 & NGC6791-R19 & 3904 & 1.95 & 0.42 & c\\
  944 & M71\_A9 & 4404 & 1.41 & -0.78 & c\\
  945 & M71\_1-109 & 4759 & 2.33 & -0.78 & c\\
  946 & M71\_1-95 & 4570 & 1.63 & -0.78 & c\\
  947 & M71\_1-107 & 4848 & 2.07 & -0.84 & b\\
  948 & M71\_1-75 & 4809 & 2.52 & -0.78 & c\\
  949 & M71\_1-73 & 4800 & 2.22 & -0.78 & c\\
  951 & M71\_1-71 & 4397 & 1.80 & -0.78 & c\\
  952 & M71\_KC-263 & 4883 & 2.61 & -0.84 & a\\
  953 & M71\_1-87 & 4988 & 2.19 & -0.84 & b\\
  954 & M71\_1-66 & 4250 & 1.65 & -0.78 & c\\
  955 & M71\_1-65 & 4664 & 1.98 & -0.78 & c\\
  956 & M71\_1-64 & 4504 & 1.64 & -0.78 & c\\
  958 & M71\_1-21 & 4486 & 1.40 & -0.78 & c\\
  959 & M71\_1-37 & 4576 & 2.12 & -0.78 & c\\
  960 & M71\_1-41 & 5020 & 2.26 & -0.84 & b\\
  962 & M71\_1-39 & 5020 & 2.26 & -0.84 & b\\
  964 & M71\_1-53 & 4167 & 1.51 & -0.84 & a\\
  965 & M71\_KC-169 & 5083 & 2.22 & -0.84 & b\\
  966 & M71\_A2 & 4679 & 2.21 & -0.84 & b\\
  968 & M71\_1-78 & 3955 & 1.48 & -0.78 & c\\
  969 & M71\_S & 4244 & 1.28 & -0.78 & c\\
  971 & M71\_I & 4222 & 1.43 & -0.78 & c\\
  972 & NGC7789-329(491) & 4527 & 2.14 & -0.13 & a\\
  973 & NGC7789-468(589) & 4167 & 1.75 & 0.01 & c\\
  974 & NGC7789-342(502) & 10968 & 3.85 & -0.13 & b\\
  975 & NGC7789-353(509) & 4538 & 2.30 & 0.01 & c\\
  976 & NGC7789-415(550) & 3815 & 1.16 & 0.01 & c\\
  977 & NGC7789-461(583) & 4123 & 1.75 & 0.01 & c\\
  978 & NGC7789-501(614) & 4057 & 1.69 & 0.01 & c\\
  979 & NGC7789-575(671) & 4547 & 2.18 & 0.01 & c\\
  980 & NGC7789-637(723) & 4857 & 2.54 & 0.01 & c\\
  981 & NGC7789-765(804) & 4397 & 2.07 & 0.01 & c\\
  982 & NGC7789-859(853) & 4666 & 2.53 & 0.01 & c\\
  983 & NGC7789-875(873) & 4861 & 2.56 & -0.13 & b\\
  984 & NGC7789-897(881) & 4918 & 2.50 & -0.13 & b\\
  985 & NGC7789-971(946) & 3746 & 1.22 & 0.01 & c\\

\end{longtable}

\begin{longtable}{clrrrcl}
\caption{\label{tab_discarded}MILES stars which were not used in the stellar population models. Column (f) gives the reference for the source of the stellar atmospheric parameters: 
$^a$\citet{MILES2}; 
$^b$\citet{prugniel+11};
$^c$\citet{sharma+16}; 
$^d$ Same as $^a$ but [Fe/H] is unknown, assumed to the solar. 
Column (g) gives the code identifying the reason(s) why each star was discarded (see discussion in Section 2 of paper): 
$^1$Excessive noise or corrupted spectrum;
$^2$Visible continuum distortions;
$^3$Visible emission lines;
$^4$Peculiar features;
$^5$E(B-V) > 0.3 from \citet{prugniel+11};
$^6$Removed by the cut in $\chi2$;
$^7$Removed by the cut in $\tilde\Delta$.}\\

\toprule\bfseries 
\bfseries MILES ID & \bfseries Star & \bfseries T$_{\rm eff}$ & 
\bfseries log g & 
\bfseries [Fe/H] & 
\bfseries Reference & \bfseries Removed by\\
(a) & (b) & (c) & (d) & (e) & (f) & (g) \\ 
\midrule 
\endfirsthead
\caption{continued.}\\
\hline
\bfseries MILES ID & \bfseries Star & \bfseries T$_{\rm eff}$ & \bfseries log g & \bfseries \feh & \bfseries Reference & \bfseries Removed by\\ 
(a) & (b) & (c) & (d) & (e) & (f) & (g) \\ 
\hline
\endhead
\bottomrule \endfoot   

%    \csvreader[
%    head to column names,
%    late after line=\\,
%    before reading={\catcode`\#=12},after reading={\catcode`\#=6}
%    ]{discarded_stars.csv}{}
%    {\milesid & \expUScore{\star} & \teff & \logg & \feh & \ref & \removedby}

  29 & HD004395 & 5444 & 3.43 & -0.27 & b & 1, 6\\
  44 & HD006474 & 6781 & 0.49 & 0.26 & b & 6, 7\\
  45 & HD006497 & 4401 & 2.55 & 0.00 & c & 1, 6\\
  62 & HD009408 & 4814 & 2.46 & -0.31 & b & 1\\
  64 & HD009919 & 6860 & 4.00 & -0.35 & b & 1\\
  68 & BD+720094 & 6131 & 4.09 & -1.68 & b & 7\\
  74 & HD012014 & 4371 & 0.66 & 0.04 & b & 1, 7\\
  87 & HD015596 & 4811 & 2.75 & -0.71 & c & 7\\
  104 & HD018391 & 5750 & 1.20 & -0.13 & b & 5, 6, 7\\
  107 & HD019510 & 6108 & 3.91 & -2.13 & b & 6\\
  140 & HD281679 & 8542 & 2.50 & -1.43 & a & 6\\
  142 & BD+060648 & 4645 & 1.38 & -1.94 & c & 6, 7\\
  181 & HD035620 & 4184 & 2.00 & 0.10 & c & 7\\
  204 & HD041117 & 20000 & 2.40 & -0.12 & b & 3\\
  210 & HD042474 & 3719 & 0.62 & -0.13 & c & 3\\
  212 & HD043042 & 6480 & 4.18 & 0.06 & b & 2\\
  220 & HD044889 & 4006 & 1.51 & -0.32 & c & 7\\
  221 & HD044691A & 7777 & 3.88 & 0.27 & b & 7\\
  246 & HD055496 & 4858 & 2.05 & -1.48 & b & 4\\
  251 & HD057060 & 33215 & 3.28 & -0.03 & b & 3\\
  252 & HD057061 & 34303 & 3.46 & 0.10 & b & 6\\
  256 & HD059612 & 8409 & 1.56 & -0.05 & b & 1\\
  330 & HD076813 & 5065 & 2.63 & -0.06 & b & 6, 7\\
  347 & HD081797 & 4171 & 1.65 & 0.01 & c & 2, 7\\
  365 & HD237846 & 4675 & 1.20 & -3.15 & b & 6, 7\\
  376 & HD088609 & 4417 & 0.91 & -2.82 & c & 1, 7\\
  459 & HD113285 & 2902 & 0.21 & -0.33 & c & 6, 7\\
  477 & HD118055 & 4391 & 0.78 & -1.86 & c & 2\\
  478 & HD118100 & 4277 & 4.48 & -0.14 & c & 3\\
  501 & HD124897 & 4245 & 1.94 & -0.70 & c & 2\\
  508 & HD126327 & 2908 & 0.37 & -0.33 & c & 6, 7\\
  624 & HD156283 & 4206 & 1.58 & 0.02 & c & 7\\
  643 & HD161796 & 7000 & 0.44 & -0.30 & b & 6, 7\\
  678 & HD169985 & 6249 & 3.84 & 0.36 & b & 6, 7\\
  680 & HD234677 & 4184 & 4.33 & -0.01 & c & 3\\
  720 & HD187216 & 3950 & 0.75 & -1.70 & b & 4\\
  731 & HD232078 & 3965 & 0.64 & -1.63 & c & 4\\
  736 & HD187111 & 4473 & 1.13 & -1.71 & c & 7\\
  740 & HD188041 & 9506 & 3.91 & 1.00 & b & 1\\
  744 & HD188727 & 5685 & 1.60 & 0.00 & a & 4\\
  751 & HD190603 & 19500 & 2.36 & 0.07 & b & 3\\
  758 & HD192577 & 4126 & 1.05 & -0.07 & c & 6\\
  774 & HD197964 & 4762 & 2.93 & 0.15 & b & 7\\
  780 & HD199478 & 11200 & 1.90 & 0.00 & d & 3, 6\\
  797 & HD204075 & 5397 & 1.48 & -0.14 & b & 7\\
  814 & HD206778 & 4196 & 0.67 & 0.04 & c & 7\\
  815 & HD207260 & 9911 & 1.57 & 0.27 & b & 5, 6, 7\\
  819 & HD207673 & 10482 & 1.87 & 0.16 & b & 5, 6, 7\\
  820 & HD208501 & 16477 & 2.80 & 0.05 & b & 6, 7\\
  834 & HD213307 & 7800 & 2.00 & 0.20 & b & 6, 7\\
  836 & HD213470 & 8943 & 1.36 & 0.11 & b & 5, 6, 7\\
  838 & G156-031 & 2805 & 5.13 & -0.04 & c & 3\\
  842 & BD+394926 & 7261 & 0.85 & -2.52 & a & 7\\
  853 & HD216916 & 21500 & 3.75 & -0.12 & b & 1, 6\\
  865 & HD219116 & 4790 & 1.79 & -0.79 & b & 7\\
  873 & HD219978 & 3910 & 0.19 & 0.18 & c & 5\\
  884 & BD+195116B & 3259 & 4.82 & -0.26 & c & 3\\
  890 & G171-010 & 2894 & 5.04 & 0.09 & c & 3\\
  892 & HD223385 & 10023 & 1.59 & 0.29 & b & 3\\
  915 & M79\_153 & 4269 & 0.77 & -1.60 & c & 1\\
  916 & M79\_160 & 4264 & 0.67 & -1.60 & c & 1, 7\\
  927 & M5 II-51 & 5718 & 1.98 & -1.29 & c & 6, 7\\
  929 & M5\_II-53 & 9441 & 2.43 & -1.11 & a & 1, 6\\
  931 & M5\_IV-86 & 5576 & 2.44 & -1.11 & a & 1, 6\\
  932 & M5\_IV-87 & 5965 & 3.80 & -1.11 & b & 1, 6\\
  950 & M71\_KC-147 & 4819 & 2.57 & -0.84 & b & 1, 7\\
  957 & M71\_1-63 & 4706 & 1.62 & -0.78 & c & 6, 7\\
  961 & M71\_1-09 & 4784 & 2.01 & -0.78 & c & 6, 7\\
  963 & M71\_1-34 & 5114 & 2.35 & -0.84 & b & 1, 7\\
  967 & M71\_1-77 & 4261 & 1.87 & -0.22 & c & 7\\
  970 & M71\_X & 3980 & 1.43 & -0.84 & b & 1\\

\end{longtable}